\documentclass{statsoc}

\usepackage[a4paper]{geometry}
\usepackage{graphicx}
\usepackage[textwidth=8em,textsize=small]{todonotes}
\usepackage{natbib}
\usepackage{url}

\usepackage{amsthm,amsmath,amssymb}
\usepackage{xcolor}
\usepackage{float}
\usepackage{subcaption}

\usepackage{setspace}
\title[\vspace*{-1cm}
Sparse microbiome clustering]{Sparse tree-based clustering of microbiome data 
	to characterize microbiome heterogeneity in pancreatic cancer
{\large \normalfont{Accepted to \textit{Journal of the Royal Statistical Society: Series C}}} \vspace*{-0.5cm}
	}

\author[Y.\ Shi et al.]{Yushu Shi}
\address{Department of Statistics, University of Missouri, Columbia, Columbia, MO, USA}
\author{Liangliang Zhang}
\address{\color{black}Department of Population and Quantitative Health Sciences, Case Western Reserve University, Cleveland, OH, USA}
\author{Kim-Anh Do}
\address{Department of Biostatistics, The University of Texas MD Anderson Cancer Center, Houston, TX, USA}
\author{Robert Jenq}
\address{Department of Genomic Medicine, The University of Texas MD Anderson Cancer Center, Houston, TX, USA}
\author[Y.\ Shi et al.]{Christine B.\ Peterson}
\address{Department of Biostatistics, The University of Texas MD Anderson Cancer Center, Houston, TX, USA}
\email{cbpeterson@mdanderson.org}
\begin{document}
	
	\begin{abstract}
		There is a keen interest in characterizing variation in the microbiome across cancer patients, given increasing evidence of its important role in determining treatment outcomes.
		Here our goal is to discover subgroups of patients with similar microbiome profiles.
		We propose a novel unsupervised clustering approach in the Bayesian framework that innovates over existing model-based clustering approaches, such as the Dirichlet multinomial mixture model, in three key respects: we incorporate feature selection, learn the appropriate number of clusters from the data, and  integrate information on the  tree structure relating the observed features.
		We compare the performance of our proposed method to existing methods on simulated data designed to mimic real microbiome data. We then illustrate results obtained for our motivating data set, a clinical study aimed at characterizing the tumor microbiome of pancreatic cancer patients.
	\end{abstract}
	
	\setstretch{1.7}
	
	\section{Introduction}
	Pancreatic cancer remains one of the hardest cancers to treat, with a 5 year survival rate of only 10\% \citep{Mizrahi2020}.
	In recent years, increasing evidence has shown that the microbiome plays an important role in shaping both pancreatic cancer risk and treatment outcomes \citep{Fan2018, Wei2019}. We are motivated by a recent multi-site study examining microbiome composition in pancreatic cancer patients \citep{RiquZhan19}, and propose a novel Bayesian clustering approach to discover groups of subjects with similar microbiome profiles from this data. Our method offers key advantages over existing methods for clustering of microbiome data: 1) it identifies specific features that are relevant, 2) it allows the appropriate number of clusters to be learned from the data, and 3) it integrates information encoded in the phylogenetic tree structure relating the observed features. Importantly, our case study findings have implications for the development of future microbiome interventions aimed at improving pancreatic cancer outcomes.
	
	We now briefly review microbiome data collection and 
	existing statistical methods for clustering of microbiome samples.
	In the past decade, advances in next-generation sequencing have enabled researchers to cheaply and comprehensively analyze microbial communities across various sites in the human body \citep{KnigCall17}.
	To date, most large scale microbiome studies rely on sequencing of the 16S ribosomal RNA marker gene. The observed sequences are grouped based on similarity into operational taxonomic units (OTUs) using specially-designed processing pipelines.
	These software pipelines also enable assignment to known taxonomic classifications based on similarity to sequences in a reference database, and the construction of a phylogenetic tree describing evolutionary relationships among the OTUs. After these preprocessing steps, the observed data from a microbiome study consist of an $N \times d$ matrix of counts, where $N$ is the number of observations and $d$ is the number of features (OTUs).
	
	Microbiome data pose a number of challenges to downstream statistical analysis. First, the data are high-dimensional, with thousands of features quantified in each sample. This represents a challenge both in terms of identifying relevant features and ensuring that methods are computationally scalable. Second, the data are noisy, with wide variability in microbial abundances across subjects. This challenge necessitates the use of sparse modeling approaches to focus on the most informative features. Next, the data are compositional, which means that the counts within each sample have a fixed sum constraint, and can only be interpreted on a relative scale.
	Finally, there is a question of how to best leverage the information encoded in microbiome tree structures. Taxonomic
	trees reflect the traditional classification of microorganisms into a hierarchy with the levels kingdom, phylum, class, order, family, genus, and species. Phylogenetic trees reflect evolutionary history, with branch points corresponding to events that gave rise to differences in the genomic sequences. Phylogenetic trees are potentially useful in analysis because they encode rich information on sequence similarity, which drives phenotypic and functional similarity. Without accounting for phylogenetic tree structures, genetically distinct but functionally similar OTUs are treated as independent features, which can make inference more challenging.

	Both machine learning and model-based approaches have been proposed for clustering of microbiome samples. Most machine learning methods require first determining pairwise distances between samples, using metrics such as Bray-Curtis dissimilarity \citep{BrayCurt57}, unweighted UniFrac distance \citep{LozuKnig05}, or weighted UniFrac distance \citep{LozuHama07}. The pairwise distances are then taken as input to an algorithm such as K-means \citep{MacQ67} or PAM \citep{KaufRous08} to identify an appropriate partition of the samples into groups.
	
	As an alternative to distance-based clustering methods,
	\citet{HolmHarr12} proposed the model-based  Dirichlet multinomial mixtures (DMM) approach. The DMM method can be applied to a few hundred variables, and may therefore be used to analyze counts grouped by genus. However, it does not perform feature selection, and therefore does not scale well to the thousands of features in data defined at the finer level of OTUs.
	Beyond the limited scalability of DMM, the machine-learning and model-based clustering methods listed above share a common limitation: the number of clusters needs to be either taken as known, or chosen based on post-hoc criteria. For example, to apply K-means or PAM when the number of clusters is not known \textit{a priori}, one can calculate the silhouette width  for a range of possible cluster numbers, and adopt the cluster number which achieves the highest value \citep{Rous87}.
	Other commonly-used methods for determining the number of clusters, such as the gap statistic, can only be applied to Euclidean distances, and are therefore not suitable for microbiome data.
	For the DMM model, \cite{HolmHarr12}  proposed determining the number of clusters by calculating model evidence via the Laplace approximation over a range of possible values and choosing the maximum.

	In this paper, we adopt a mixture of finite mixtures (MFM) model, which puts a prior on the number of clusters. To efficiently handle the high dimensionality of  microbiome data, in the choice of mixture component distributions, we exploit the conjugacy between the Dirichlet and Dirichlet tree distributions and the multinomial distribution. 
	Also, we hypothesize that microbiome datasets often contain ``noise" OTUs that mask signal from informative OTUs and hinder successful clustering. We therefore select informative features during the clustering processes.
	We refer to our proposed modeling approaches based on the Dirichlet and Dirichlet tree distributions as MFM 
	Dirichlet multinomial (MFMDM) and  MFM Dirichlet tree multinomial (MFMDTM), respectively.
	
	
	To illustrate the real-world utility of our proposed microbiome clustering approaches,
	we apply these methods to characterize tumor microbiome heterogeneity in a study of 
	pancreatic ductal adenocarcinoma (PDAC) patients \citep{RiquZhan19}.
	This dataset consists of 68 tumor samples from PDAC patients collected at two hospitals, where each patient contributed one tumor sample.  Of the patients included in the study, 36 survived longer than 5 years after surgery (long-term survivors), while the rest died within 5 years after surgery (short-term survivors). \cite{RiquZhan19} reported that higher microbiome diversity was associated with better outcomes in these patients and identified a microbiome signature based on differentially abundant microbiome features between the long- and short-term survivors. However, their results indicated that long-term survivors had more consistent tumor microbiome profiles than short-term survivors, which suggests that two-group comparisons might not appropriately capture the heterogeneity among short term survivors. Here, we provide further insight into this dataset, through the use of unsupervised clustering to characterize naturally occurring sample groups and identify relevant features.
	
	The paper is structured as follows. In Section \ref{sec:method}, we present the formulation of the clustering and feature selection methods. In Section \ref{sec:mcmc}, we describe the implementation of the proposed method. In Section \ref{sec:sim}, we provide an extensive comparison of the performance of our proposed method to competing methods on simulated data. Finally, we provide results from our proposed method on the motivating pancreatic cancer dataset in Section \ref{sec:application}, and 
	we conclude with a brief discussion in Section  \ref{sec:discussion}.
	
	\section{Formulation of the MFMDM and MFMDTM methods} \label{sec:method}
	In a model-based clustering approach, samples are assumed to come from various subpopulations. The observations within each subpopulation, or mixture component, are assumed to follow a parametric distribution with parameters specific to that mixture component. In the current work, we adopt this model-based framework. In this section, we first describe the likelihood of the data within each mixture component. We consider both the basic Dirichlet multinomial distribution, which underlies the existing DMM approach, as well as the Dirichlet tree multinomial, which allows us to integrate information on the taxonomic or phylogenetic tree structure. We then
	develop the Bayesian prior formulation that allows us to achieve both feature selection and clustering into a flexible number of mixture components, using an MFM model. We therefore refer to our method based on the Dirichlet multinomial as the MFMDM method and the approach based on the Dirichlet tree multinomial as the MFMDTM method.
	
	\subsection{Dirichlet multinomial and Dirichlet tree multinomial distributions}
	
	The simplest form of a microbiome dataset involves an $N \times d$ matrix $\mathbf{Y}$, where $N$ is the number of observations, and $d$ is the number of features. The entry $y_{i,j}$ represents the number of counts observed for the $j$th feature in the $i$th observation. Under the simplest model for count data, the vector of counts for the $i$th observation could be modeled as $\mathbf{y}_i \sim \mathrm{Multinomial}(q_1, q_2, \dots, q_d).$
	In practice, microbiome data tend to have higher variation than captured by the multinomial distribution, which assumes fixed proportions $\boldsymbol{q} = q_1, q_2, \dots, q_d$. For this reason, it is helpful to treat the proportions $\boldsymbol{q}$ as random variables.
	If we assume $\boldsymbol{q}  \sim \mathrm{Dirichlet}(\alpha,\alpha,\dots,\alpha)$, we can integrate out $\boldsymbol{q}$ to obtain the Dirichlet multinomial distribution: 
	
	\begin{equation}
		P(\mathbf{y}_i|\alpha)=\frac{y_{i,.}!}{y_{i,1}!y_{i,2}!\dots y_{i,d}!}\frac{\Gamma(d\alpha)}{\Gamma(\alpha)^d}{\color{black}\frac{\Gamma(y_{i,1}+\alpha)\Gamma(y_{i,2}+\alpha)\dots \Gamma(y_{i,d}+\alpha)}{\Gamma(y_{i,.}+d\alpha)}},
	\end{equation}
	\noindent
	{\color{black}where $y_{i,.}$ is the summation of counts for the $i$th observation.}  The Dirichlet multinomial distribution is frequently used to model microbiome data because it better captures overdispersion than the simple multinomial likelihood \citep{HolmHarr12, ChenLi13}.  In particular, several regression models for microbiome data have been proposed that rely on the Dirichlet multinomial distribution \citep{ChenLi13, Wadsworth2017}. 
	
	
	
\begin{figure}[ht]
		\centering
		\includegraphics*[width=0.37\linewidth]{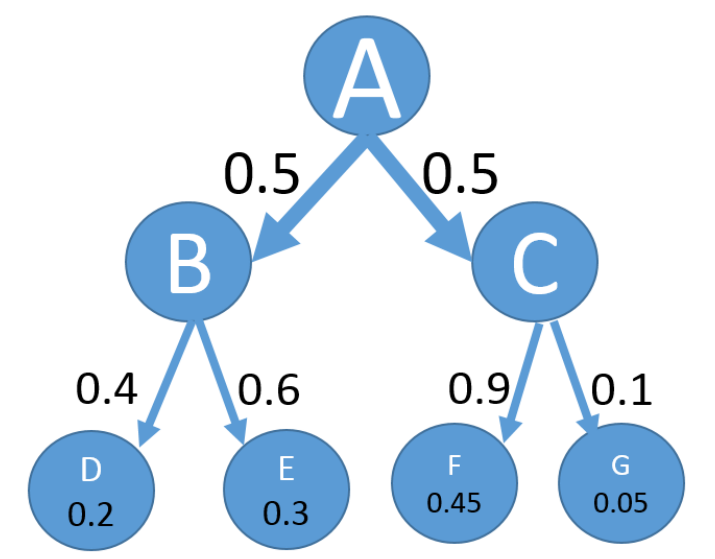}
		\caption{An example of the Dirichlet tree distribution.}
		\label{Dirichlettree}
	\end{figure}

	Since microorganisms that are closely related often have similar functions, we can potentially better capture microbiome variation among samples by recognizing the structure among the observed features, which can be described using a taxonomic or phylogenetic tree structure. An extension of the Dirichlet distribution, the Dirichlet tree distribution \citep{Denn91}, can help us achieve this goal while maintaining conjugacy to the multinomial distribution. Figure \ref{Dirichlettree} gives an example of the Dirichlet tree distribution, where the probability of a count being allocated to a leaf is the product of the branch probabilities leading to that leaf. {\color{black} When applied to model microbiome data, each tree node represents a taxonomy unit, and nodes closer to the top correspond to more generic or broad groupings. In the toy example shown in Figure \ref{Dirichlettree}, node A (the root node) could represent the kingdom Bacteria. We could then classify a random sequence as belonging to either phylum B or phylum C, each with 50\% probability. If the sequence corresponds to phylum B, it could be further classified as belonging to class D with 40\% probability or class E with 60\% probability. Finer taxonomic or phylogenetic classification could be reflected by a tree with additional levels. }
	While the  Dirichlet multinomial distribution remains more commonly used to model microbiome data, the Dirichlet tree multinomial distribution has also been proposed in settings such as differential abundance testing and regression modeling  \citep{WangZhao17,TangNico17, TangLi18}. To the best of our knowledge, we are the first to propose using the Dirichlet tree multinomial in the context of clustering.
	
	
	
	We now describe some useful properties of the Dirichlet tree distribution in more detail. In particular, the Dirichlet tree distribution is conjugate to the multinomial distribution.
	To demonstrate this property, we can represent a multinomial sample as the outcome of a finite stochastic process. {\color{black} The probability of a count assigned to tree node $j$ being further classified to its child node $k$ is $b_{jk}$.} Given the tree structure $T$ and the branch probabilities $B$ between nodes and their children, the probability of a single count $x$ can be written as $P(x|B,T)=\prod_{j \in J} \prod_{k \in K_j} b_{jk}^{\delta_{jk}(x)}$, where $J$ is the set of parent nodes, {\color{black} $K_j$ is the set of child nodes directly descending from node $j$ (not including any grandchildren or further descendants), and $\delta_{jk}(x)$ is the indicator of whether the count $x$ passes through the branch linking node $j$ and node $k$. For the branch probabilities $b_{jk} \in B,$ $b_{jk}>0$ for $j\in J,$ and $\sum_{k\in K_j}b_{jk}=1$. In this paper, we use $\mathbf{y}_i$ to denote the count vector for the $i$th observation, and $\mathbf{Y}$ to denote the corresponding count matrix for all the observations.
		Given the tree structure $T$ and the vector $\mathbf{y}_i$, one can compute the $|J|\times (|J|+d)$ matrix $X_i$ for the $i$th observation, where $|J|$ is the number of internal nodes of the tree and $d$ is the number of OTUs, i.e., leaf nodes of the tree. The matrix $X_i$ describes the number of counts traveling from parent nodes to children nodes. The collection of these matrices is denoted as $\mathbf{X}.$} The probability of the matrix of counts for the $i$th sample is:
	\begin{equation}
		P(X_i|B,T)=\prod_{j \in J}\frac{n_{j,.}(X_i)!}{n_{j,1}(X_i)! n_{j,2}(X_i)! \dots n_{j,|K_j|}(X_i)!}\prod_{k \in K_j} b_{jk}^{n_{j,k}(X_i)},
	\end{equation}
	where $n_{j,.}(X_i)$ is the sum of the counts for all the nodes descending from node $j$ for the $i$th observation,  $n_{j,k}(X_i)$ is the number of counts descending from node $j$ to node $k$ for the $i$th observation, and $|K_j|$ is the number of children of node $j$.

	A Dirichlet tree distribution can be expressed as a product of Dirichlet densities, $P(B)=\prod_{j \in J} p(\mathbf{b}_j)$. If each $\mathbf{b}_j$ is given a Dirichlet prior, $P(\mathbf{b}_j|\alpha)=\mathrm{Dirichlet}(\alpha,\alpha,\dots,\alpha)$, one can obtain a Dirichlet tree multinomial distribution for observation $i$, after integrating out $B$: \hspace*{-1cm} \begin{equation} 
		\begin{split}
			P(X_i|T)=&\prod_{j \in J}\left[\frac{n_{j,.}(X_i)!}{n_{j,1}(X_i)!n_{j,2}(X_i)!\dots n_{j,|K_j|}(X_i)!} \cdot \right.\\
			&\left.\prod_{k \in K_j} \frac{\Gamma(|K_j| \alpha)}{\{\Gamma(\alpha)\}^{|K_j|}}\frac{\{\prod_{k \in K_j}\Gamma(\alpha+n_{j,k}(X_i))\}}{\Gamma(|K_j| \alpha+\sum_{k \in K_j} n_{j,k}(X_i))}\right].
		\end{split}
	\end{equation}
	\subsection{Feature selection} \label{sec:sel}
	In the current work, feature selection is achieved by identifying a parsimonious set of OTUs or tree nodes that exhibit differential abundance across groups. The rest of the features are not informative regarding the sample clustering, meaning that the corresponding parameters describing their relative abundance  are not cluster-specific. To obtain a sparse model that selects informative OTUs or tree nodes, we introduce a binary vector $\boldsymbol{\gamma}$, whose entries indicate whether the corresponding features are useful in discriminating between sample groups, where $\gamma_j=1$ if $j$ is an informative feature and $0$ otherwise.
	For simplicity, we use a Bernoulli prior for $\boldsymbol{\gamma}$: $p(\boldsymbol{\gamma}) \propto \prod_{j=1}^d w^{\gamma_j}(1-w)^{1-\gamma_j}$ \citep{GeorMcCu93,MadiYork95}, where $d$ is the total number of features, and the prior probability of a feature being informative is $w$. {\color{black}
		In both the MFMDM and MFMDTM models, we select features that are relevant to clustering. We consider features not contributing to clustering as ``noise" features, and features facilitating clustering as informative features.}
	
	\subsubsection{Feature selection in the MFMDM model}
	In the proposed parsimonious model, the parameters describing the $i$th observation can be divided into a common part shared across all observations $p_1, p_2, ..., p_{d-d_{\boldsymbol{\gamma}}}$, and a part specific to the cluster the $i$th observation belongs to, $w_{c_i}$, $q_{{c_i},1}, q_{{c_i},2}, ..., q_{{c_i},d_{\boldsymbol{\gamma}}}$, where $c_i$ denotes the index of the cluster to which observation $i$ belongs, $d_{\boldsymbol{\gamma}}$ denotes the number of informative features, and $w_{c_i}$ controls the proportion of counts belonging to the {\color{black} informative} features in cluster ${c_i}$. Each observation is modeled as a multinomial variable with parameters {\color{black} $\mathbf{y}_i \sim \mathrm{Multinomial}((1-w_{c_i}) p_1,(1-w_{c_i}) p_2,$ $\dots,$ $(1-w_{c_i}) p_{d-d_{\boldsymbol{\gamma}}} , w_{c_i} q_{{c_i},1}, w_{c_i} q_{{c_i},2},\dots,$ $w_{c_i} q_{{c_i},d_{\boldsymbol{\gamma}}})$. In our notation, we reorder the elements of $\mathbf{y}_i$  such that the first $d-d_{\boldsymbol{\gamma}}$ elements are noisy and the remaining are informative, i.e., $\mathbf{y}_i=(\mathbf{y}_{ni},\mathbf{y}_{ei}).$}
	
	The length of the vectors $\boldsymbol{p}=(p_1, p_2, ..., p_{d-d_{\boldsymbol{\gamma}}})$ and $\boldsymbol{q}_{c_i}=(q_{{c_i},1}, q_{{c_i},2}, ..., q_{{c_i},d_{\boldsymbol{\gamma}}})$ will change with the number of OTUs selected as informative, but both $\boldsymbol{p}$ and $\boldsymbol{q}_{c_i}$ sum to $1$, thus guaranteeing the mean vector of the multinomial distribution will sum to $1$ as well. Finally, the likelihood for observation $i$ can be written as: 
	\begin{equation}
		\begin{split}
			P(\mathbf{y}_i|c_i,\boldsymbol{p},\boldsymbol{q}_{c_i},w_{c_i},\boldsymbol{\gamma})=&\frac{\textcolor{black}{y_{i,.}!}}{y_{n,i,1}!\dots y_{n,i,d-d_{\boldsymbol{\gamma}}}! y_{e,i,1}!\dots y_{e,i,d_{\boldsymbol{\gamma}}}!}\\
			& {\color{black}\times\{(1-w_{c_i}) p_1\}^{y_{n,i,1}}\dots \{(1-w_{c_i}) p_{d-d_{\boldsymbol{\gamma}}}\}^{y_{n,i,d-d_{\boldsymbol{\gamma}}}}}\\
			&{\color{black} \times (w_{c_i} q_{{c_i},1})^{y_{e,i,1}}\dots (w_{c_i} q_{{c_i},d_{\boldsymbol{\gamma}}})^{y_{e,i,d_{\boldsymbol{\gamma}}}}.}
		\end{split}
	\end{equation}
	
	For observation $i$, $y_{n,i,j}$ is the number of counts of ``noise" OTU $j$, $y_{e,i,l}$ is the number of counts of the informative OTU $l$, and $y_{i,.}=y_{n,i,.} + \ y_{e,i,.}$ is the total number of counts. The likelihood for all the observations is $P(\mathbf{Y}|\boldsymbol{p},\boldsymbol{q},\mathbf{w},\boldsymbol{\gamma},\mathbf{c}) = \prod_{c\in C}\prod_{c_i=c} P(\mathbf{y}_i|c_i,\boldsymbol{p},\boldsymbol{q}_{c_i},w_{c_i},\boldsymbol{\gamma})$, where $C$ denotes the set of distinct cluster indices.

	%

	%
	
	In order to obtain a more tractable posterior distribution, we use the simplest conjugate priors by setting:
	\begin{equation}
		\begin{split}
			p_1, p_2, \dots, p_{d-d_{\boldsymbol{\gamma}}} &\sim \mathrm{Dirichlet}( \alpha , \alpha , \dots, \alpha )\\  
			q_{c,1}, q_{c,2}, \dots, q_{c,d_{\boldsymbol{\gamma}}} &\sim \mathrm{Dirichlet}( \alpha , \alpha , \dots , \alpha)\\
			w_c &\sim \mathrm{Beta}( \beta_1 , \beta_2).
		\end{split}
	\end{equation}
	For simplicity and objectivity, we assume all the parameters in the Dirichlet distribution are equal. Since our primary interest is in learning the cluster assignments and identifying discriminating features, we integrate out the remaining parameters to speed up computation. The resulting marginal likelihood is:
	
	\begin{equation}
		\begin{split}
			P(\mathbf{Y}|\boldsymbol{\gamma},\mathbf{c})=&\prod_{i=1}^{N}\frac{y_{i,.}!}{y_{n,i,1}!\dots y_{n,i,d-d_{\boldsymbol{\gamma}}}! y_{e,i,1}! \dots y_{e,i,d_{\boldsymbol{\gamma}}}\textcolor{black}{!}}  \frac{\Gamma((d-d_{\boldsymbol{\gamma}})\alpha)}{\{\Gamma(\alpha)\}^{d-d_{\boldsymbol{\gamma}}}} \\
			&\times \frac{\Gamma(\sum_{i=1}^N y_{n,i,1}+\alpha)\dots\Gamma(\sum_{i=1}^N y_{n,i,d-d_{\boldsymbol{\gamma}}}+\alpha)}{\Gamma(y_{n,.,.}+(d-d_{\boldsymbol{\gamma}})\alpha)}\\
			&\times \prod_{c\in C} \frac{\Gamma(\beta_1+\beta_2)}{\Gamma(\beta_1)\Gamma(\beta_2)} \frac{ {\color{black}\Gamma(\beta_1+\sum_{c_i=c} y_{e,i,.})\Gamma(\beta_2+\sum_{c_i=c} y_{n,i,.})}}{\Gamma(\beta_1+\beta_2+\sum_{c_i=c} y_{i,.})}\\
			&\times \frac{\Gamma(d_{\boldsymbol{\gamma}}\alpha)}{\{\Gamma(\alpha)\}^{d_{\boldsymbol{\gamma}}}}\frac{\Gamma(\sum_{c_i=c} y_{e,i,1}+\alpha) \dots \Gamma(\sum_{c_i=c} y_{e,i,d_{\boldsymbol{\gamma}}}+\alpha)}{\Gamma(\sum_{c_i=c} y_{e,i,.}+d_{\boldsymbol{\gamma}}\alpha)}.
		\end{split}
	\end{equation}
	
	\subsubsection{Feature selection in the MFMDTM model}
	{\color{black} In the MFMDTM model, ``noise" nodes' allocation probabilities are the same across clusters. In contrast, informative nodes have cluster-specific allocation probabilities. We denote the set of informative parent nodes $J^e$, and the set of ``noise" parent nodes $J^n$, with $J = J^e \cup J^n$.} The MFMDTM method can therefore highlight the level within the tree structure where the sample composition begins to differentiate between clusters. This could be potentially advantageous when groups of microorganisms close together in the tree, which are typically functionally related, are up- or down-regulated in a coordinated manner.
	
	
	
	We denote all the parameters associated with  observation $i$ as $B_i$, with the part belonging to the ``noise" nodes as $\mathbf{b}_{j}$, $j \in J^n$, and the part shared only by the observations in the same cluster $\mathbf{b}_{j}(c_i)$, $j \in J^e$, $c_i \in \{1,2,\dots,C\}$. The likelihood of the vector of counts for the $i$th observation $X_i$ is:  \begin{equation}
		\begin{split}
			P(X_i|B,T,\boldsymbol{\gamma},c_i)=&\prod_j\frac{n_{j,.}(X_i)!}{n_{j,1}(X_i)!n_{j,2}(X_i)!\dots n_{j,|K_j|}(X_i)!}\\
			&\prod_{j \in J^n} \prod_{k \in K_j} b_{jk}^{n_{j,k}(X_i)}\prod_{j \in J^e} \prod_{k \in K_j} b_{jk} (c_i)^{ n_{j,k}(X_i)}.
		\end{split}
	\end{equation}
	The likelihood for the matrix of counts across all observations $\mathbf{X}$ is then:\begin{equation}
		\begin{split}
			P(\mathbf{X}|B,T,\boldsymbol{\gamma},C)=&\prod_i \left\{\prod_j\frac{n_{j,.}(X_i)!}{n_{j,1}(X_i)!n_{j,2}(X_i)!\dots n_{j,|K_j|}(X_i)!} \right\}\\
			& \prod_{j \in J^n} \prod_{k \in K_j} b_{jk}^{n_{j,k}(\mathbf{X})}\prod_{c \in C} \prod_{j \in J^e}\prod_{k \in K_j} b_{jk} (c)^{ n_{j,k}(\mathbf{X}^c)}. 
		\end{split}
	\end{equation}
	{\color{black} Here $n_{j,k}(\mathbf{X})$ is the total number of counts descending from node $j$ to the $k$th child node for all the observations, $n_{j,k}(\mathbf{X}^c)$ is the total number of counts descending from node $j$ to the $k$th child node for the observations in cluster $c$, $b_{j,k}$ is the probability of assigning count from non-informative node $j$ to its $k$th descending node, and $b_{j,k}(c)$ is the probability of assigning count from informative node $j$ to its $k$th descending node in cluster $c$.} 
	
	Again, for simplicity of computation, we use the strategy of integrating out parameters. We consider the simplest conjugate prior $\mathbf{b}_{j}|\boldsymbol{\alpha} \sim \mathrm{Dirichlet}(\alpha, \alpha, \ldots, \alpha)$, and integrate out $B$. The marginal likelihood is:
	\begin{equation}
		\begin{split}
			P(\mathbf{X}|T,\boldsymbol{\gamma},\mathbf{c})=& \prod_i \left\{\prod_j\frac{n_{j,.}(X_i)!}{n_{j,1}(X_i)!n_{j,2}(X_i)!\dots n_{j,|K_j|}(X_i)!} \right\}\\
			&\prod_{j \in J^n} \frac{\Gamma(|K_j| \alpha)}{\{\Gamma(\alpha)\}^{|K_j|}}
			\frac{(\prod_{k \in K_j}\Gamma(\alpha+n_{j,k}(\mathbf{X})))}{\Gamma(|K_j| \alpha+\sum_{k \in K_j} n_{j,k}(\mathbf{X}))}\\
			&\prod_{c \in C}\prod_{j \in J^e} \frac{\Gamma(|K_j| \alpha)}{\{\Gamma(\alpha)\}^{|K_j|}}
			\frac{(\prod_{k \in K_j}\Gamma(\alpha+n_{j,k}(\mathbf{X}^c)))}{\Gamma(|K_j| \alpha+\sum_{k \in K_j} n_{j,k}(\mathbf{X}^c))}.
		\end{split}
	\end{equation}

	\subsection{Mixture of finite mixtures}
	
	When the number of clusters is not prespecified, one can treat the data as arising from an infinite mixture of distributions, as in a Bayesian nonparametric approach such as the Dirichlet process mixture model. However, it has been shown that in a Dirichlet process mixture model, the number of clusters will grow with the number of observations \citep{MillHarr14}. Alternatively, one can treat the data as arising from a finite mixture of a given distribution and use methods such as reversible jump Markov chain Monte Carlo (RJMCMC)  to learn the number of clusters from the data \citep{RichGree97, TadeSha05}. \citet{MillHarr18} proved that the MFM can consistently estimate the number of clusters, while the number of clusters in Dirichlet process mixtures will increase with sample size. They also demonstrated that efficient algorithms designed for the Dirichlet process mixture context can be applied for MFMs, avoiding the need for RJMCMC, which is notorious for being difficult to implement and computationally intensive. For this reason, we adopt the MFM in our modeling approach.
	The hierarchical formulation of our proposed model using the MFM framework is:
	
	{\color{black}
		\begin{align*}
			M&\sim p_m, &\text{where }p_m\text{ is a p.m.f on }\{1,2,\dots\},\\
			(\pi_1,\dots,\pi_M)|M=m &\sim\mathrm{Dirichlet}_m(\eta,\dots,\eta),& \\
			c_1,\dots,c_N |\boldsymbol{\pi} & \sim \mathrm{Categorical}(\boldsymbol{\pi}),&\\
			\boldsymbol{\gamma} &\sim \prod_{j=1}^d \text{Bernoulli}(w),&\\
			\boldsymbol{\theta} \sim G_{00},& \quad\boldsymbol{\theta}_1,\dots,\boldsymbol{\theta}_M \sim G_0, &\\
			\text{MFMDM model: }&\\
			G_0|\boldsymbol{\gamma}&= \text{Dirichlet}(\boldsymbol{\alpha}_e),&\\
			G_{00}|\boldsymbol{\gamma}&= \text{Dirichlet}(\boldsymbol{\alpha}_n),\\
			\text{MFMDTM model:}&\\
			G_0|\boldsymbol{\gamma}&= \prod_{\gamma_k=1} \text{Dirichlet}(\boldsymbol{\alpha}_k),&\\
			G_{00}|\boldsymbol{\gamma}& = \prod_{\gamma_l=0} \text{Dirichlet}(\boldsymbol{\alpha}_l),\\
			\mathbf{X}_i |\boldsymbol{\theta}_{1:M},c_{1:N}, \boldsymbol{\theta}&\sim F(\boldsymbol{\theta}_{c_i},\boldsymbol{\theta}),& \text{ for } i=1,\dots,N .
		\end{align*}
	}
	Here, $M$ is the underlying number of components in the population. The vector $\boldsymbol\pi=(\pi_1,\dots,\pi_M)$ is the probability of a random sample belonging to a component. $c_1,\dots,c_N$ are the indices of the cluster to which each sample belongs. The vector $\boldsymbol{\gamma}$ is the feature selection indicator introduced in Section \ref{sec:sel}. 
	$\boldsymbol{\theta}$ includes the parameters of the ``noise" features, which are shared across all clusters, while $\boldsymbol{\theta}_i$ is the set of corresponding parameters for informative features of cluster $i$. \textcolor{black}{For the MFMDM model, the distribution for $\boldsymbol{\theta}$ --- $G_{00}$ and the base distribution for $\boldsymbol{\theta}_m$s --- $G_{0}$, are two Dirichlet distributions, and the length of the parameter vectors correspond to the number of $0$ and $1$ elements in $\boldsymbol{\gamma}$ respectively. For the MFMDTM model, $G_{00}$ and $G_{0}$ are products of Dirichlet distributions, where the length of vector $\boldsymbol{\alpha}_j$ depends on the number of children nodes for node $j$. Without knowledge suggesting that informative features allocate counts more evenly (or less evenly) to their children than noisy features, we adopt the same prior for noisy features and informative features, with the purpose of facilitating a smooth transition between noisy and informative in the feature selection process. For simplicity, we let the parameter vectors of the Dirichlet distributions, including $\boldsymbol{\alpha}_e$, $\boldsymbol{\alpha}_n$ in the MFMDM model and $\boldsymbol{\alpha}_j$ in the MFMDTM model be vectors of $1$s for both $G_0$ and $G_{00}$.
		$F$ is the mixing kernel introduced in the above subsections, i.e., the multinomial distribution.} 
	
	Similar to the Dirichlet process mixture model, the underlying discrete measure of the MFM model has a P\'{o}lya urn scheme representation, which enables sampling of the parameters for each observation sequentially. This close parallelism makes most sampling algorithms designed for the Dirichlet process mixture model directly applicable \citep{MillHarr18}. The parameter set for observation $i$, $\boldsymbol{\theta}_i$, will either take the identical value of an existing parameter, or a newly generated value from the base distribution $G_0$ with the following probability: $
	\boldsymbol{\theta}_i|\boldsymbol{\theta}_{-i} \sim \sum_{c \in C} (n_{c,-i}+\eta)\delta(\boldsymbol{\theta}_c^*)+ \frac{V_N(|C|+1)}{V_N(|C|)}\eta G_0,$
	where $\boldsymbol{\theta}_{-i}$ are the parameters for all the observations except for the $i$th observation; $\boldsymbol{\theta}_c^*$s are the distinct values of $\boldsymbol{\theta}_{-i}$, and $n_{c,-i}$s are the corresponding numbers of observations having the parameter $\boldsymbol{\theta}_c^*$, except for the $i$th observation. The function $V_N(R)$ is defined as $
	V_N(R)=\sum_{m=R}^{\infty}\frac{\Gamma(m+1) \Gamma(\eta m)}{\Gamma(m-R+1) \Gamma(\eta m+N)}p_M(m).$
	This completes the specification of the MFMDM and MFMDTM models.
	
	\section{Method implementation} \label{sec:mcmc}
	Obtaining a sample from the posterior distribution of either model requires the use of Markov chain Monte Carlo (MCMC).
	In each MCMC iteration, we first select OTUs or tree nodes by updating the latent indicator  $\boldsymbol{\gamma}$ given the current cluster assignments, then fix  $\boldsymbol{\gamma}$ and apply the split-and-merge algorithm to assign observations into clusters. Here we provide a high-level description of the algorithm, with additional details provided in the Supplementary Material.
	
	
	\subsection{Updates to feature selection indicators}
	The latent selection indicator $\boldsymbol{\gamma}$ is updated by repeating the following Metropolis step $t$ times, where $t=20$ following the suggestion of \citet{KimTade06}. A new candidate $\boldsymbol{\gamma}^{new}$ is generated by randomly choosing one of the two transition moves:
	\begin{enumerate}
		\item \textbf{add/delete} by randomly picking one of the $d$ indices in $\boldsymbol{\gamma}^{old}$ and changing its value (from $0$ to $1$ or from $1$ to $0$);
		\item \textbf{swap} by randomly drawing a 0 and a 1 in $\boldsymbol{\gamma}^{old}$ and switching their values.
	\end{enumerate} The new candidate is accepted with probability $
	\min\left\{1,\frac{f(\boldsymbol{\gamma}^{new}|\mathbf{X},\mathbf{c})}{f(\boldsymbol{\gamma}^{old}|\mathbf{X},\mathbf{c})}\right\}
	$, where $\boldsymbol{c}$ is the cluster assignment vector. As $f(\boldsymbol{\gamma}|\mathbf{X},\boldsymbol{c})\propto f(\mathbf{X}|\boldsymbol{\gamma},\boldsymbol{c}) \mathrm{P}(\boldsymbol{\gamma})$, the proposed acceptance probability can be calculated by: 
	\begin{equation}
		\frac{f(\boldsymbol{\gamma}^{new}|\mathbf{X},\boldsymbol{c})}{f(\boldsymbol{\gamma}^{old}|\mathbf{X},\boldsymbol{c})} = \frac{f(\mathbf{X}|\boldsymbol{\gamma}^{new},\boldsymbol{c}) \mathrm{P}(\boldsymbol{\gamma}^{new}) }{ f(\mathbf{X}|\boldsymbol{\gamma}^{old},\mathbf{c}) \mathrm{P}(\boldsymbol{\gamma}^{old})}.
	\end{equation}
	\textcolor{black}{With the saved MCMC samples, one can calculate the marginal posterior inclusion probability vector $\boldsymbol{\pi}$ for all the features. For feature $i$, the posterior inclusion probability $\pi_i$ is the number of times feature $i$ was selected divided by the number of saved MCMC iterations.} {\color{black} One could then rely on a pre-specified threshold as the cutoff for selection; a threshold of $0.5$ is a common choice, as it was shown to be optimal in the context of regression modeling \citep{Barbieri2004}.  An alternative approach is to calculate the expected false discovery rate (FDR) from $\boldsymbol{\pi}$ and control the FDR to a target level. Further discussion is given in Section S4.1 of the Supplementary Material.}
	\subsection{Updates to cluster assignments}
	We update the latent sample allocation vector $\boldsymbol{c}$ using \citet{JainNeal04}'s split-and-merge algorithm by first selecting two distinct observations, $i$ and $l$ uniformly at random. Let $\mathcal{C}$ denote the set of other observations that are in the same cluster with $i$ or $l$.
	
	
	If $\mathcal{C}$ is empty, we use the simple random split-merge algorithm. Otherwise, we use the restricted Gibbs sampling split-merge algorithm. Both involve a Metropolis-Hastings sampling step, with acceptance probability: 
	\begin{equation}
		a(\boldsymbol{c}^{merge},\mathbf{c})=\min\left\{1, \frac{q(\boldsymbol{c}|\boldsymbol{c}^{merge})\mathrm{P}(\boldsymbol{c}^{merge})L(\boldsymbol{c}^{merge}|\mathbf{X},\boldsymbol{\gamma})}{q(\boldsymbol{c}^{merge}|\boldsymbol{c})\boldsymbol{P}(\boldsymbol{c})L(\boldsymbol{c}|\mathbf{X},\boldsymbol{\gamma})}
		\right\}\end{equation} if $c_i \neq c_l$, and \begin{equation}
		a(\mathbf{c}^{split}|\mathbf{c})=\min\left\{ 
		1, \frac{q(\boldsymbol{c}|\boldsymbol{c}^{split})\mathrm{P}(\mathbf{c}^{split})L(\mathbf{c}^{split}|\mathbf{X},\boldsymbol{\gamma})}{q(\boldsymbol{c}^{split}|\boldsymbol{c})\mathrm{P}(\boldsymbol{c})L(\boldsymbol{c}|\mathbf{X},\boldsymbol{\gamma})}
		\right\} \end{equation} if $c_i = c_l$.
	
	For the simple random split-merge algorithm, $\frac{q(\mathbf{c}|\mathbf{c}^{merge})}{q(\mathbf{c}^{merge}|\mathbf{c})} =1$, $\frac{q(\mathbf{c}|\mathbf{c}^{split})}{ q(\mathbf{c}^{split}|\mathbf{c})} = 1$. For the restricted Gibbs sampling, we first randomly create a launch state. This launch state is modified by a series of ``intermediate" restricted Gibbs sampling steps to achieve a reasonable split of the observations. The last launch state is used for the calculation of the transient probabilities. Details on the split-and-merge algorithm and computing times are provided in the Supplementary Material.
	
	
	The prior ratio, $\mathrm{P}(\mathbf{c}^{merge})/\mathrm{P}(\mathbf{c})$ or $\mathrm{P}(\mathbf{c}^{split})/\mathrm{P}(\mathbf{c})$, relies on the partition distribution $\mathrm{P}(\mathbf{c})$. In an MFM model, the probability function of $\mathbf{c}$ is  $\mathrm{P}(\mathbf{c})=V_N(|C|)\prod_{c \in C}\eta^{(n_c)}$.
	%
	%
	%
%
%
The probabilities of splitting a cluster and combining two clusters are: 
\begin{align}
	\frac{\mathrm{P}(\mathbf{c}^{split})}{\mathrm{P}(\mathbf{c})}&=\frac{V_N(|C|+1)}{V_N(|C|)}\frac{\Gamma(n_{c_1}+\eta)\Gamma(n_{c_2}+\eta)}{\Gamma(n_{c_1}+n_{c_2}+\eta)\Gamma(\eta)};\\
	\frac{\mathrm{P}(\mathbf{c}^{merge})}{\mathrm{P}(\mathbf{c})}&=\frac{V_N(|C|-1)}{V_N(|C|)}\frac{\Gamma(n_{c_1}+n_{c_2}+\eta)\Gamma(\eta)}{\Gamma(n_{c_1}+\eta)\Gamma(n_{c_2}+\eta)},\end{align}
where $n_{c_1}$ and $n_{c_2}$ are the number of observations in the two clusters.

\subsection{Post-processing of MCMC samples}
The sampled values of the cluster indices can only describe whether two observations belong to the same cluster, but are not comparable between iterations, as the same index value may represent different clusters due to the ``label switching" issue.
In this paper, we adopt \cite{FritIcks09}'s method for summarizing posterior cluster labels from MCMC samples. We denote the proposed clustering estimate as $c^*$, and estimate the probability that samples $i$ and $j$ belong to the same cluster from $M$ MCMC samples by $\zeta_{ij}=\frac{1}{M}\sum_{m=1}^M I(c_i^{(m)}=c_j^{(m)})$. A posterior cluster assignment can be obtained by maximizing the adjusted Rand index: 

\begin{equation}
	AR(c^*,\zeta)=\frac{\sum_{i<j}\mathrm{I}_{\{c_i^*=c_j^*\}}\zeta_{ij}-\sum_{i<j}\mathrm{I}_{\{c_i^*=c_j^*\}}\sum_{i<j}\zeta_{ij}/ {\color{black}{N \choose 2}}}
	{\frac{1}{2}[\sum_{i<j}\mathrm{I}_{\{c_i^*=c_j^*\}}+\sum_{i<j}\zeta_{ij}]-\sum_{i<j}\mathrm{I}_{\{c_i^*=c_j^*\}}\sum_{i<j}\zeta_{ij}/ {\color{black}{N \choose 2}}}.
\end{equation}

This method can handle the label-switching issue, and can be simply implemented using the R package ``mcclust" \citep{Frit12}.

\section{Simulation studies} \label{sec:sim}

In this section, we first introduce the simulation setup, and then compare the performance of the proposed MFMDM and MFMDTM approaches with those from existing distance-based clustering methods, including PAM and hierarchical clustering (i.e., hcut) with complete linkage using the Euclidean, Bray Curtis, unweighted UniFrac, and weighted UniFrac distance metrics. {\color{black} For the completeness of comparison, the performances from the Dirichlet process mixture of Dirichlet (tree) multinomials (DPDM and DPDTM) are also included.}

To construct the simulated data, we generated observations with structure similar to the dataset described in \citet{DeFiCava10}. {\color{black} This study included two groups of samples, 14 from Africa and 15 from Italy. The original dataset has 2,803 OTUs with a median sequencing depth of 13,523. Due to the geographic distance and lifestyle difference between the two sample groups, we expect the microbiome profiles for the two groups to be well separated.
	We chose this dataset as the basis for our simulation study
	since their sequencing data is publicly available and the samples are well annotated. Also, it has a large number of OTUs with only a moderate number of observations, which is typical in microbiome data analysis. By relying on an existing microbiome dataset as the basis for our simulation study, we ensure that aspects of the simulated data such as the distribution of counts and shape of the tree structure resemble those in real data.
} 
In our simulation design, 
each group has 15 observations, and each observation has 15,000 total counts. We simulated 5 scenarios, with decreasing levels of complexity, and generate the OTU counts for the $z$th scenario, $z=1,2,\dots,5$, in the following way.
\begin{enumerate}
	\item {\color{black} Choose} two non-overlapping subsets of OTUs, $\Psi$ and $\Lambda$. In our simulation set-up, one subset $\Psi$ accounts for $13\%$ of the counts and $356$ OTUs, while the other subset $\Lambda$ accounts for $15\%$ of the counts and $595$ OTUs.
	\item {\color{black} Set the expected abundance of the two groups using the marginal distributions. Let $\mathbf{p}_{\Psi}$ and $\mathbf{p}_{\Lambda}$ represent the vectors of marginal probabilities for the subsets $\Psi$ and $\Lambda$, respectively. For group A, we set the marginal probabilities for OTUs in subset $\Psi$ to $\mathbf{p}_{\Psi}^A=(1-z/5)\mathbf{p}_{\Psi}$, and correspondingly change the marginal probabilities for OTUs in subset $\Lambda$ to $\mathbf{p}_{\Lambda}^A=\mathbf{p}_{\Lambda}(\sum_{i \in \Lambda}p_i+z/5\sum_{i \in \Psi}p_i)/\sum_{i \in \Lambda}p_i$.  For group B, we change the marginal probabilities in the opposite direction, $\mathbf{p}_{\Psi}^B=(1+z/5)\mathbf{p}_{\Psi}$ and $\mathbf{p}_{\Lambda}^B=\mathbf{p}_{\Lambda}(\sum_{i \in \Lambda}p_i-z/5\sum_{i \in \Psi}p_i)/\sum_{i \in \Lambda}p_i$.}
	\item Generate the count vectors from a Dirichlet multinomial with the sum of parameters to be $200$.
	The tree used in the MFMDTM model is the phylogenetic tree of the De Filippo dataset.
	\item Repeat the above steps 200 times to generate 200 simulated data sets.
\end{enumerate}

We now describe the parameter settings using in applying the proposed MFMDM and MFMDTM models.
The parameters of the Beta distribution in the MFMDM model are set to be $\beta_1=\beta_2=1$, which corresponds to a uniform prior on the number of counts which are informative for clustering. Similarly, the parameters of the Dirichlet distribution for both the MFMDM and MFMDTM are set to be $\alpha=1$, which is a uniform prior in the multinomial case. The prior probabilities of OTUs being informative is $50\%$, i.e., $w=0.5$ for the MFMDM model. We give $M-1$ a $\mathrm{Poisson}(1)$ distribution, which expresses a preference for a small number of clusters. {\color{black} Sensitivity analyses regarding the choice of priors are included in the supplentary material S.3.}


In both models, a large number of observed sequences inflates the factorial terms in the likelihood, which tends to support finer clusters. Unlike scale invariant mixtures \citep{MalsFruh14}, such as Gaussian mixtures, the likelihood of the Dirichlet (tree) multinomial is dependent on the number of sequences, which reflects both sequencing depth and rarefaction. To temper this effect and better achieve meaningful clustering, we take an approach similar to that of \citet{GrieMcDa18} who ``normalize" the data by first dividing the observed counts by a scaling parameter. For the simulated data, we found $50$ to be a reasonable scaling parameter. \textcolor{black}{Based on our experiments with both simulated and real data, we found that the maximum sequencing depth divided by $300$ worked well for the choice of scaling parameter across all settings considered.}

\begin{figure}
	\includegraphics*[width=\linewidth]{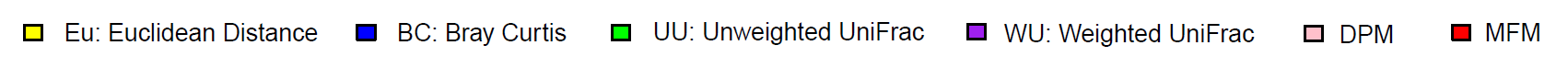}
	\begin{subfigure}{0.32\linewidth}
		\caption{Scenario 2}
		\includegraphics*[width=\textwidth]{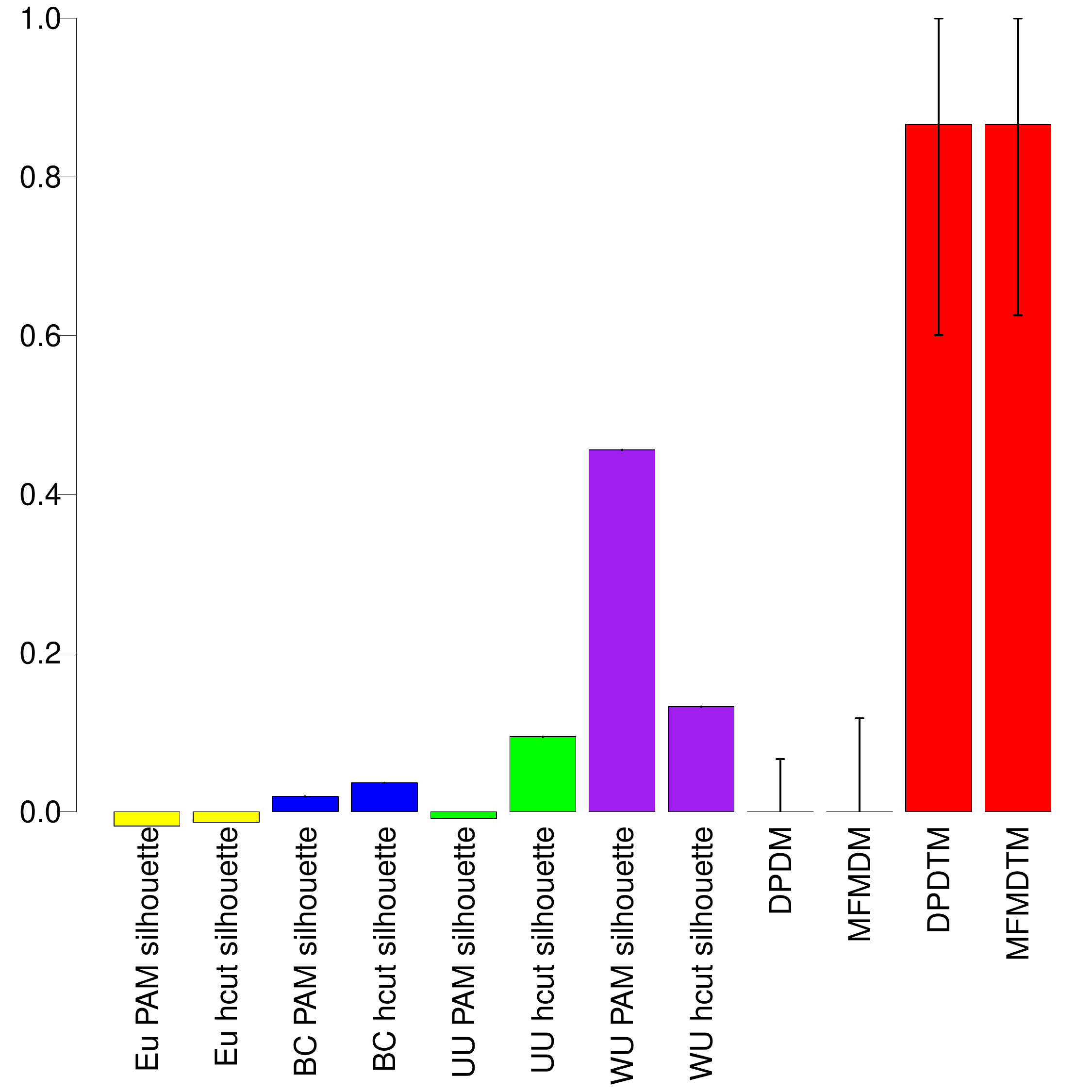}
	\end{subfigure}
	%
	\begin{subfigure}{0.32\linewidth}
		\caption{Scenario 4}
		\includegraphics*[width=\textwidth]{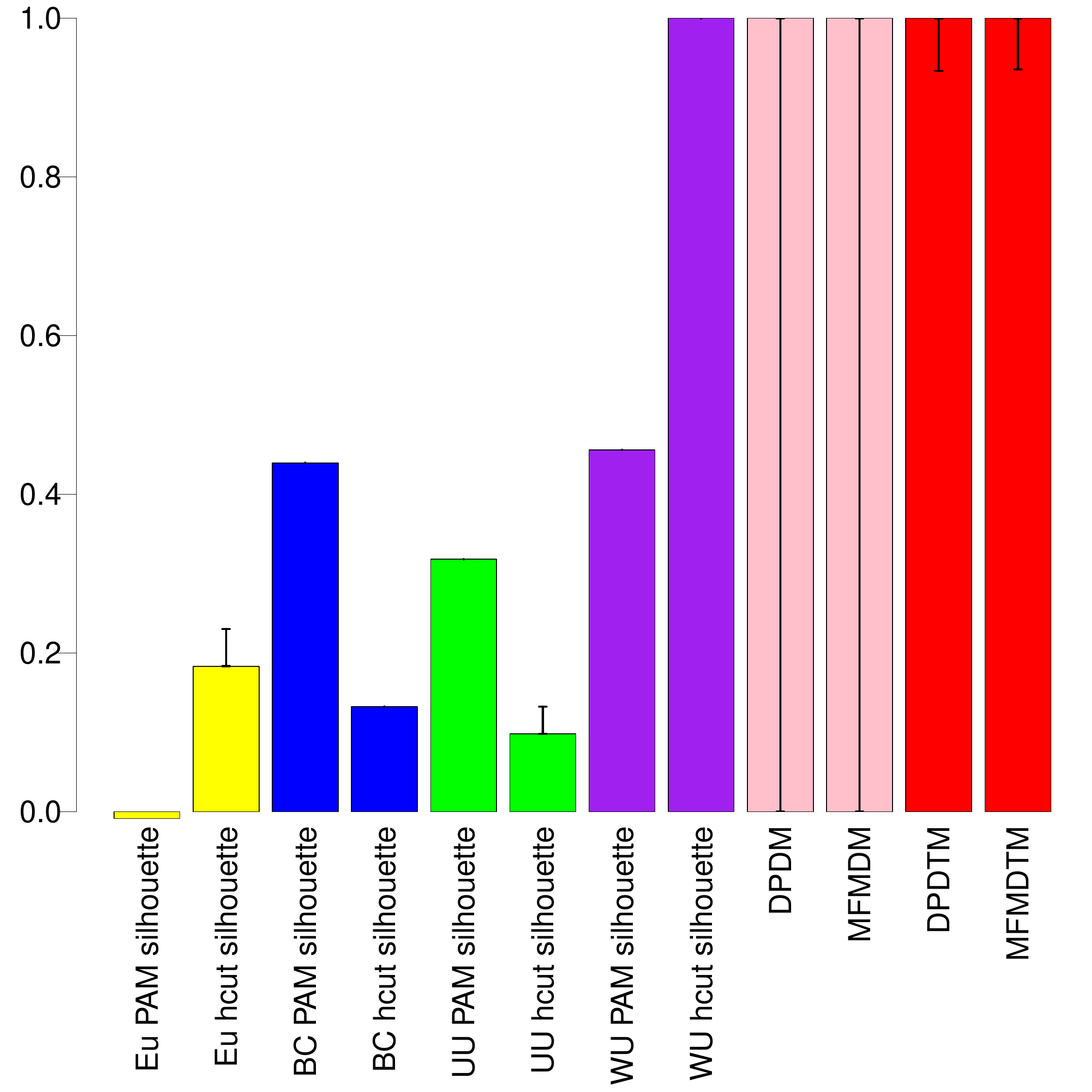}
	\end{subfigure}
	\begin{subfigure}{0.32\linewidth}
		\caption{OTU Selection AUC}
		\includegraphics*[width=\textwidth]{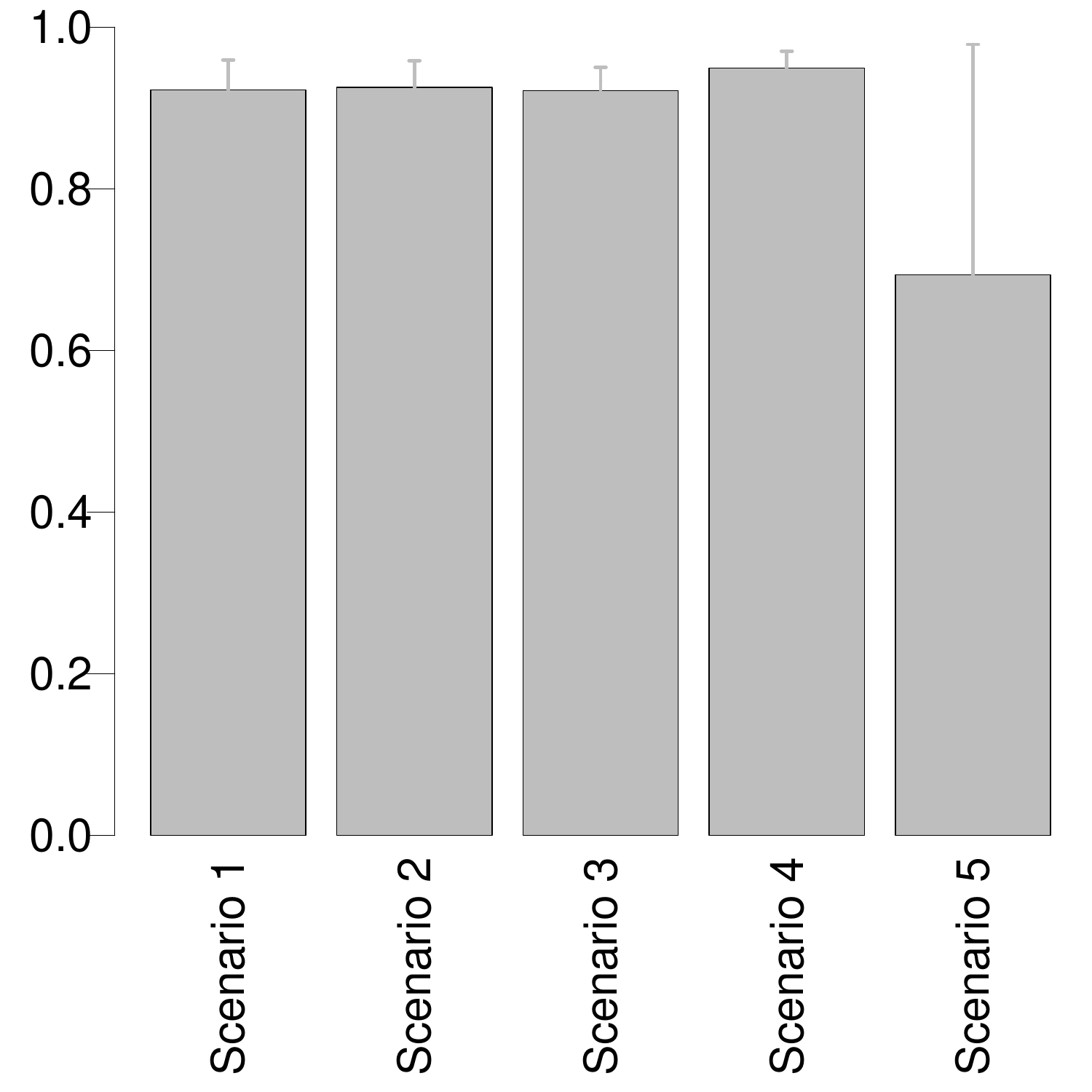}
	\end{subfigure}
	\caption{Comparison with machine learning methods in terms of Rand indices, and the area under the curve (AUC) of the high abundance OTUs. The bar heights represent the median over 200 datasets, and the black intervals represent the empirical estimate of the $95\%$ confidence interval.}
	\label{randAUC}
\end{figure}

For each dataset, we run 20,000 iterations, the first 10,000 of which are discarded as burn-in, and then apply a thinning of $10$ and keep 1,000 samples for inference. For machine learning methods, the number of clusters is determined by the silhouette width, which is appropriate for non-Euclidean distances. We measure clustering performance using the adjusted Rand index. The expected value of the adjusted Rand index is 0 when clustering is done at random, while 1 reflects perfect recovery of the true underlying clusters in the data. Figure \ref{randAUC} (a) and (b) show the performance of our proposed methods, compared to some distance-based clustering methods for scenarios 2 and 4. Barplots for scenarios 1, 3, and 5 can be found in Supplementary Figure S1. \textcolor{black}{Compared with the Dirichlet process mixture (DPM) model, the mixture of finite mixture (MFM) model can estimate the number of clusters consistently, i.e., the estimated number of clusters will not inflate with increased sample size. However, we found from our simulation studies that the performance of the DPM model and the MFM model is similar, which is in alignment with the empirical comparison in \cite{MillHarr18}, who observed that the two methods perform similarly on simulated datasets with moderate sample sizes. }{\color{black}The main difference is the underlying belief: DPM assumes there are infinite number of mixture components, whereas MFM assumes finite number of mixture components.} When the separation between the two clusters is relatively small, the proposed MFMDTM method, which performs variable selection accounting for the tree structure among features, shows a sizeable advantage over the distance-based methods including PAM using UniFrac distances, which incorporate phylogenetic information  (Figure \ref{randAUC} (a)). When the separation between the clusters becomes larger, the performance of MFMDTM still shows significant improvement over that of competing methods that do not account for phylogeny (Figure \ref{randAUC} (b)). \textcolor{black}{Though the empirical confidence interval is wide, the performance of the MFMDM model also improves significantly in this more separated scenario, achieving a median Rand index of 1.} In general, the methods that incorporate tree information outperform those that do not, with MFMDTM achieving the highest adjusted Rand indices across all methods considered.

An advantage of the proposed methods over existing alternatives is that they enable  the selection of informative features. The inference about informative vs.\ noisy features is based on the marginal posterior distribution of the latent indicator $\boldsymbol{\gamma}$, which is estimated from the selection frequencies in the MCMC output \citep{KimTade06}. It is worth mentioning that $\Psi$ and $\Lambda$ contain OTUs with low abundance, whose effects are negligible compared with the simulation noise. To set a meaningful goal for selection, we consider the $37$ high abundance OTUs that differ across groups, from among the $197$ high abundance OTUs in the dataset, as the true discriminatory features, where ``high abundance" is defined as marginal abundance greater than $0.001$. \textcolor{black}{For more details regarding feature selection under different thresholds, readers can refer to Supplementary Material Figure S9.}
As shown in Figure \ref{randAUC} (c), the area under the curve (AUC) values for the receiver operating curve (ROC) describing the accuracy of feature selection suggest that MFMDM's selections are successful even when the Rand index is low.
\textcolor{black}{To provide intuition for this result, we note that achieving a high Rand index is a more challenging task than identifying influential features. To give an extreme example, the true cluster assignment is: two clusters with 15 observations each, however, the clustering algorithm concludes that there are two clusters, one with one observation while the other has 29 observations. The Rand index in this example is 0, but the features that separate one observation from the rest can still be the correct features that distinguish the two true clusters.}
The decrease in the AUC for Scenario 5 is due to the fact that the larger separation enables detection of informative OTUs that have marginal abundance below $0.001$, reducing the specificity. \textcolor{black}{For five simulation scenarios, we show the ROC curves generated by varying the threshold on the posterior probability of feature inclusion of the 200 datasets in  Supplementary Material Figure S10.} These results show that the proposed method is able to accurately recover the informative features.

{\color{black} The proposed methods are fairly computationally intensive, but still feasible to run on a desktop computer. More specifically, for the simulated data described above,
	on a computer with an Intel Core i9-10900K 3.70GHz processor and 64GB memory, it takes Rcpp 14.58 minutes for 1000 MCMC iterations with the MFMDM model, and MATLAB R2020a 21.87 minutes for 1000 MCMC iterations with the MFMDTM model. We adopted a MATLAB implementation for the MFMDTM due to its simplicity in handling multi-dimensional arrays. The run time for the DPMDM model is similar to that of the MFMDM model (14.53 minutes). The  difference in computation time between the MFM model and the DPM model is within $1\%$ of the total run time. Increasing the number of samples to 200 results in a run time of less than 2 hours for 1000 MCMC iterations.
}

\section{Case study: tumor microbiome heterogeneity in pancreatic cancer} \label{sec:application}
The goal of this case study is to provide insight into heterogeneity of the microbiome across pancreatic cancer patients.
We rely on the data set described by \citet{RiquZhan19}, which consists of microbiome profiles for 68 surgically resected PDAC samples. Among these 68 samples, 36 were obtained from long-term survivors, with survival times greater than 5 years, and 32 were obtained from stage-matched short-term survivors, who survived 5 years or less. Subjects were recruited from two cancer centers: The University of Texas MD Anderson Cancer Center (MDA, $n$ = 43) and Johns Hopkins University (JHU, $n$ = 25). The microbiome profiling data includes 2,410 OTUs corresponding to \textcolor{black}{1,095 taxonomic units at or above the species level.}

\begin{figure}[ht!]
	\centering
		\includegraphics*[width=0.75\textwidth]{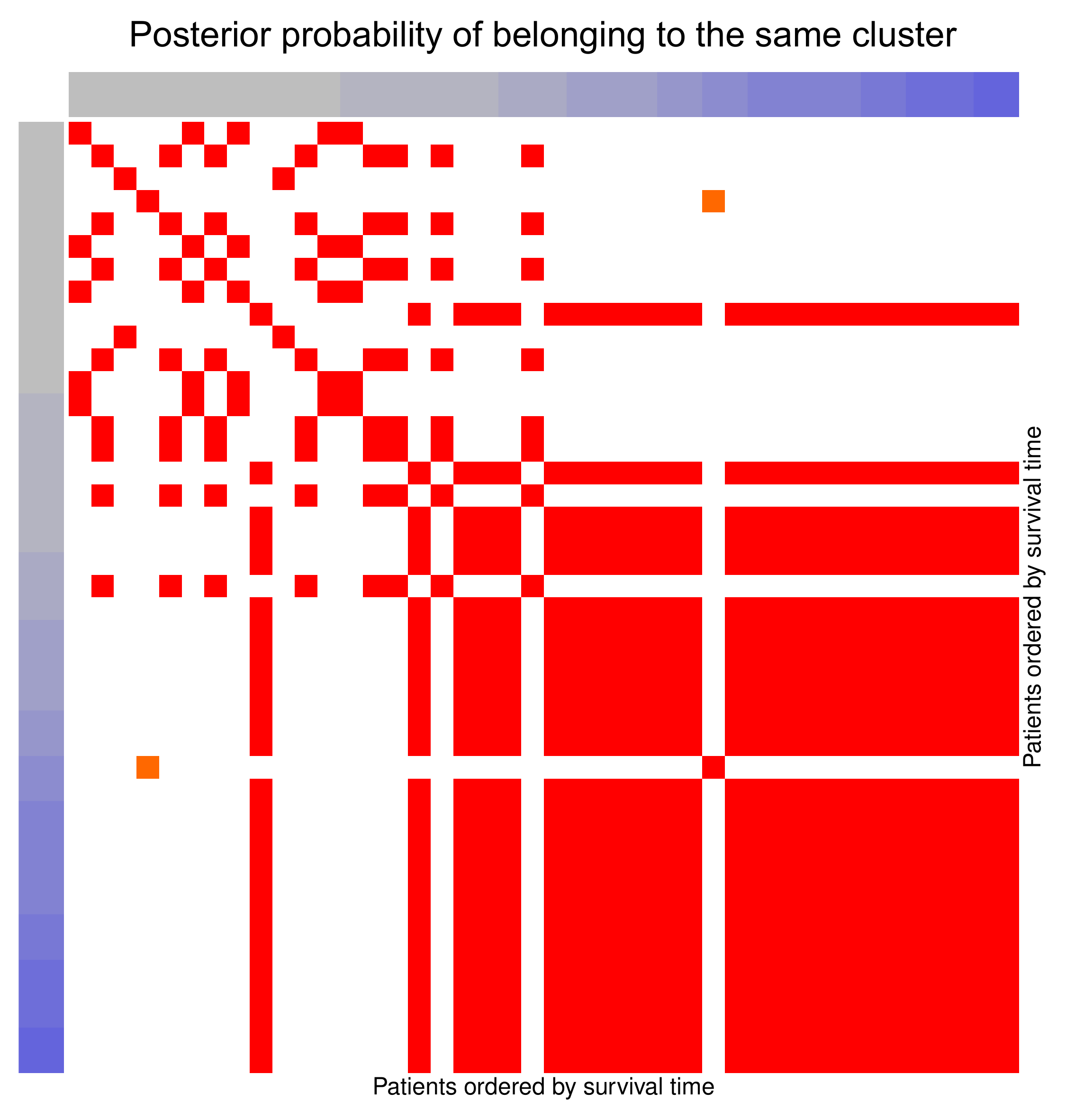}
		\includegraphics*[width=\textwidth]{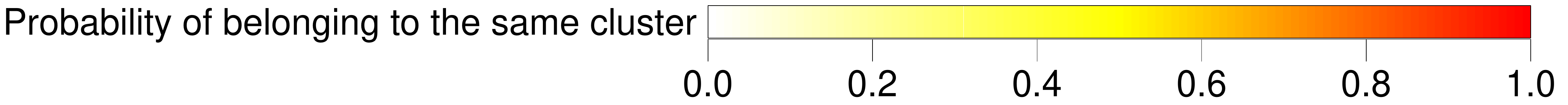}	
		\includegraphics*[width=\textwidth]{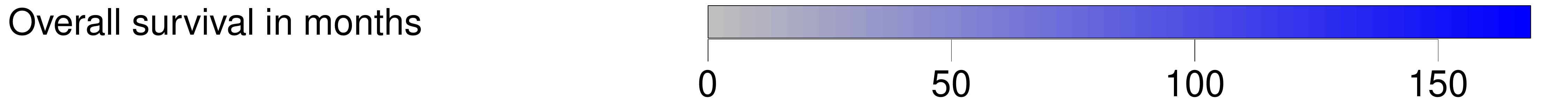}	
		\caption{Heatmap showing the posterior probability of being assigned to the same cluster for patients with exact survival time for \citet{RiquZhan19} data. {\color{black} Patients are ordered by their survival times.}}
		\label{RiquHeatmap}
\end{figure}

In order to uncover the natural sample groups and features driving these sample clusters,
\textcolor{black}{we applied the MFMDTM method to this data set}, using the same parameter settings as in the simulation.
For the MDA cohort, exact survival times were reported for 42 subjects, while for the JHU cohort, only a binary indicator of whether a patient survived more than 5 years was provided.
For the patients with exact survival times, we plot a heatmap showing the posterior probability of any two observations belonging to the same cluster, where samples are sorted by survival time (Figure \ref{RiquHeatmap}). The heatmap shows that long-term survivors cluster together with high probability, suggesting that their
microbiome profiles are more consistent than that of short-term survivors.
This finding suggests that it may be fruitful for researchers to investigate the specific bacteria present in long-term survivors, as reflecting a distinctive protective microbiome state.
A similar conclusion can be drawn from the heatmap of all the samples using both the  MFMDM and MFMDTM methods (Supplemental Figure S2). Our finding that long-term survivors consistently cluster together is particular interesting, since the long-term survivors came from two distinct geographic locations. Though some pre-clinical models \citep{PushHund18,AykuPush19} have suggested that certain microbial species are positively associated with tumor progression, our finding is aligned with that of \citet{RiquZhan19}, who concluded that a protective microbiome induces anti-tumor immunity in long-term survivors and that those protective species are the key for future interventions.



\begin{figure}[ht!]
		\centering
		\includegraphics*[width=0.8\textwidth]{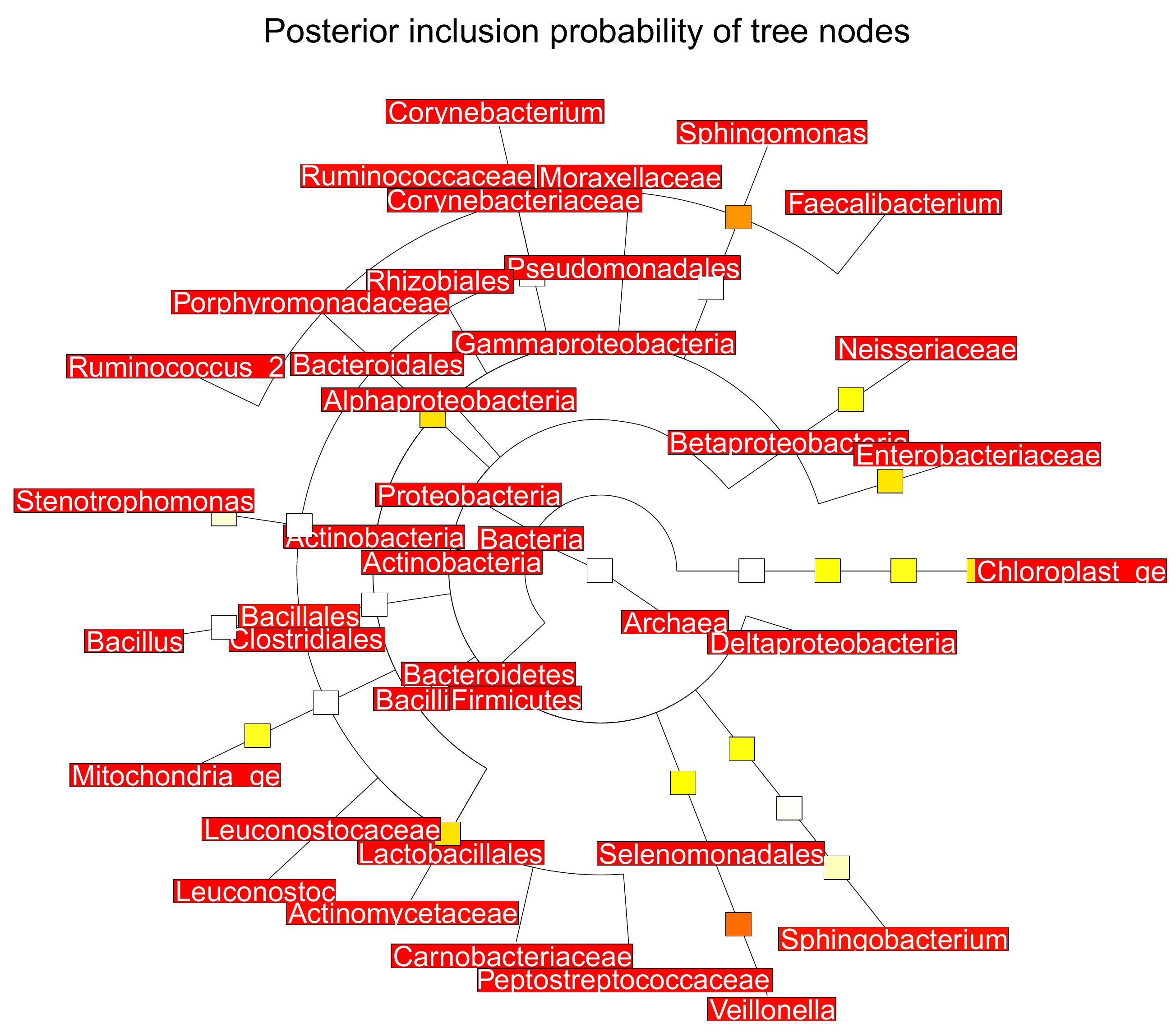}
		\includegraphics*[width=\textwidth]{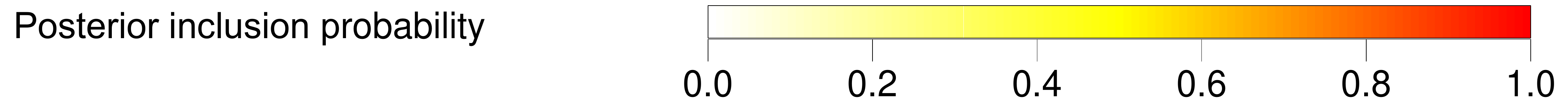}	
		\caption{\textcolor{black}{Posterior inclusion probabilities $\pi_i$} of the tree nodes for the \citet{RiquZhan19} data}
		\label{featureSelection}
\end{figure}

\textcolor{black}{Our model can also identify nodes in the taxonomic tree that drive the clustering of the samples. Figure \ref{featureSelection} shows the nodes with high posterior inclusion probabilities in red, the majority of which were also identified in \citet{RiquZhan19}.} The original paper showed the predominance of Clostridia in short-term survivors and Alphaproteobacteria in long-term survivors at the class level, while our method shows that the two corresponding phyla, Firmicutes and Proteobacteria, are differential across clusters with posterior probability greater than $95\%$. 
\citet{RiquZhan19} identified the species \textit{Bacillus clausii} as predictive of survivorship, and our method selects the genus it belongs to as a relevant feature. Some of the taxa identified our analysis using MFMDTM and the analysis by \citet{RiquZhan19} were discovered by previous research on PDAC. For example, \cite{FarrZhan12} found that the abundance of the genus \textit{Corynebacterium} is lower in PDAC patients than in healthy individuals, while our method specifically points out that the allocation probability for the family Corynebacteriaceae is differential across clusters. \cite{GellBarz17} discovered that Proteobacteria producing cytidine deaminase are most associated with pancreatic cancer and our model identified several features under the phylum Proteobacteria, including the families \textit{Porphyromonadaceae} and \textit{Enterobacteriaceae}. In addition, MFMDTM identifies features that have not been thoroughly discussed in the pancreatic cancer literature before, such as the order Rhizobiales, which was found in higher abundance among patients with Helicobacter pylori-negative intestinal metaplasia than those with Helicobacter pylori-negative chronic superficial gastritis  or cancer \citep{ParkLee19}.

Compared with the LEfSe method used in the original paper for differential abundance analysis \cite{lefse}, which tends to select nested features, our method can identify the exact taxonomic level at which clusters are different. For completeness, we plot a heatmap showing the conditional probability of allocating a count to each child node for each selected parent node (Supplementary Figure S3) and a principal coordinate analysis (PCoA) plot of the samples colored by the MFMDTM cluster assignment (Supplement Figure S4).

\section{Discussion} \label{sec:discussion}
We have proposed two novel approaches for clustering of microbiome samples.
Unlike existing approaches for microbiome clustering, our methods perform variable selection, which enables biological understanding of features that differentiate clusters present in the data, and does not require pre-specification of the number of clusters.
The simulation results demonstrate that our methods can outperform commonly used unsupervised clustering algorithms in terms of the adjusted Rand index, suggesting that our sparse models are not only more interpretable but also more robust to noise.

Our application to tumor microbiome profiling of pancreatic cancer patients enhances the originally published analysis of this data set: while \cite{RiquZhan19} applied LefSe to identify microbiome features assuming known group membership, our approach shows that the long-term survivors make up a more natural cluster, while the short-term survivors are more heterogeneous, and our approach identifies additional features that are differential across the inferred clusters. These findings could guide the development of future microbiome interventions to improve cancer outcomes, which is an active and exciting area of current medical research \citep{RetiEngl19, McQuDani19}. The Dirichlet multinomial mixture model code is included in the R package BayesianMicrobiome, while the Dirichlet tree multinomial mixture is implemented using Matlab. Both are available at \url{https://github.com/YushuShi/sparseMbClust}.

In this manuscript, we have largely focused on data obtained from 16S profiling. 
However, there is increasing interest in metagenomic whole genome sequencing (WGS) approaches, which allow for the comprehensive sequencing of all DNA present in a sample. While there are some differences between these two sequencing approaches, our methods are applicable to WGS data as well. More specifically,  WGS can allow additional characterization of functional metabolic pathways by assigning the observed genetic sequences to biological roles, using tools such as HUMAnN2 or FMAP \citep{Franzosa2018, Kim2016}. Interestingly, these pathways can be organized into an ontological hierarchy \citep{Caspi2013}, making our tree-based clustering approach applicable in this context as well.   \textcolor{black}{    The proposed methods can be scaled to hundreds of observations with several thousand features, as they exploit the conjugacy between the Dirichlet (tree) priors and multinomial distributions and rely on efficient Rcpp/Matlab implementations.
	If greater computational scalability is needed (for example, when applying the method to amplicon sequence variants obtained from WGS),
	a faster approach would be using variational inference to approximate the posterior. Variational inference for unsupervised clustering with simultaneous feature selection is an area that we would like to explore in our future research.}

{\color{black} Finally, the proposed MFMDTM model assumes a fixed tree,  but in reality, there may be uncertainty regarding the tree structure. One potential approach to incorporate uncertainty in the tree structure is to summarize the tree as a matrix of pairwise distances between the features  \citep{ZhanShi21} and express the uncertainty of the tree structure through the variance of this matrix. The application of our methods to functional data derived from WGS and to settings with uncertainty regarding the tree structure are potential topics of interest in future work. }



\section*{Data availability}
The case study data was originally described in the publication \cite{RiquZhan19}. Sequencing data and patient survival times can be accessed through the NCBI BioProject under Accession Number PRJNA542615. A processed version of the 16S data is available at \url{https://github.com/YushuShi/sparseMbClust/}, along with an R package
implementing the  Dirichlet multinomial mixture model code and Matlab code for the 
the Dirichlet tree multinomial mixture.

\section*{Acknowledgements}

This work was supported by the National Institutes of Health [P30CA016672 to K.A.D. and C.P., P50CA140388, UL1TR003167 and 5R01GM122775 to K.A.D., HL124112 to R.J., R01HL158796 to C.P. and R.J.]; MD Anderson Moon Shot Programs [to K.A.D. and C.P.]; and Cancer Prevention \& Research Institute of Texas [RR160089 to R.J. and RP160693 to K.A.D.].

\bibliographystyle{rss}
\bibliography{ref2} 

\end{document}


\date{}
	\maketitle
	
	\noindent
	Here we provide details on the MCMC algorithm described in Section 3 of the main manuscript, including the steps in the split-merge algorithm.
	We provide additional plots from the simulation studies and pancreatic cancer application.
	{\color{black} We also show the results from sensitivity analyses, estimates of the expected false discovery rate,
		and assessments of the impact of varying the 
		threshold for being considered a ``high abundance" feature.}

	\section*{Contents}
	\begin{enumerate}
		\item[S1] Details on the computational approach
		\begin{enumerate}
			\item[S1.1] Simple random split-merge algorithm
			\item[S1.2] Restricted Gibbs sampling split-merge
		\end{enumerate}
		\item[S2] Additional plots
		\begin{enumerate}
			\item[S2.1] Additional plots with simulated data
			\item[S2.2] Additional plots for \citet{RiquZhan19} pancreatic cancer dataset
		\end{enumerate}
		{\color{black}
			\item[S3] Sensitivity analysis
			\begin{enumerate}
				\item[S3.1] Sensitivity analysis on $\beta_1$ and $\beta_2$ for MFMDM model
				\item[S3.2] Sensitivity analysis on $\alpha$
				\item[S3.3] Sensitivity analysis on $w$, $\eta$ and $p_m$
			\end{enumerate}
			\item[S4] Additional analysis on feature selection
			\begin{enumerate}
				\item[S4.1] Expected false discovery rate
				\item[S4.2] Impact of threshold for ``high abundance" OTUs
		\end{enumerate}}
	\end{enumerate}
	\section{Details on the computational approach}
	
	Here we provide additional details on the MCMC steps to update the cluster assignments, described at a high level in Section 3.2 of the main manuscript. This algorithm is based on the one proposed in \cite{JainNeal04}. We also include a note on the computing times.
	
	\subsection{Simple random split-merge algorithm}  
	
	\begin{enumerate}
		\item If $c_i=c_l$, then
		\begin{enumerate}
			\item a new cluster not equal to $\{c_1, . . . , c_n\}$ is created, and the allocations for other observations remain unchanged. In the proposal, $c_i$ is allocated to this new cluster. The new allocation with $i$ and $l$ in different clusters is called $\mathbf{c}^{split}$;
			\item the proposal is accepted with probability $$
			a(\mathbf{c}^{split}|\mathbf{c})=\min\left\{ 
			1, \frac{q(\mathbf{c}|\mathbf{c}^{split})\mathrm{P}(\mathbf{c}^{split})L(\mathbf{c}^{split}|\mathbf{X},\boldsymbol{\gamma})}{q(\mathbf{c}^{split}|\mathbf{c})\mathrm{P}(\mathbf{c})L(\mathbf{c}|\mathbf{X},\boldsymbol{\gamma})}
			\right\},
			$$
			
			where $\frac{q(\mathbf{c}|\mathbf{c}^{split})}{q(\mathbf{c}^{split}|\mathbf{c})}=1$, and 
			$$
			\frac{L(\mathbf{c}^{split}|\mathbf{X},\boldsymbol{\gamma})}{L(\mathbf{c}|\mathbf{X},\boldsymbol{\gamma})}=\frac{\int F(\mathbf{X}_i; \boldsymbol{\theta},\boldsymbol{\gamma}) dG_0(\boldsymbol{\theta};\boldsymbol{\gamma})\int F(\mathbf{X}_l; \boldsymbol{\theta},\boldsymbol{\gamma}) dG_0(\boldsymbol{\theta};\boldsymbol{\gamma})}{\int F(\mathbf{X}_i; \boldsymbol{\theta},\boldsymbol{\gamma}) F(\mathbf{X}_l; \boldsymbol{\theta},\boldsymbol{\gamma}) dG_0(\boldsymbol{\theta};\boldsymbol{\gamma})}.
			$$
			
			
		\end{enumerate}
		
		\item If $c_i \neq c_l$, then
		\begin{enumerate}
			\item $c_i$ and $c_l$ are merged into a single cluster, and the allocations for other observations remain unchanged. We name such an allocation $\mathbf{c}^{merge}$;
			\item the proposal is accepted with probability$$
			a(\mathbf{c}^{merge},\mathbf{c})=\min\left\{1, \frac{q(\mathbf{c}|\mathbf{c}^{merge})\mathrm{P}(\mathbf{c}^{merge})L(\mathbf{c}^{merge}|\mathbf{X},\boldsymbol{\gamma})}{q(\mathbf{c}^{merge}|\mathbf{c})\mathrm{P}(\mathbf{c})L(c|\mathbf{X},\boldsymbol{\gamma})}
			\right\}
			$$
			
			where $\frac{q(\mathbf{c}|\mathbf{c}^{merge})}{q(\mathbf{c}^{merge}|\mathbf{c})}=1$, and 
			$$
			\frac{L(\mathbf{c}^{merge}|\mathbf{X},\boldsymbol{\gamma})}{L(\mathbf{c}|\mathbf{X},\boldsymbol{\gamma})}=\frac{\int F(\mathbf{X}_i; \boldsymbol{\theta},\boldsymbol{\gamma}) F(\mathbf{X}_l; \boldsymbol{\theta},\boldsymbol{\gamma}) dG_0(\boldsymbol{\theta};\boldsymbol{\gamma})}{\int F(\mathbf{X}_i; \boldsymbol{\theta},\boldsymbol{\gamma})  dG_0(\boldsymbol{\theta};\boldsymbol{\gamma})\int F(\mathbf{X}_l; \boldsymbol{\theta},\boldsymbol{\gamma}) dG_0(\boldsymbol{\theta};\boldsymbol{\gamma})}.
			$$
		\end{enumerate}
	\end{enumerate}
	\subsection{Restricted Gibbs sampling split-merge}
	
	\begin{enumerate}
		\item Start by building a launch state as follows:
		\begin{enumerate}
			\item if $c_i=c_l$, then split the component, such that $c_i^{launch}\notin\{c_1, . . . , c_n\}$ and
			$c_l^{launch}=c_l$;
			\item if $c_i\neq c_l$, then $c_i^{launch}=c_i$ and $c_l^{launch}=c_l$;
			\item for every $s \in \mathscr{C}$, i.e., $s\neq i$, $s\neq l$ and $c_s=c_i$ or $c_s=c_l$, set $c_s^{launch}$ independently and at random with probability $0.5$
			to either $c_i^{launch}$ or $c_l^{launch}$;
			\item perform $t$ (we suggest using $t=20$) intermediate restricted Gibbs sampling scans to allocate each observation $s \in \mathscr{C}$ to either $c_i^{launch}$ or $c_l^{launch}$, such that 
			$$
			\mathrm{Pr}(c_s=c_i^{launch}|c_{-s},\mathbf{X}_s,\boldsymbol{\gamma})=\frac{Q(\mathbf{X}_s;\boldsymbol{\theta},\boldsymbol{\gamma},c_i^{launch})}{Q(\mathbf{X}_s;\boldsymbol{\theta},\boldsymbol{\gamma},c_i^{launch})+Q(\mathbf{X}_s;\boldsymbol{\theta},\boldsymbol{\gamma},c_l^{launch})}
			$$	
			Here $Q(\mathbf{X}_s;\boldsymbol{\theta},\boldsymbol{\gamma},c_i^{launch})=n_{c_i^{launch},-s}\int F(\mathbf{X}_s;\boldsymbol{\theta},\boldsymbol{\gamma})dH_{c_i^{launch},-s}(\boldsymbol{\theta};\boldsymbol{\gamma})$, where  $H_{c_i^{launch},-s}(\boldsymbol{\theta};\boldsymbol{\gamma})$ is the posterior of the parameter set $\boldsymbol{\theta}$ with all the observations in the cluster $c_i^{launch}$, except for observation $s$. Similarly, $n_{c_i^{launch},-s}$ is the number of observations in the cluster $c_i^{launch}$, except for observation $s$.
			
		\end{enumerate}
		\item If $c_i=c_l$, then
		\begin{enumerate}
			\item let $c_i^{split}=c_i^{launch}$ and $c_l^{split}=c_l^{launch}$;
			\item for every $s \in \mathscr{C}$, perform one final Gibbs sampling scan from
			$c_s^{launch}$ to set $c_s^{split}$ to either $c_i^{split}$ or $c_l^{split}$;
			\item the allocation for observations not in $\mathscr{C}$ remains unchanged; we name the proposed observation allocation $\mathbf{c}^{split}$;
			\item evaluate the proposal by the Metropolis Hastings acceptance probability $a(\mathbf{c}^{split}, \mathbf{c})$. The $q(\mathbf{c}^{split}|\mathbf{c})$ inside it is obtained by computing the Gibbs sampling
			transition probability from $\mathbf{c}^{launch}$ to $\mathbf{c}^{split}$. 
		\end{enumerate}
		\item If $c_i \neq c_l$, then
		\begin{enumerate}
			\item let $c_i^{merge}=c_l$ and $c_l^{merge}=c_l$;
			\item for every $s \in \mathscr{C}$, let $c_s^{merge}=c_l$;
			\item the allocation for observations not in $\mathscr{C}$ remains unchanged; we name the proposed observation allocation $\mathbf{c}^{merge}$;
			\item the proposal is accepted with probability $a(\mathbf{c}^{merge}, \mathbf{c})$, where $q(\mathbf{c}|\mathbf{c}^{merge})$ is the
			product over $s \in \mathscr{C}$ of the probabilities of setting each $c_s$ from the original split state to the launch state.
		\end{enumerate}
	\end{enumerate}

\section{Additional plots}

In this section, we provide some additional plots for the simulation studies (Section 4 of the main manuscript) and pancreatic cancer application (Section 5 of the main manuscript).

\subsection{Additional plots with simulated data}

{\color{black} To assess how much the use of silhouette width influences the results from the machine learning methods, we also applied these methods with the number of cluster fixed to the true value.
	Figure \ref{randAUC} provides the outcomes for all the simulation scenarios, including the performance of machine-learning based methods where the number of clusters are set to be the truth (2 clusters), machine-learning based methods where the number of clusters is determined by the silhouette width, the Dirichlet process mixture of Dirichlet multinomials (DPDM) and Dirichlet tree multinomials (DPDTM), and the mixture of finite mixtures of Dirichlet multinomials (MFMDM) and Dirichlet tree multinomials (MFMDTM). As expected, knowing the true number of clusters greatly improves the performance of the machine-learning based methods. However, such information is typically not available for real-world applications.} 

\begin{figure}[H]
	\includegraphics*[width=\linewidth]{plots/captureNew.png}
	\begin{subfigure}{0.3\linewidth}
		\caption{Scenario 1}
		\includegraphics*[width=\textwidth]{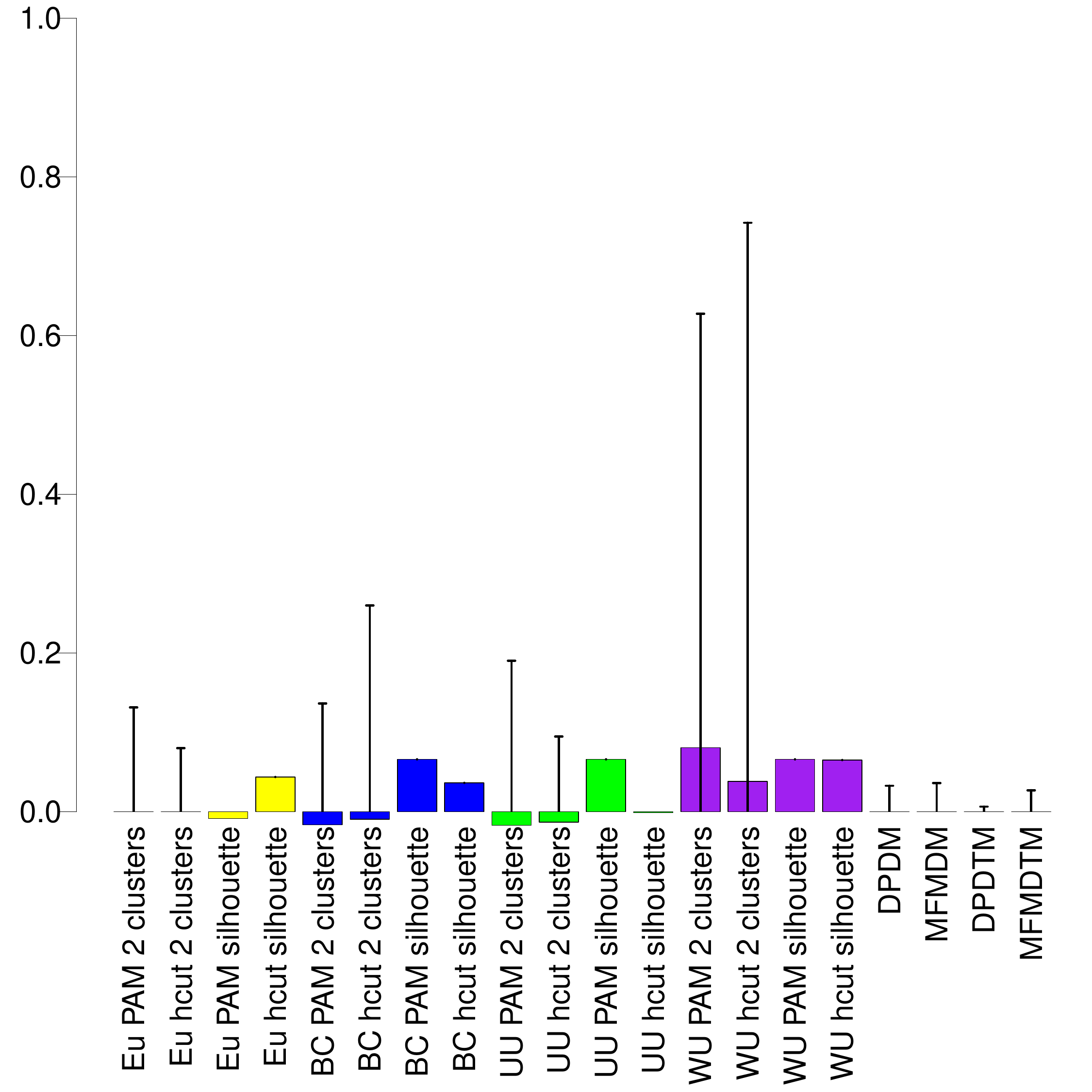}
	\end{subfigure}
	\begin{subfigure}{0.3\linewidth}
		\caption{Scenario 2}
		\includegraphics*[width=\textwidth]{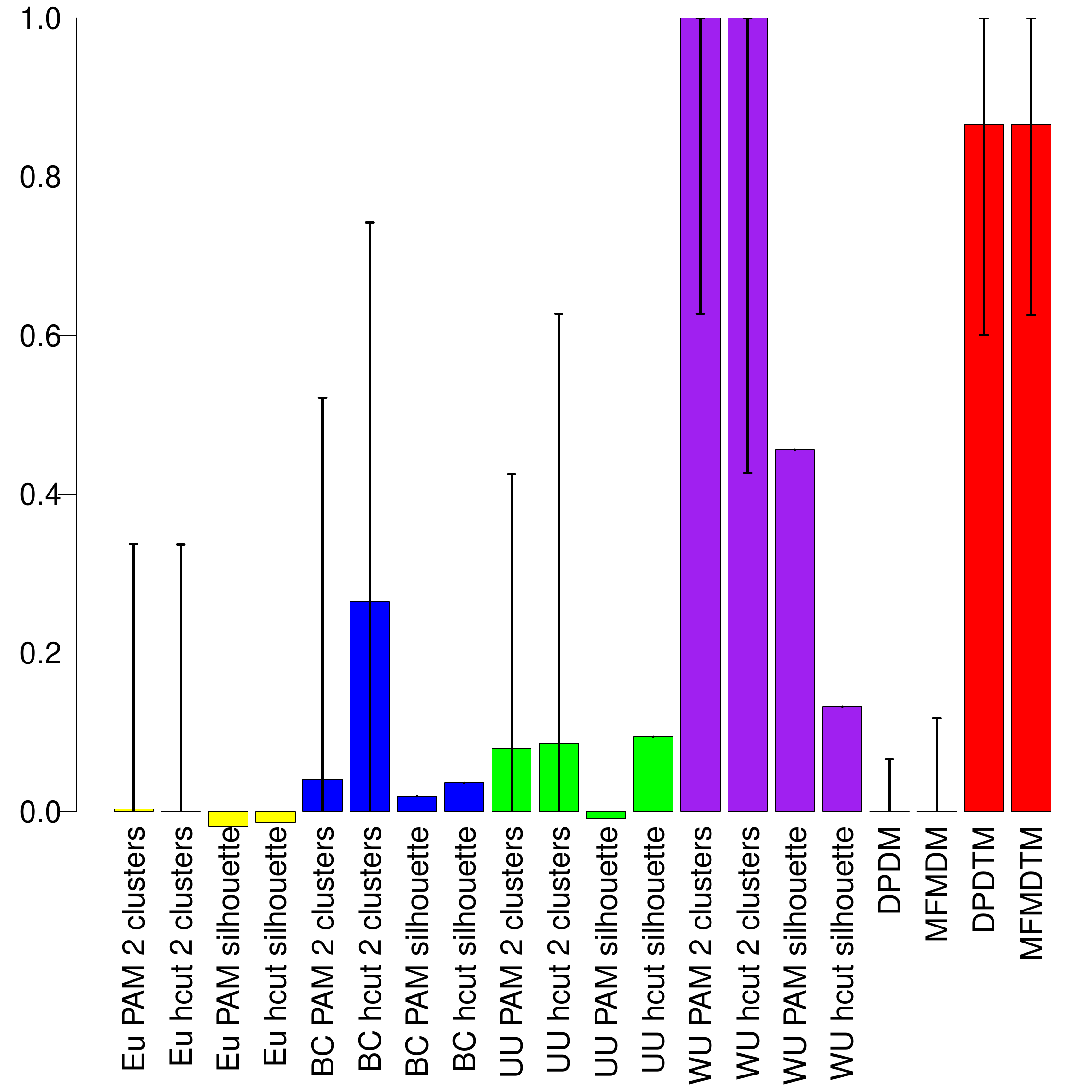}
	\end{subfigure}
	\begin{subfigure}{0.3\linewidth}
		\caption{Scenario 3}
		\includegraphics*[width=\textwidth]{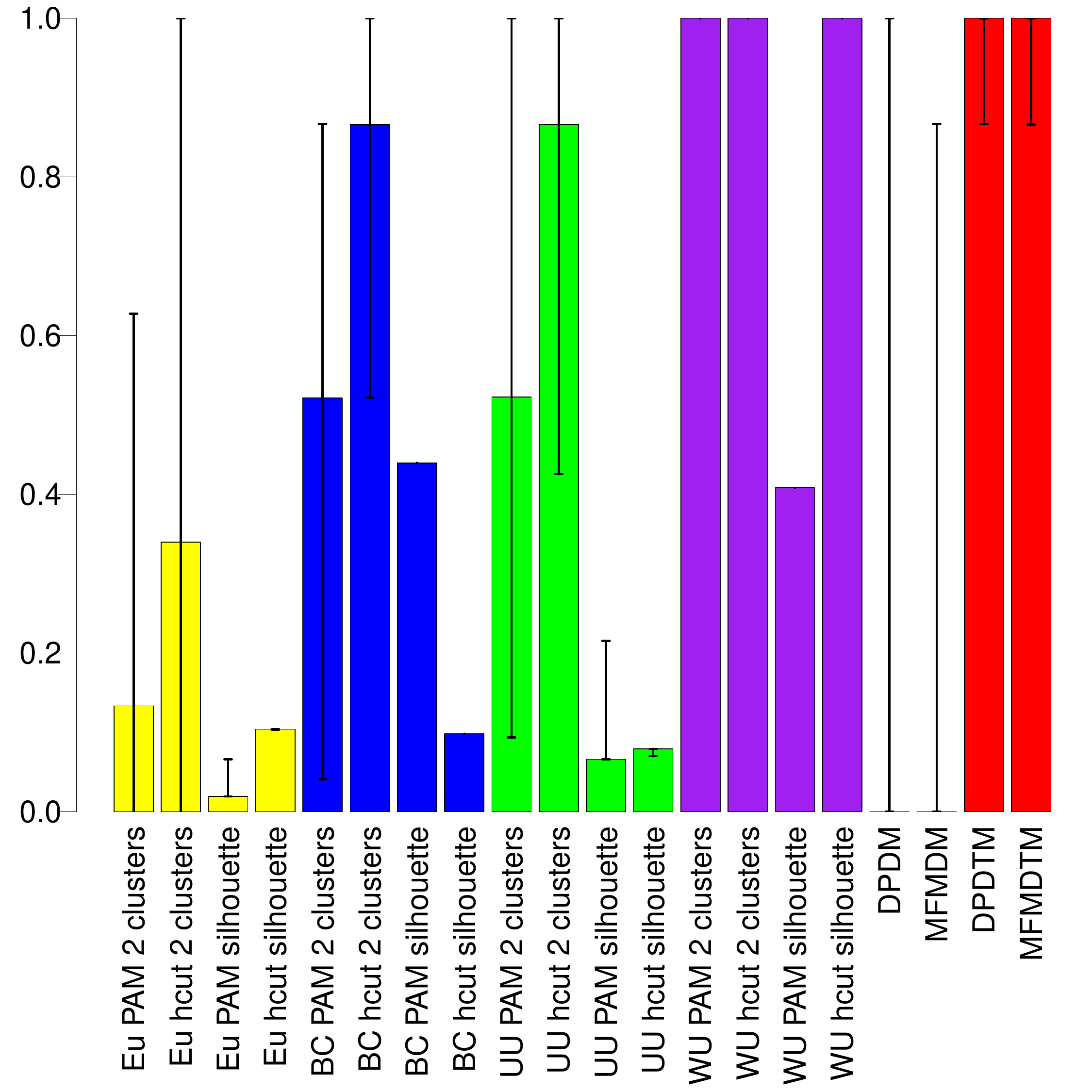}
	\end{subfigure}
\end{figure}

\begin{figure}[H]\ContinuedFloat
	\begin{subfigure}{0.3\linewidth}
		\caption{Scenario 4}
		\includegraphics*[width=\textwidth]{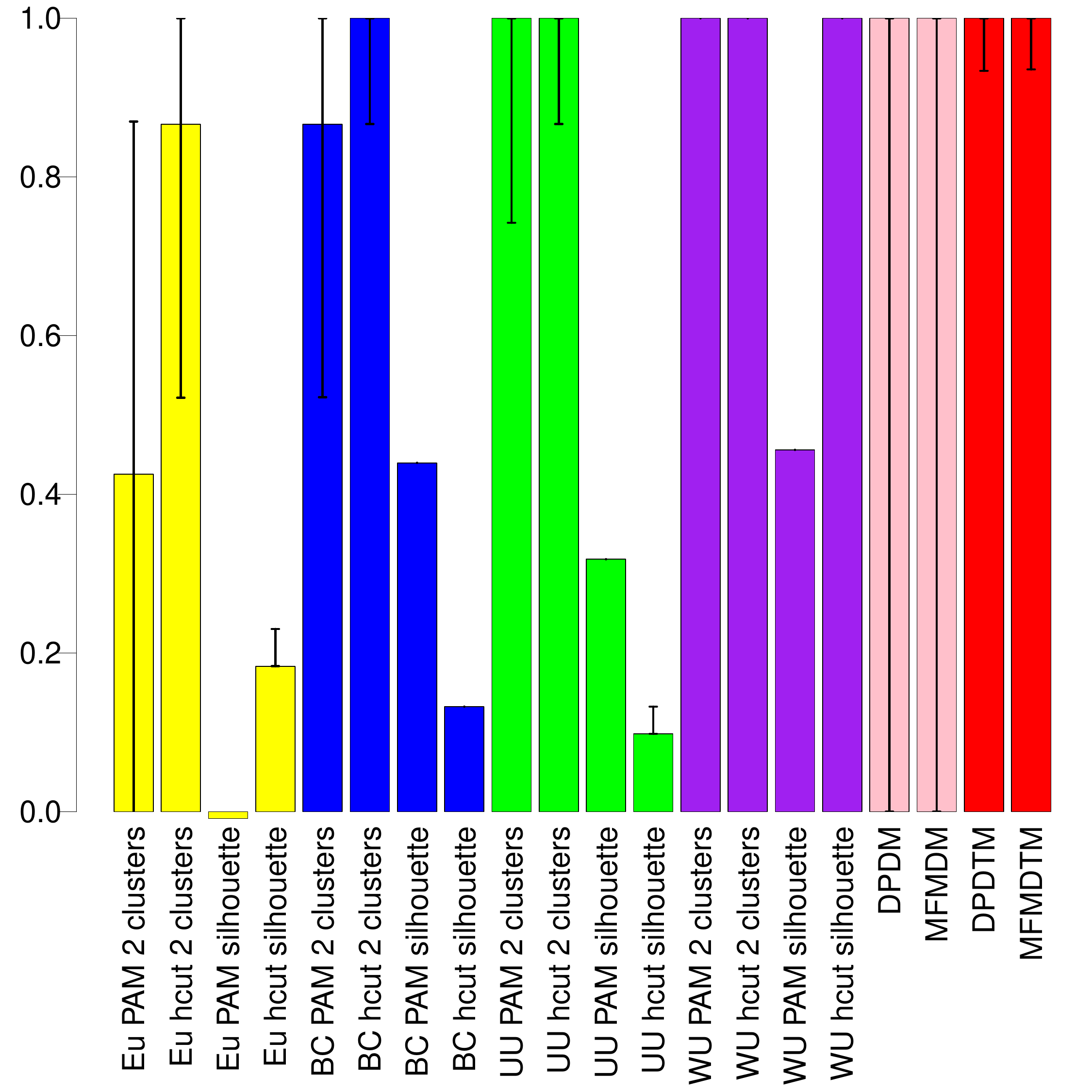}
	\end{subfigure}
	\begin{subfigure}{0.3\linewidth}
		\caption{Scenario 5}
		\includegraphics*[width=\textwidth]{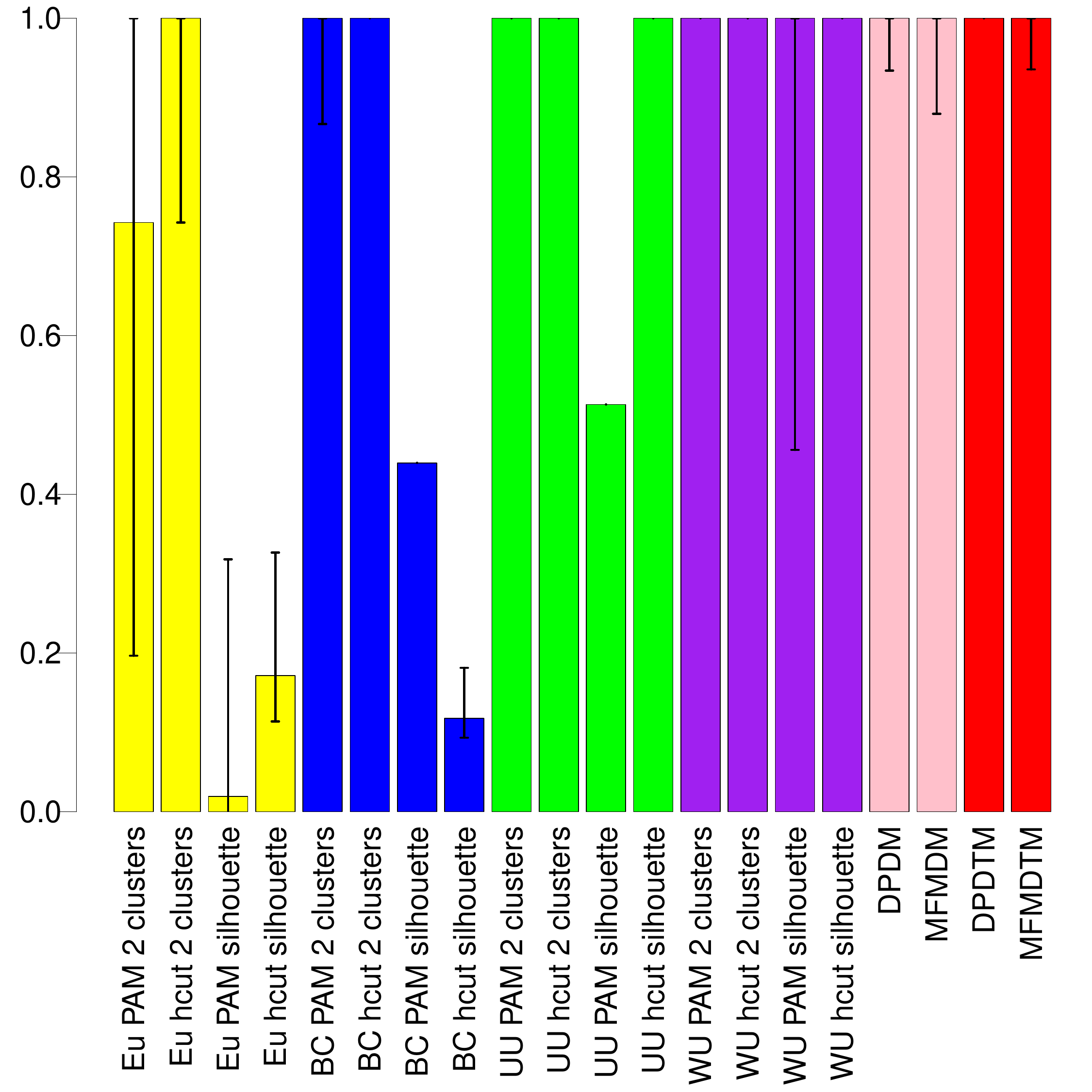}
	\end{subfigure}
	\begin{subfigure}{0.3\linewidth}
		\caption{OTU Selection AUC}
		\includegraphics*[width=\textwidth]{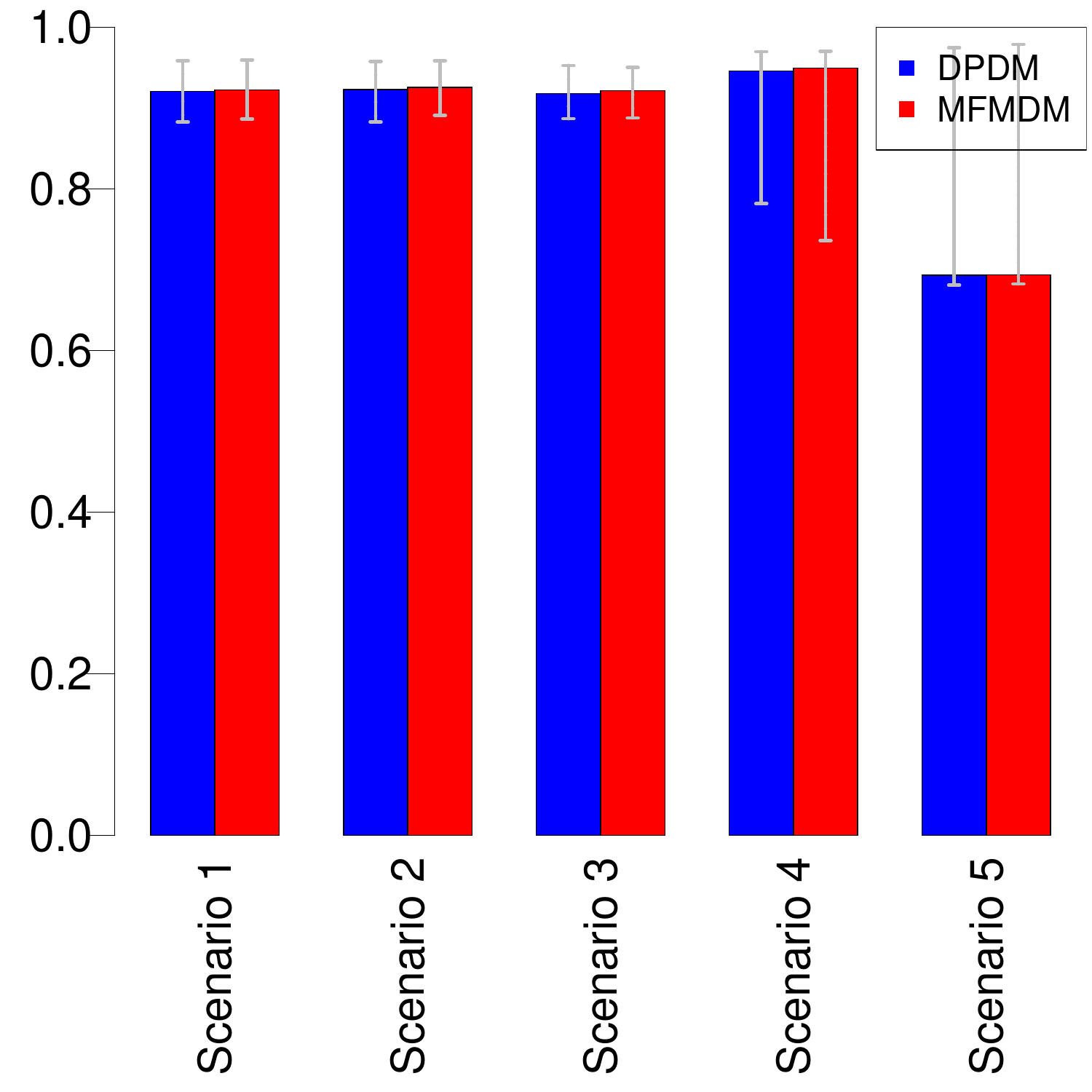}
	\end{subfigure}
	\caption{Comparison with machine learning methods in terms of Rand indices, and the area under the curve (AUC) of the high abundance OTUs using simulated data}
	\label{randAUC}
\end{figure}
\newpage
\subsection{Additional plots for the \cite{RiquZhan19} pancreatic cancer dataset}
\begin{figure}[ht]
	\centering
	\begin{subfigure}{\linewidth}
		\caption{Result using MFMDM}
		\includegraphics*[width=0.9\textwidth]{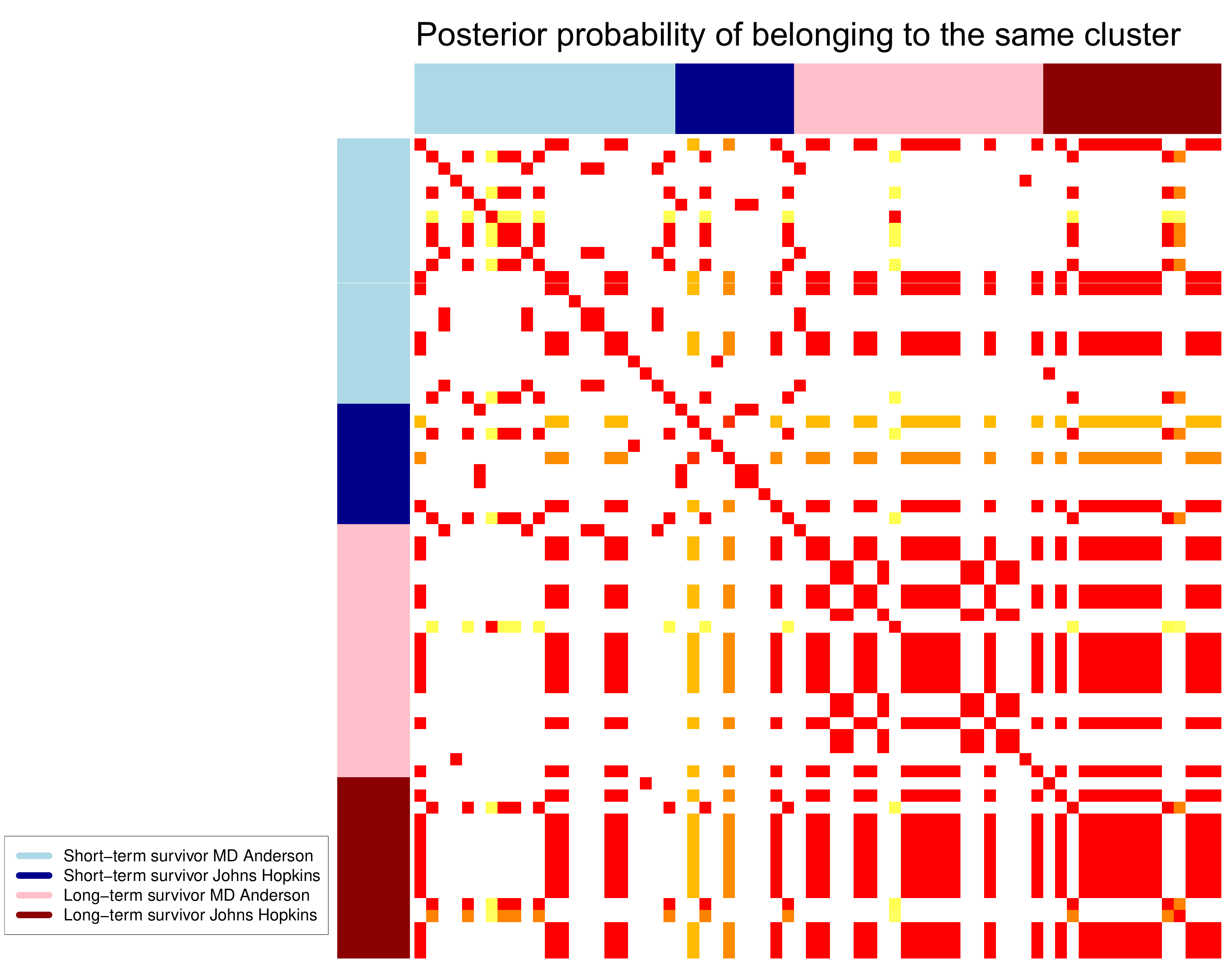}
	\end{subfigure}
\end{figure}
\newpage
\begin{figure}[H]\ContinuedFloat
	\begin{subfigure}{\linewidth}
		\caption{Result using MFMDTM}
		\includegraphics*[width=0.9\textwidth]{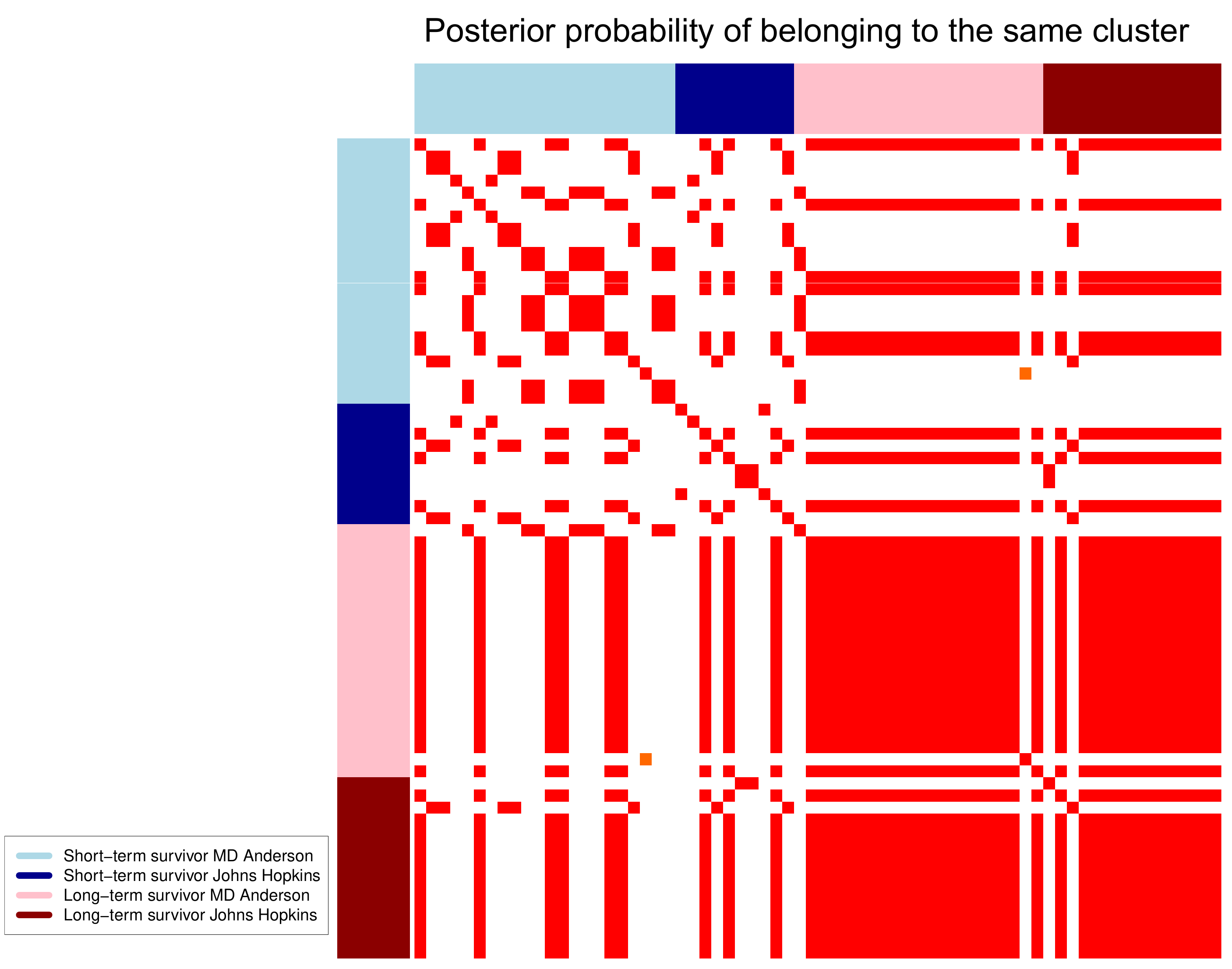}
	\end{subfigure}
	\caption{Heatmaps showing the posterior probability of being assigned to the same cluster for all the patients in the Riquelme et~al.\ (2019) data using the MFMDM and MFMDTM models}
\end{figure}

\begin{figure}[H]
	\includegraphics*[width=\textwidth]{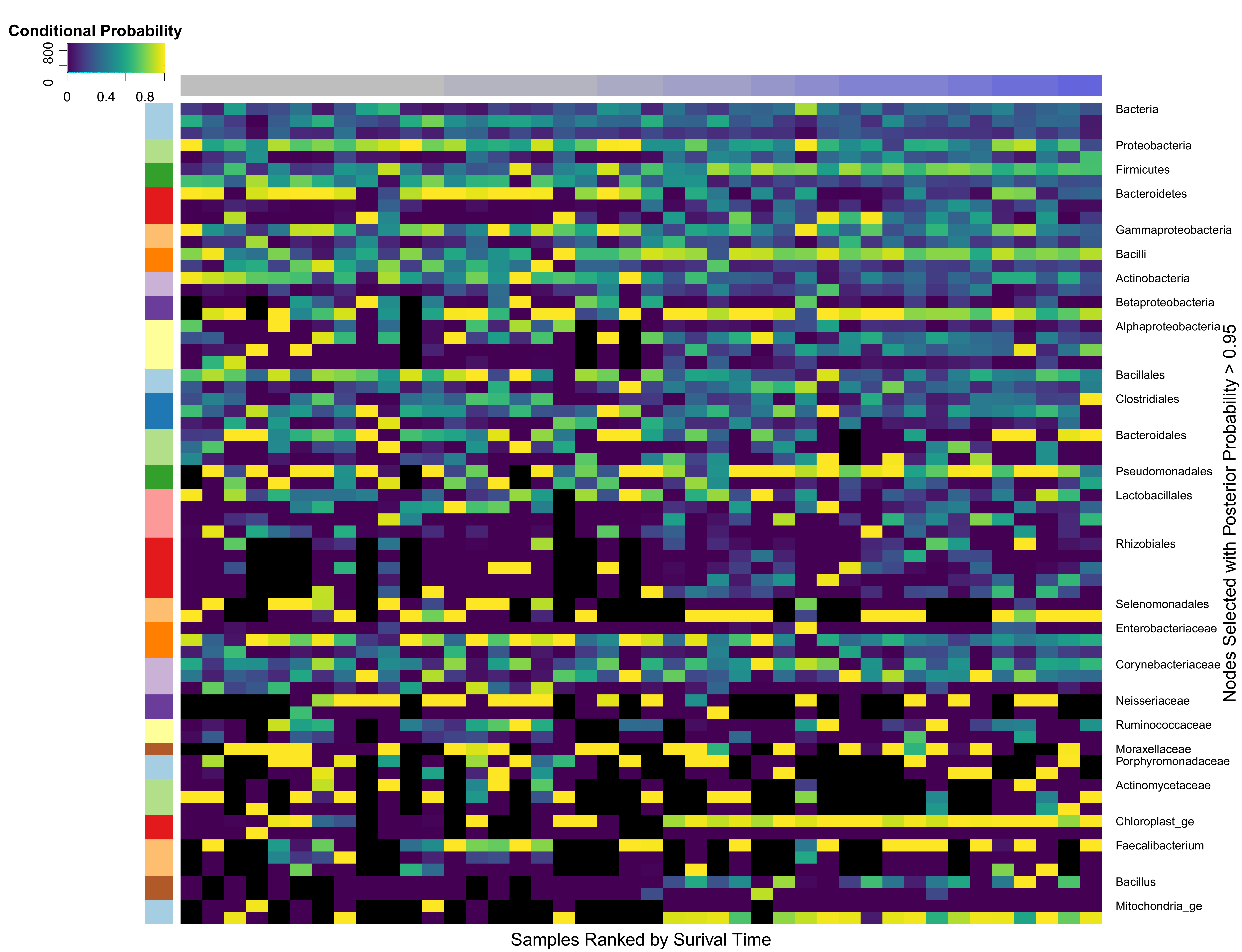}
	\caption{Conditional probability of allocating a count to child nodes for selected nodes in the \cite{RiquZhan19} data}
\end{figure}

\begin{figure}[H]
	\includegraphics*[width=0.3\textwidth]{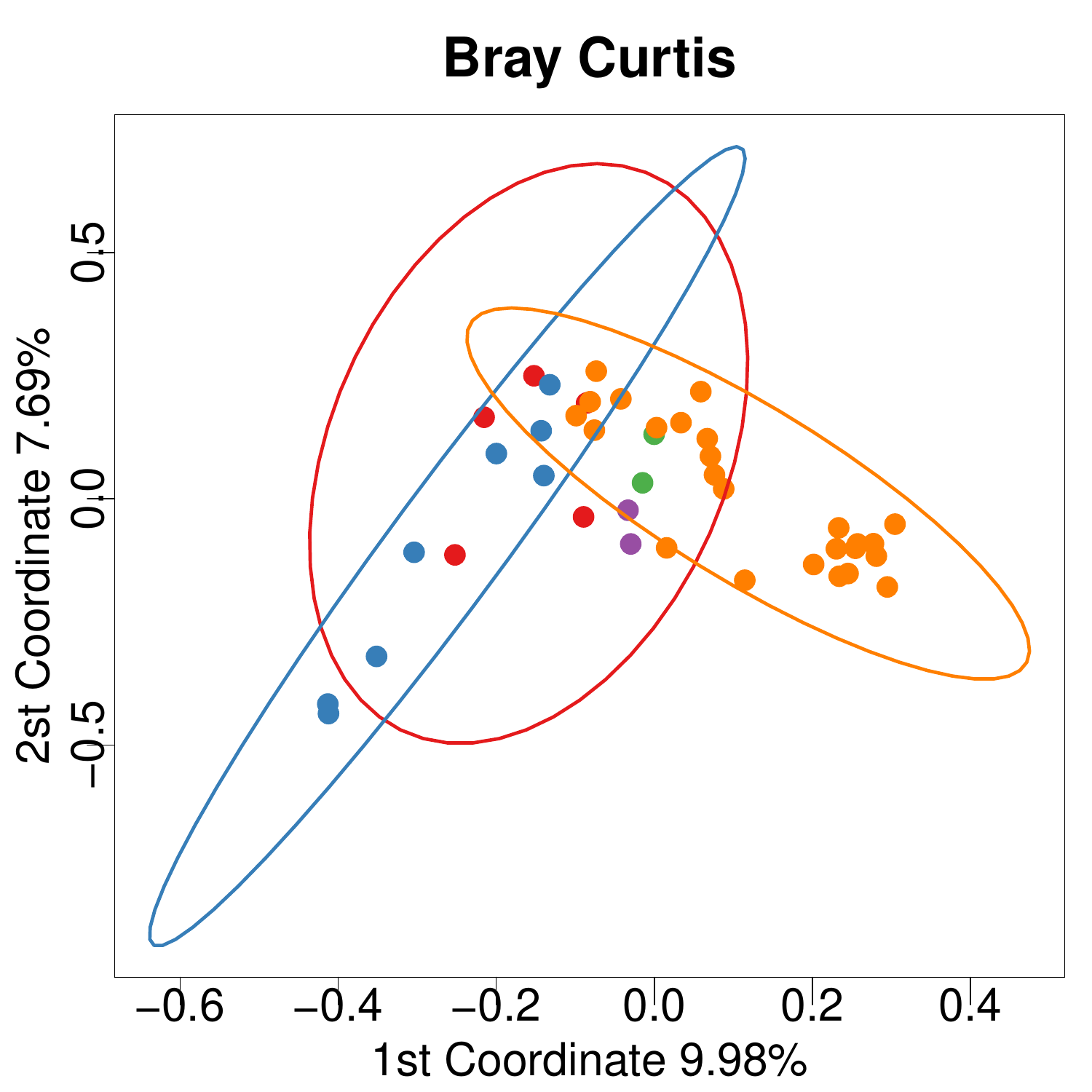}
	\includegraphics*[width=0.3\textwidth]{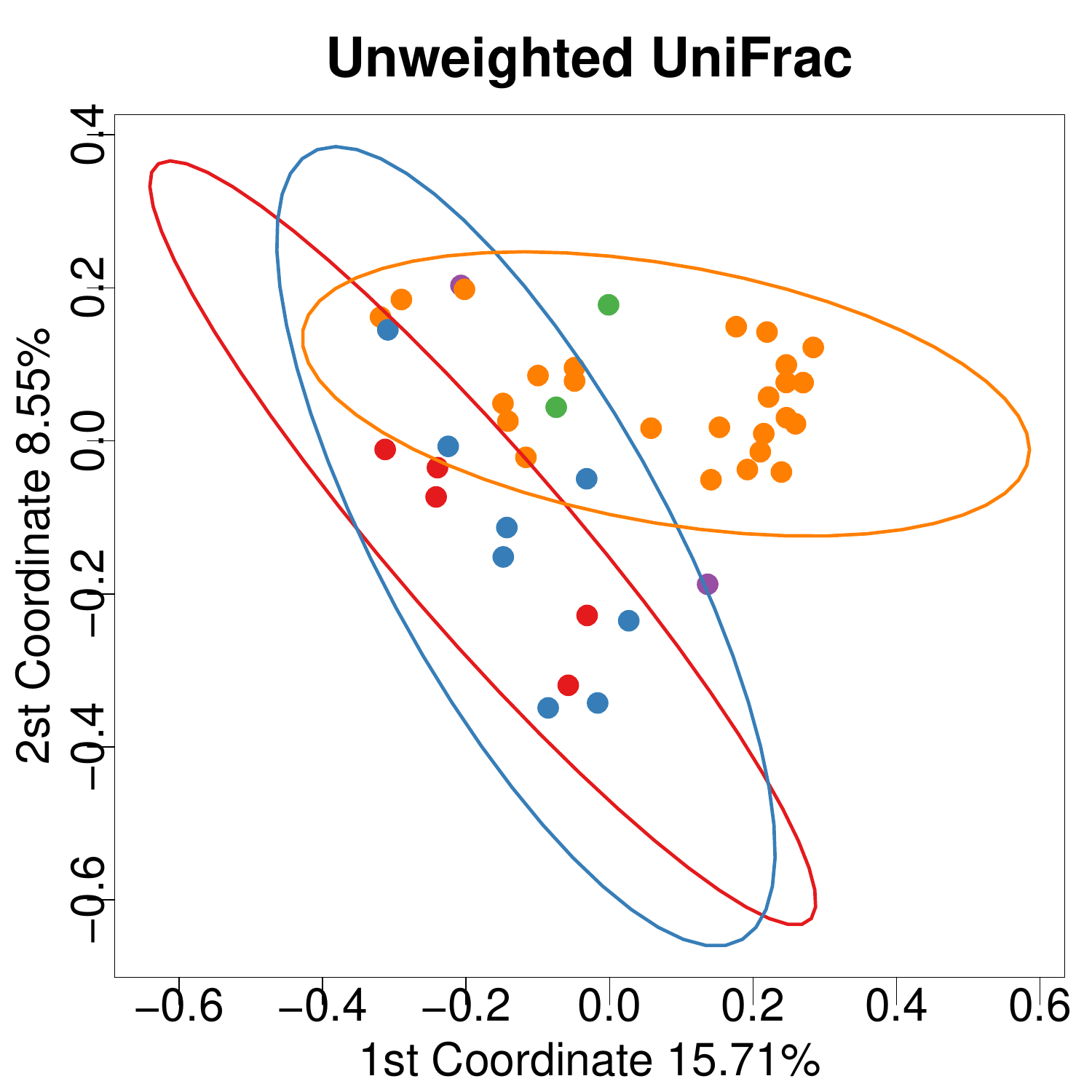}
	\includegraphics*[width=0.3\textwidth]{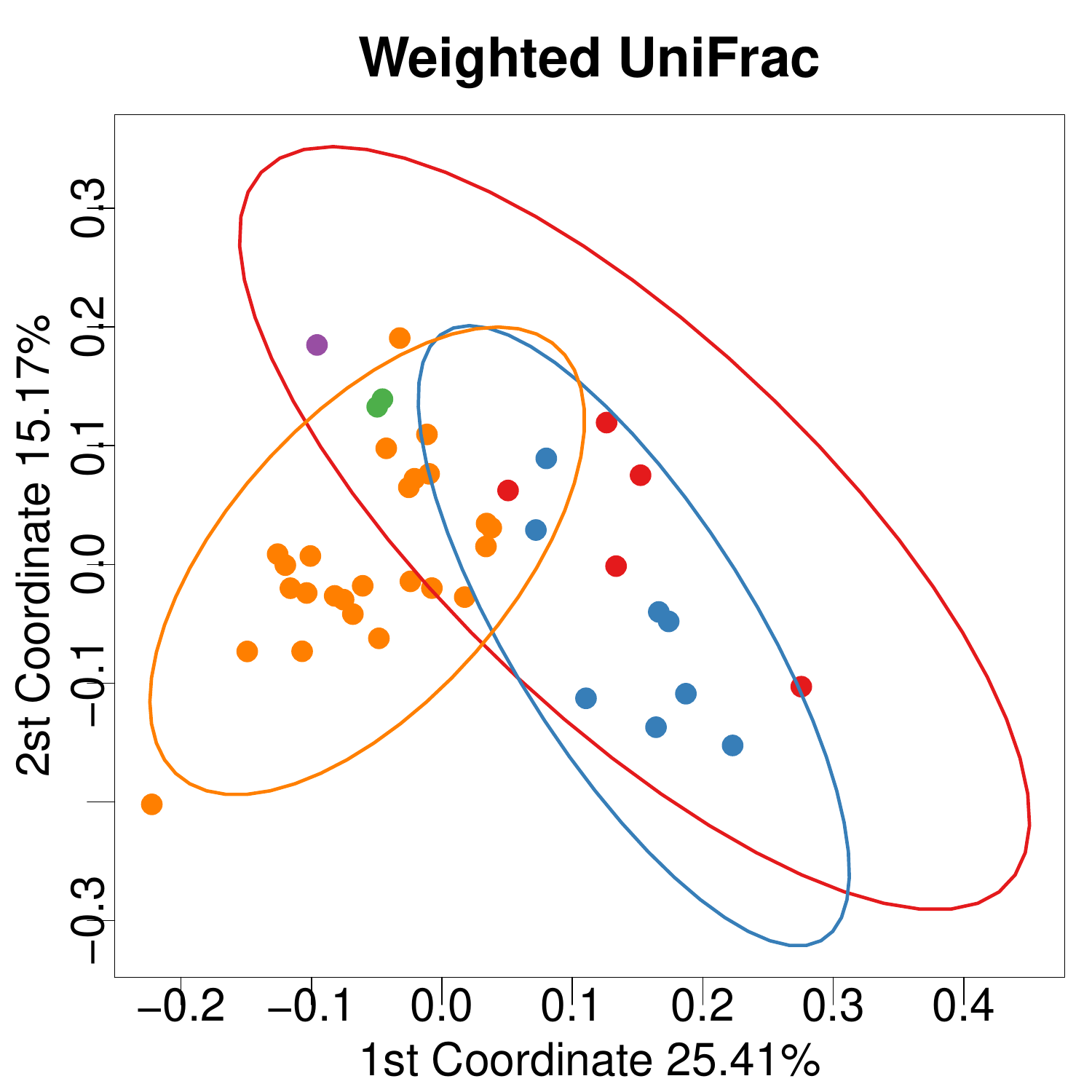}
	\caption{Principal coordinates analysis (PCoA) plots of the samples colored by the cluster assignments derived from the posterior of MFMDTM under three metrics}
\end{figure}


{\color{black}
	\section{Sensitivity analysis}
	In the plots provided in this section, the blue bars indicate the hyperparameter settings used in the main text, while the gray bars represent alternative hyperparameter settings.
	\subsection{Sensitivity analysis on $\beta_1$ and $\beta_2$ for the MFMDM model}
	In our MFMDM model, $w\sim \mathrm{Beta}(\beta_1,\beta_2).$ We conduct sensitivity analysis for different choices of $\beta_1$ and $\beta_2$.}

\begin{figure}[H]
	\includegraphics*[width=0.33\textwidth]{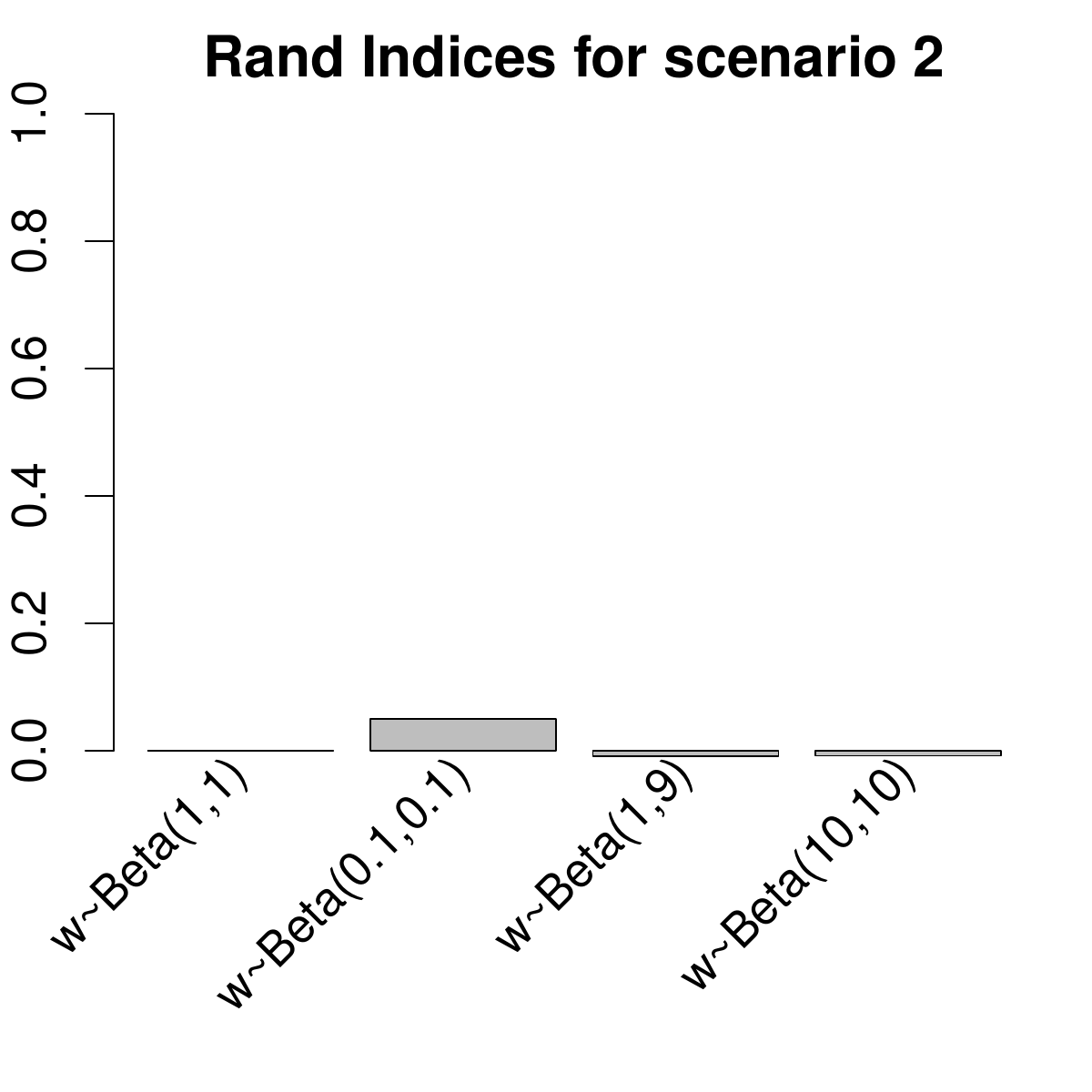}
	\includegraphics*[width=0.33\textwidth]{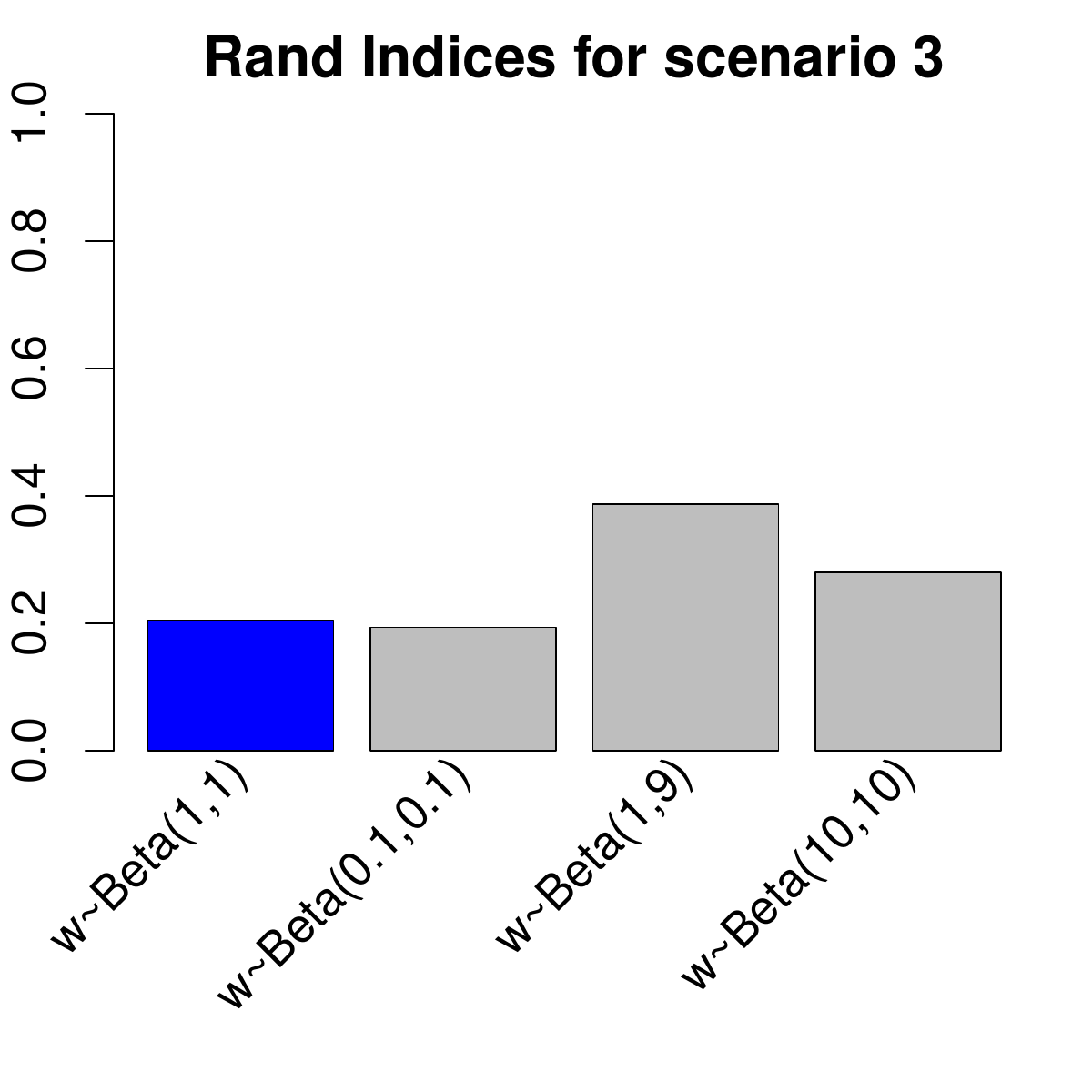}
	\includegraphics*[width=0.33\textwidth]{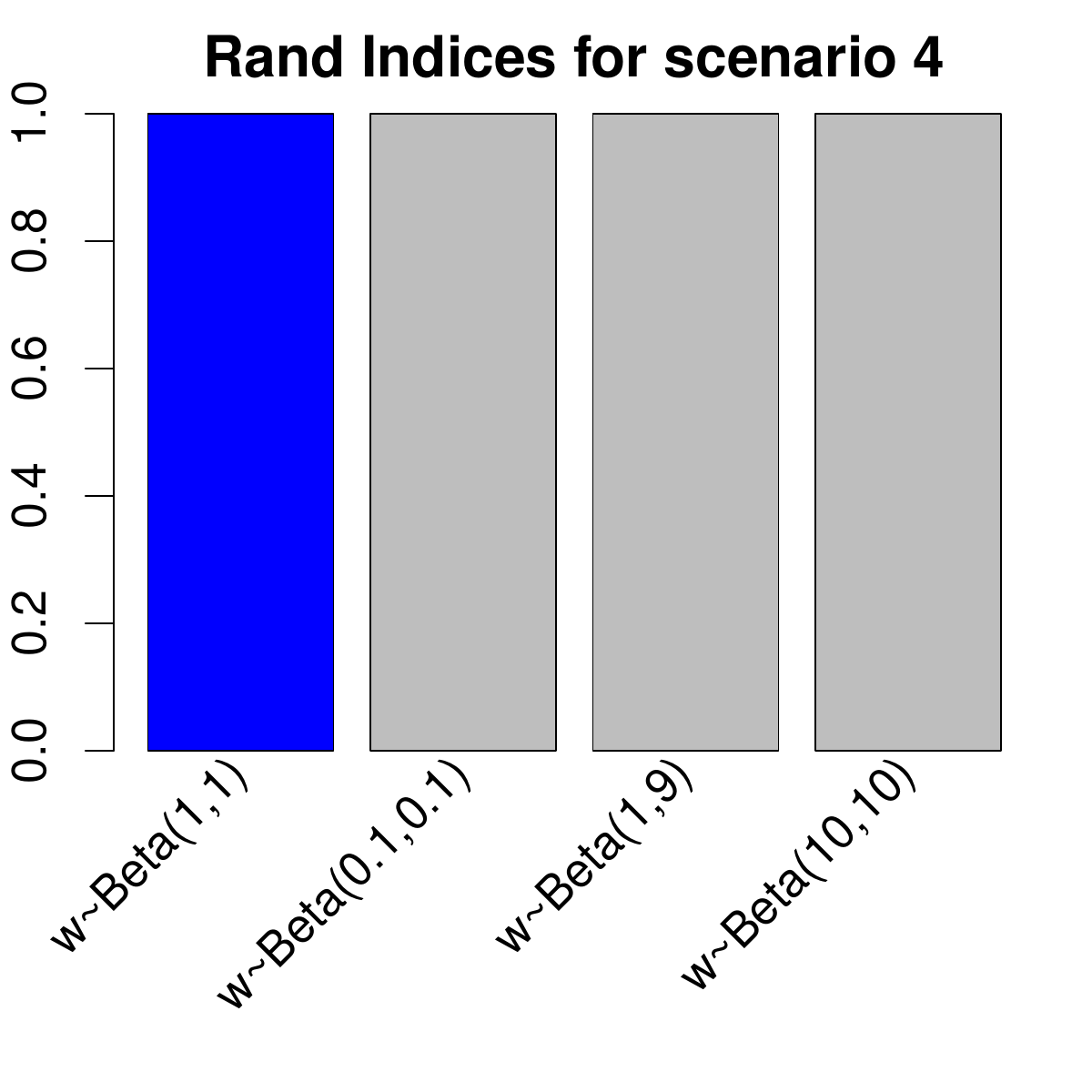}
	\caption{\textcolor{black}{Sensitivity analysis: Clustering performance using different $\beta_1$ and $\beta_2$ values}}
	\label{beta1beta2}
\end{figure}

\noindent {\color{black}As shown in Figure \ref{beta1beta2}, clustering performance is similar with different $\beta_1$ and $\beta_2$ values across simulation scenarios with varying degrees of cluster separation.}

{\color{black}
	\subsection{Sensitivity analysis on $\alpha$}
	The choice of $\alpha=1$ for the Dirichlet (tree) distribution corresponds to the frequentists' likelihood based inference, which is uniform over all the points in the support. For the Dirichlet distribution, ``$\alpha>1$" prefers evenly distributed count allocation, while ``$\alpha<1$" prefers unbalanced count allocation.  To compare the performance of different choices of $\alpha$ value, we run sensitivity analysis based on one random dataset from Scenario 2 and one random dataset from Scenario 4 using Dirichlet tree multinomial mixture (MFMDTM).
	
	As shown in Figure \ref{alphaRand}, the choice of $\alpha$ does not impact clustering performance when clusters are more separated (Scenario 4).  However, when clusters are less separated (Scenario 2), values close to $1$ give better performance. For too large (2 or 10) or too small (0.1) values of $\alpha$, the MFMDTM method clusters all the observations into the same group.
	\setcounter{figure}{6}    
	\begin{figure}[H]
		\includegraphics*[width=0.49\textwidth]{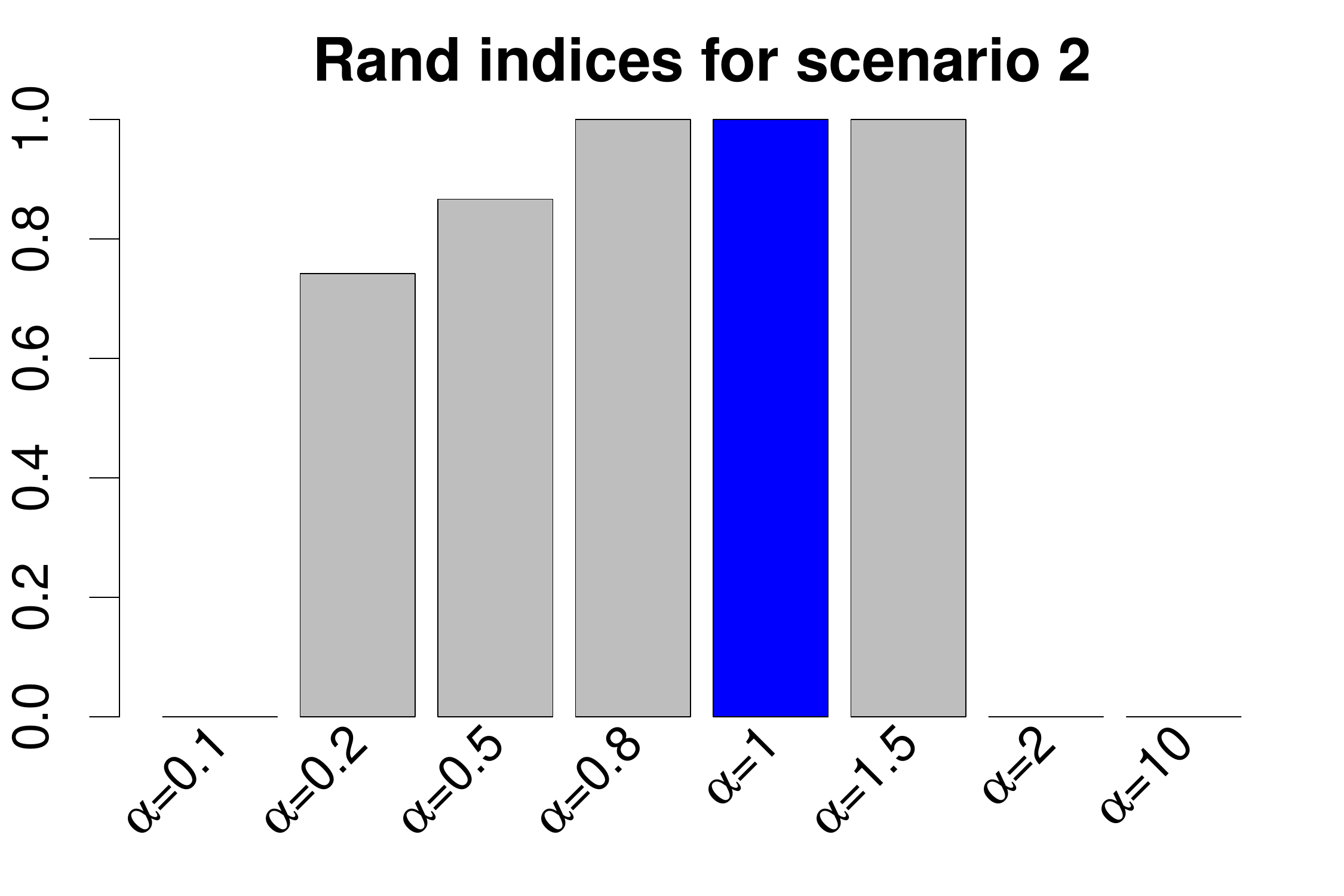}
		\includegraphics*[width=0.49\textwidth]{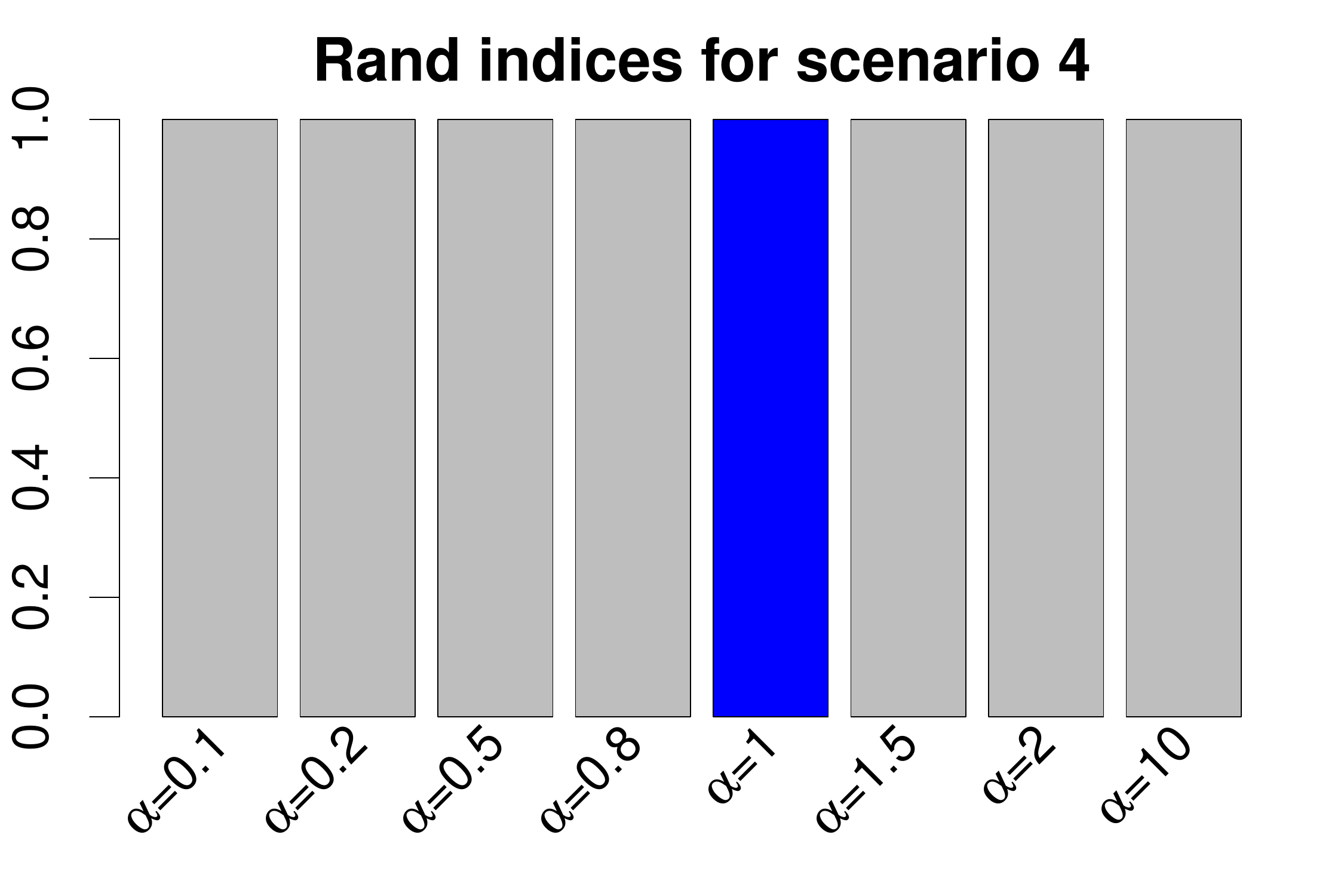}
	\end{figure}
	
	\begin{figure}[H]\ContinuedFloat
		\includegraphics*[width=0.49\textwidth]{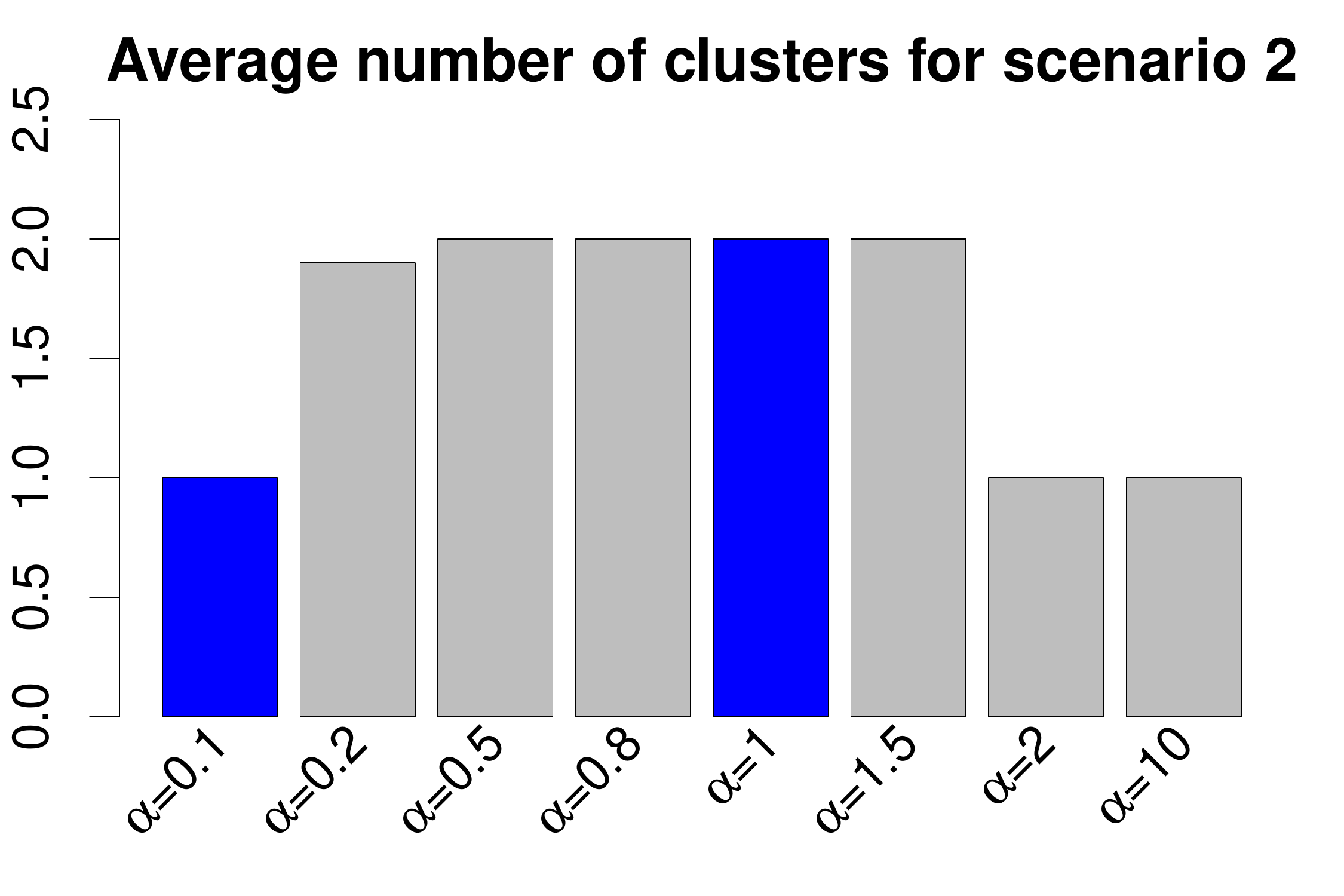}
		\includegraphics*[width=0.49\textwidth]{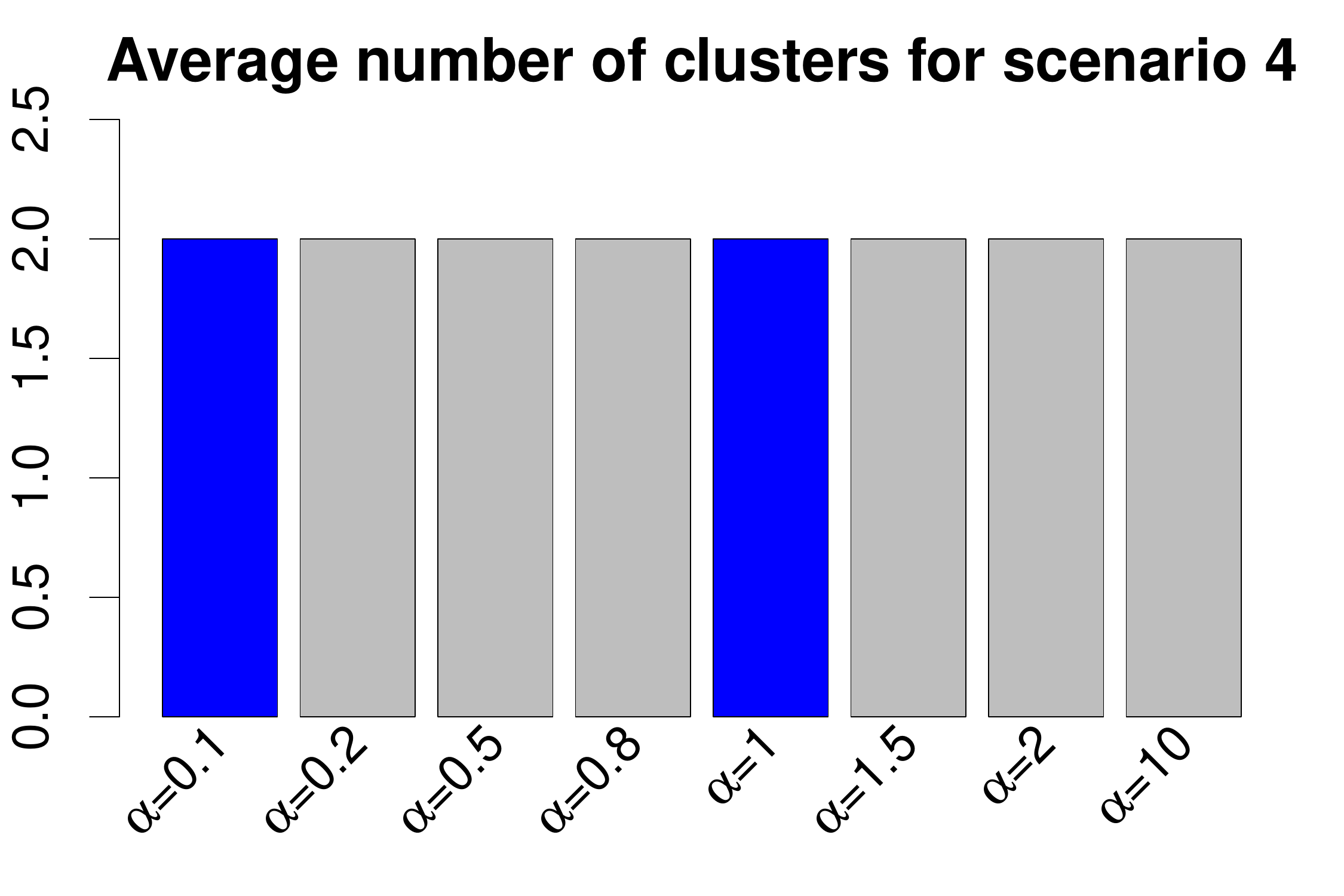}
		\caption{Sensitivity Analysis: Clustering performance using different $\alpha$ values}
		\label{alphaRand}
	\end{figure}
	
	\subsection{Sensitivity analysis on $w$, $\eta$ and $p_m$}
	We also conduct sensitivity analysis for other parameters. We vary $w$, $\eta$ and the distribution of potential number of clusters $p_m $ one at a time, while fixing other parameters to $\eta=1$, $w=0.5$ and $p_m-1\sim \mathrm{Poisson}(1)$ (the setting used in the main text). We also consider adding more flexibility on $w$ by giving it a conjugate prior $w\sim \mathrm{Beta}(a,b).$ The add-delete-swap feature selection algorithm has an easy implementation with this hyperprior.
	
	In the MFMDTM model, consider the situation where $s$ nodes are selected and $d-s$ nodes are not selected. The marginal distribution of $\boldsymbol{\gamma}$ is
	\begin{equation*}
		\begin{split}
			P(\boldsymbol{\gamma})&=\int_{0}^{1}w^s(1-w)^{d-s}\frac{\Gamma(\beta_1+\beta_2)}{\Gamma(\beta_1)\Gamma(\beta_2)}w^{\beta_1}(1-w)^{\beta_2} dw\\
			&=\int_{0}^{1}w^{(s+\beta_1)}(1-w)^{d-s+\beta_2}\frac{\Gamma(\beta_1+\beta_2)}{\Gamma(\beta_1)\Gamma(\beta_2)} dw\\
			&=\frac{\Gamma(\beta_1+\beta_2)}{\Gamma(\beta_1)\Gamma(\beta_2)}\frac{\Gamma(\beta_1+s)\Gamma(\beta_2+d-s)}{\Gamma(\beta_1+\beta_2+d)}.
		\end{split}
	\end{equation*}
	
	\noindent
	\textbf{Swap} Because the number of selected features is not changed, then 
	$$ 	 \frac{P(\boldsymbol{\gamma}^{new})}{P(\boldsymbol{\gamma}^{old})}=1.
	$$
	
	\noindent
	\textbf{Add} The number of selected features increases by $1$, then
	\begin{equation*}
		\begin{split}
			\frac{P(\boldsymbol{\gamma}^{new})}{P(\boldsymbol{\gamma}^{old})}
			&=\frac{\Gamma(\beta_1+s+1)\Gamma(\beta_2+d-s-1)}{\Gamma(\beta_1+\beta_2+d)}
			\frac{\Gamma(\beta_1+\beta_2+d)}{\Gamma(\beta_1+s)\Gamma(\beta_2+d-s)}\\
			&=\frac{\Gamma(\beta_1+s+1)\Gamma(\beta_2+d-s-1)}{\Gamma(\beta_1+s)\Gamma(\beta_2+d-s)}\\
			&=\frac{\beta_1+s}{\beta_2+d-s-1}.\\
		\end{split}
	\end{equation*}
	
	\noindent
	\textbf{Delete} The number of selected features decreases by $1$, then
	\begin{equation*}
		\begin{split}
			\frac{P(\boldsymbol{\gamma}^{new})}{P(\boldsymbol{\gamma}^{old})}
			&=\frac{\Gamma(\beta_1+s-1)\Gamma(\beta_2+d-s+1)}{\Gamma(\beta_1+\beta_2+d)}
			\frac{\Gamma(\beta_1+\beta_2+d)}{\Gamma(\beta_1+s)\Gamma(\beta_2+d-s)}\\
			&=\frac{\Gamma(\beta_1+s-1)\Gamma(\beta_2+d-s+1)}{\Gamma(\beta_1+s)\Gamma(\beta_2+d-s)}\\
			&=\frac{\beta_2+d-s}{\beta_1+s-1}.\\
		\end{split}
	\end{equation*}
	\begin{figure}[H]
		\includegraphics*[width=0.49\textwidth]{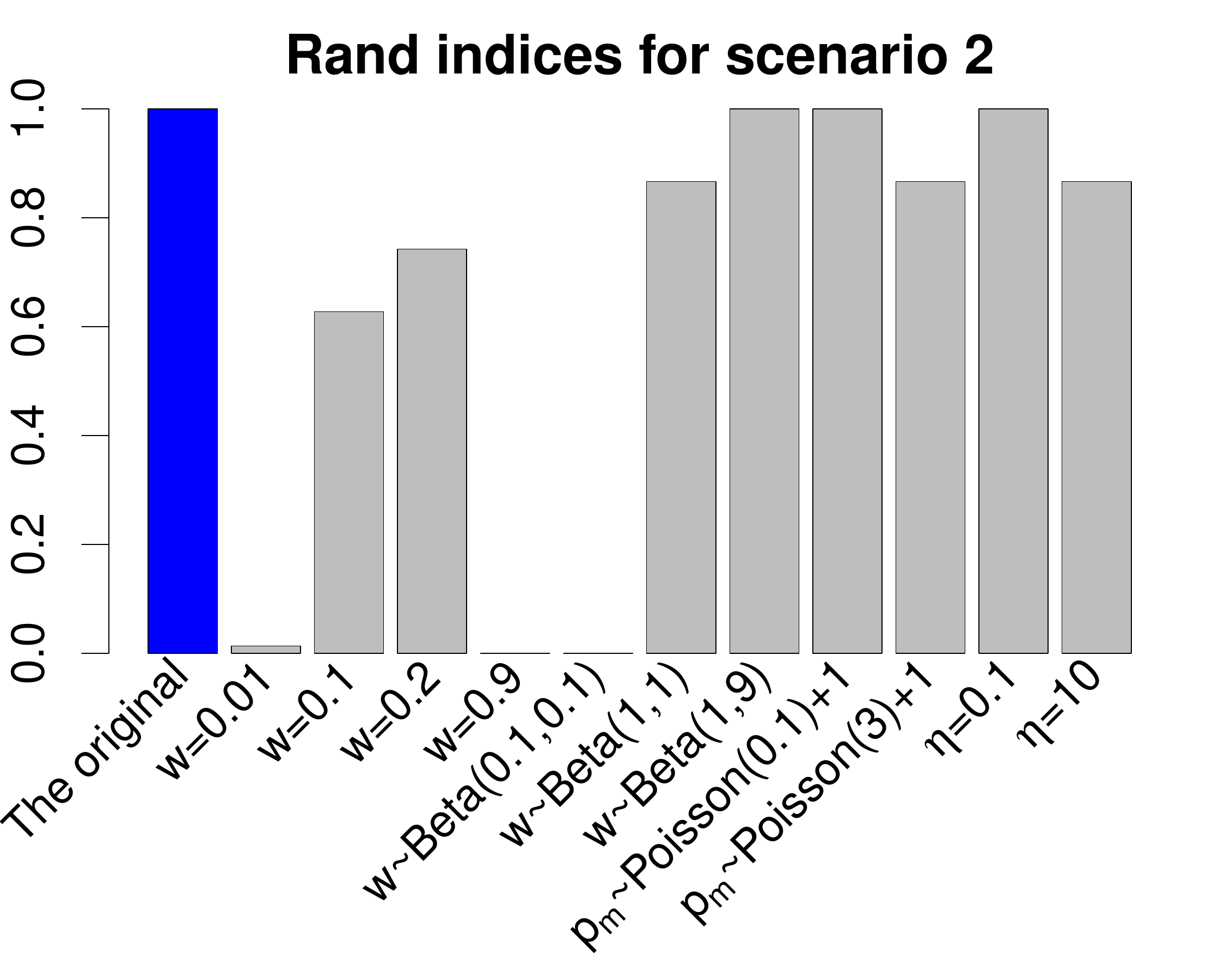}
		\includegraphics*[width=0.49\textwidth]{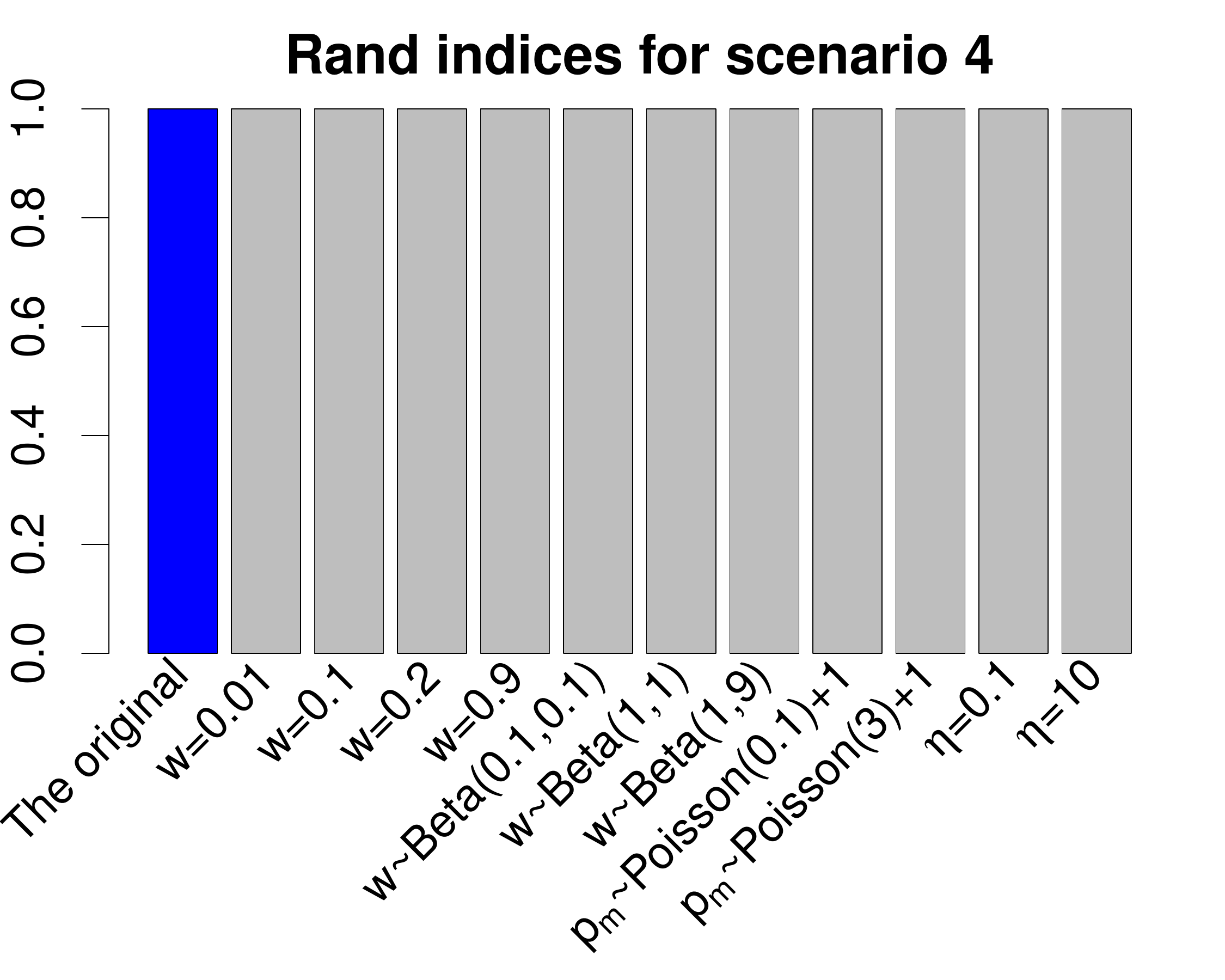}
		\includegraphics*[width=0.49\textwidth]{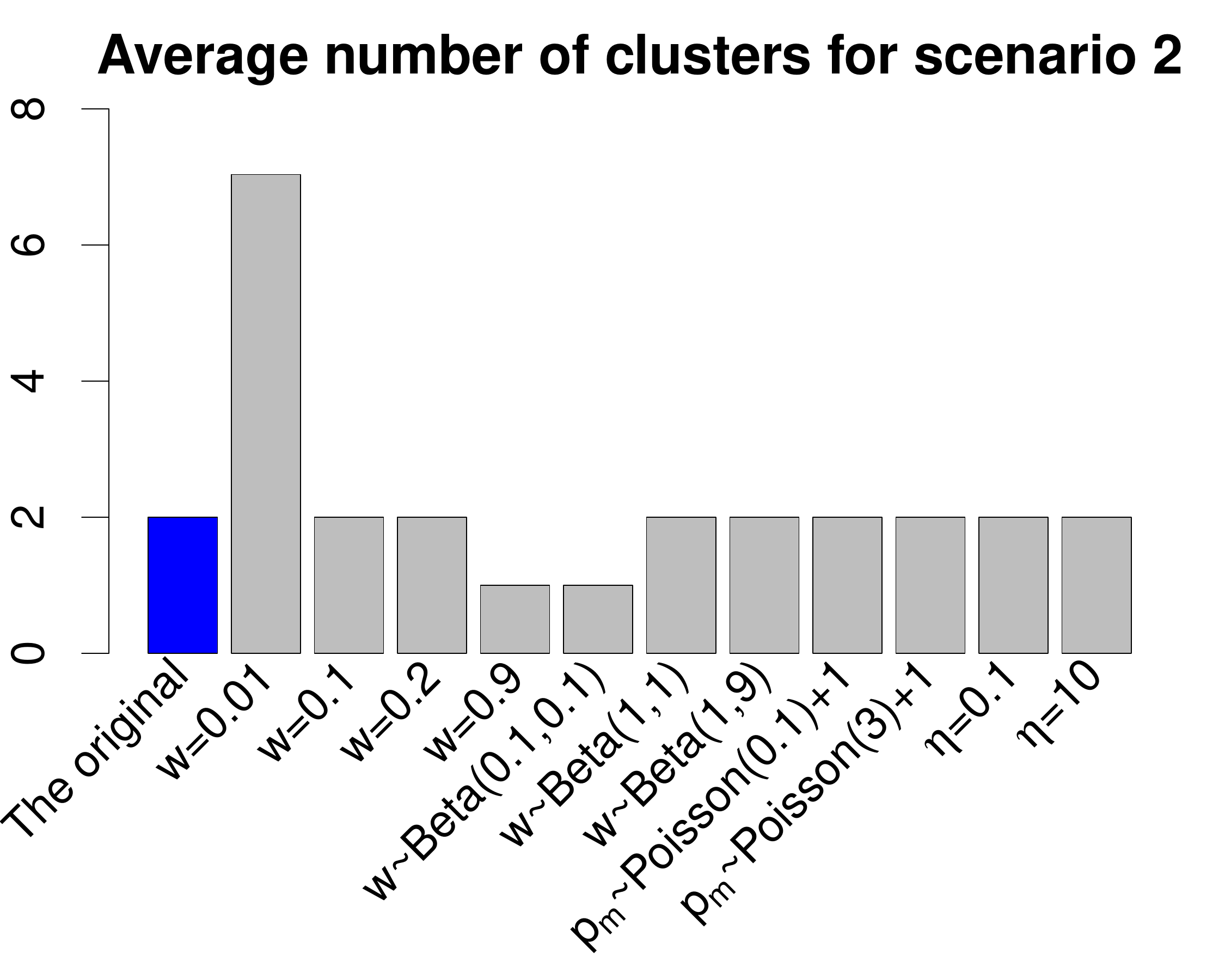}
		\includegraphics*[width=0.49\textwidth]{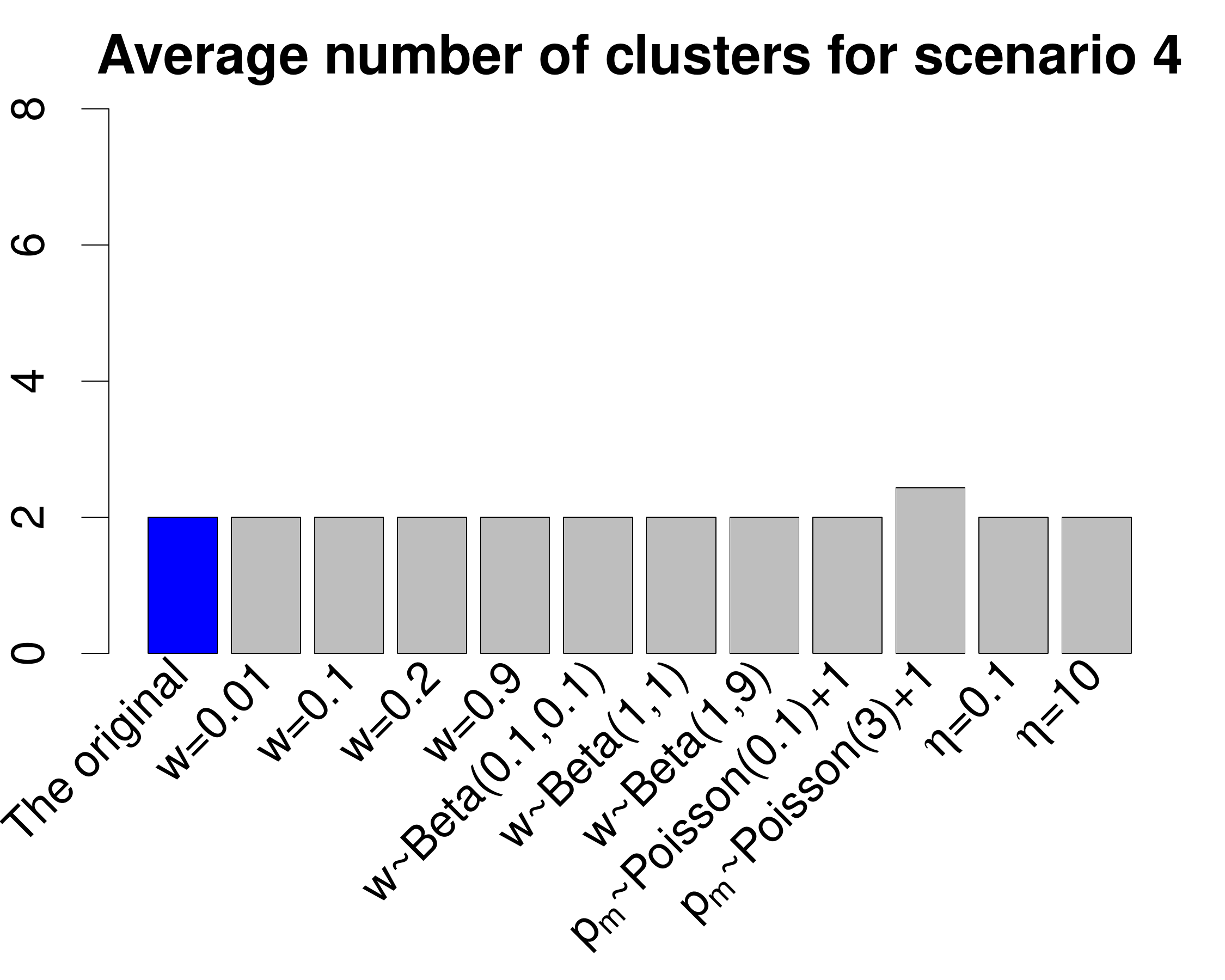}
		
		\caption{Sensitivity analysis: Clustering performance using different $w$, $\eta$ and $p_m.$}
		\label{varyRand}
	\end{figure}
	
	As shown in Figure \ref{varyRand}, when clusters are more separated (Scenario 4) the choice of parameters does not affect the clustering performance. When clusters are less separated (Scenario 2), the choices of $\eta$ and $p_m$ do not heavily influence the clustering performance. In the main text, we rely on the Poisson+1 distribution for $p_m$ due to its nice mathematical properties shown by \cite{MillHarr18}. Values of $w$ close to $0.5$ give better performance, whereas too large ($w$=0.9) or too small $w=0.01$ tend to give poor clustering performance. For the model with hierarchy on $w$, sensitivity analysis shows that less informative priors on $w$ ($\mathrm{Beta}(1,1)$, $\mathrm{Beta}(1,9)$) give similar clustering performance to the default choice, while priors favoring large or small $w$ values ($\mathrm{Beta}(0.1,0.1)$) give poor results. The default choice $w=0.5$ reflects indifference, in the sense that \textsl{a priori} a feature is considered equally likely to be an informative feature or a noisy feature.
}

{\color{black} 
	\section{Additional analysis on feature selection}
	\subsection{Expected false discovery rate}
	Using the method from \cite{NewtNoue04}, the expected FDR can be calculated given the marginal posterior probabilities of inclusion for all the features $\boldsymbol{\pi},$
	
	$$
	FDR=\frac{\sum_{i=1}^{|J|}(1-\pi_i)I(\pi_i\geqslant \kappa)}{\sum_{i=1}^{|J|} I(\pi_i	\geqslant \kappa)\vee 1},
	$$
	where $\kappa$ is the threshold for a feature to be considered as ``informative", $I$ is the indicator function, and $|J|$ is the number of features.
	\begin{figure}[H]
		\includegraphics*[width=0.49\textwidth]{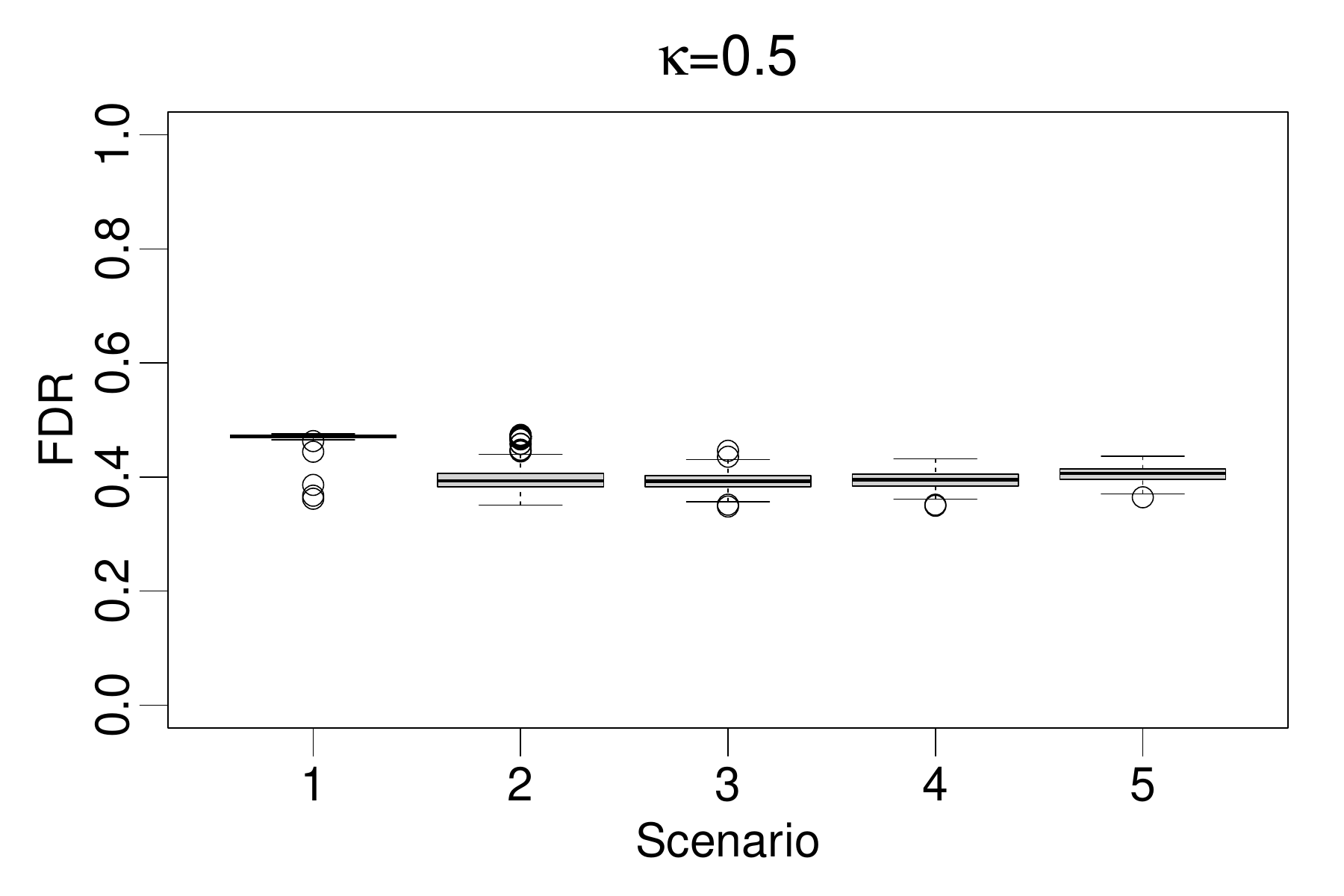}
		\includegraphics*[width=0.49\textwidth]{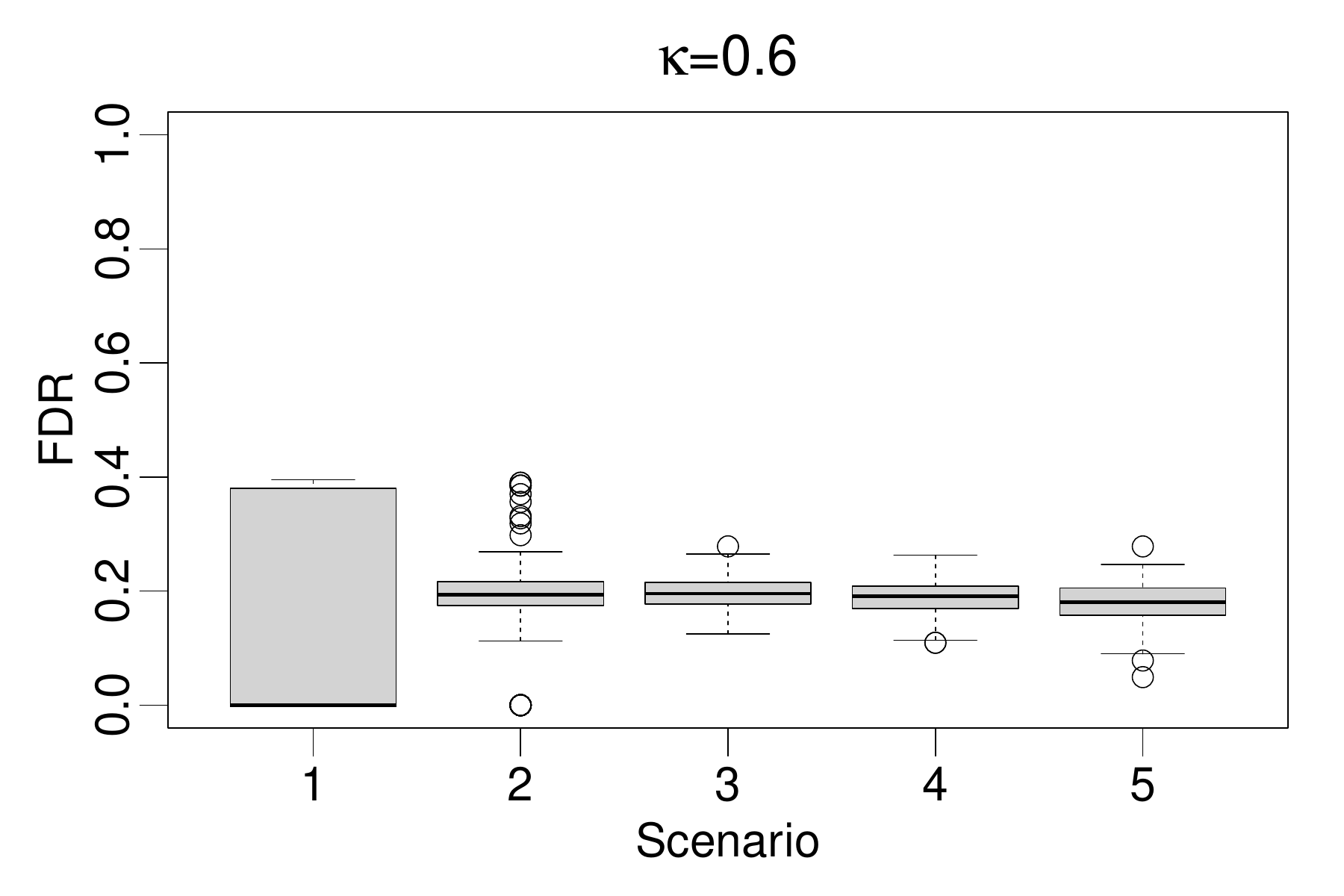}
		\includegraphics*[width=0.49\textwidth]{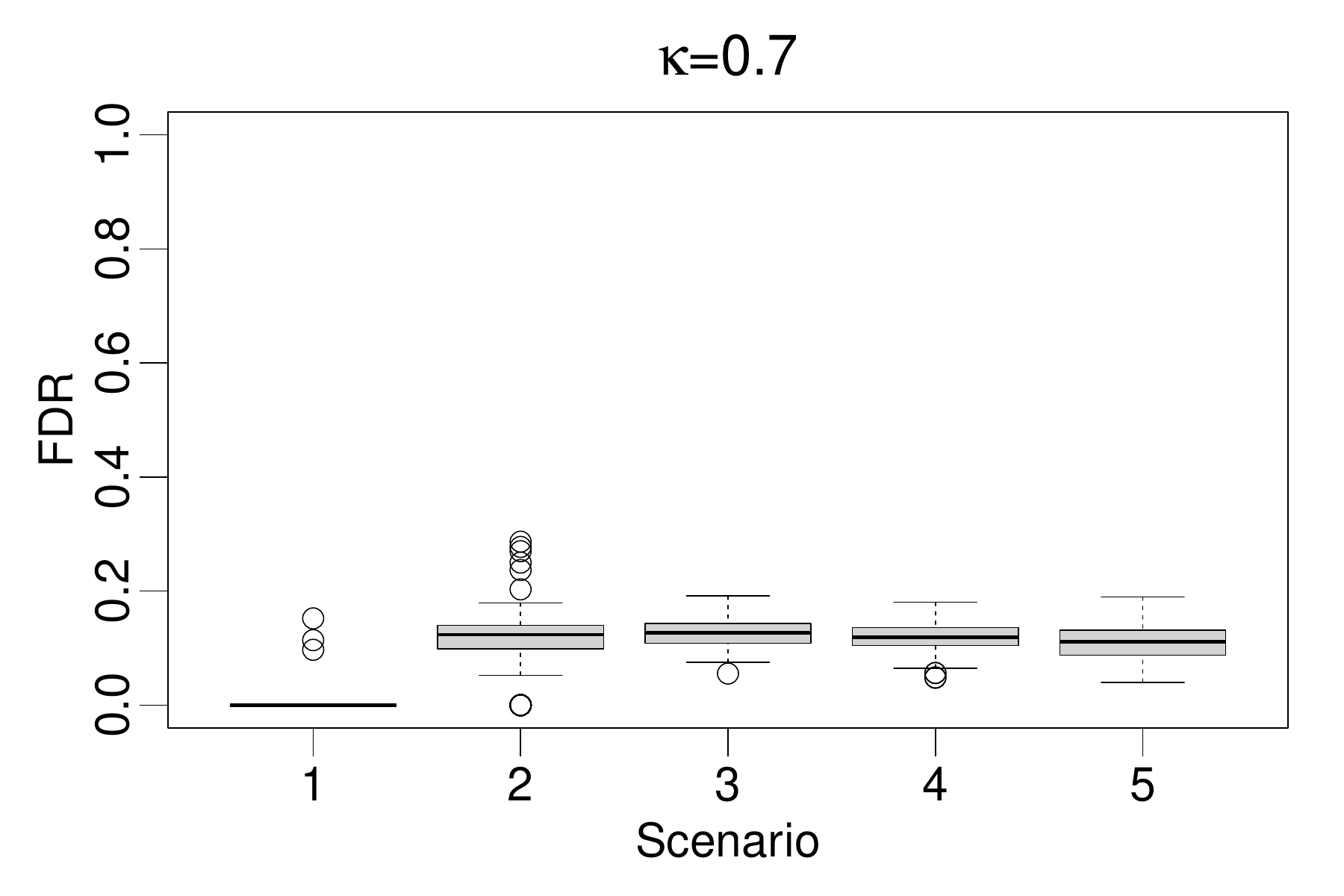}
		\includegraphics*[width=0.49\textwidth]{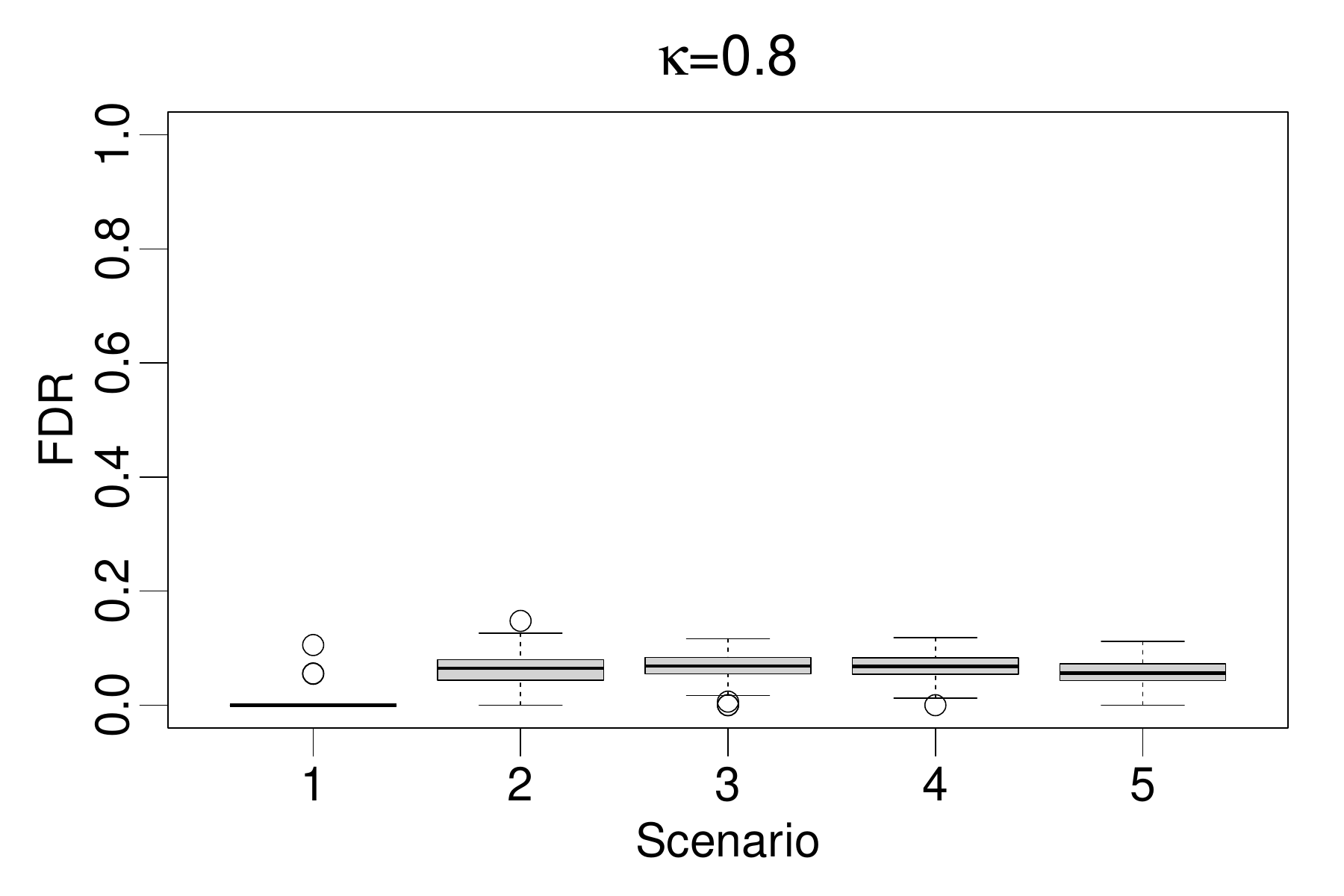}
		\includegraphics*[width=0.49\textwidth]{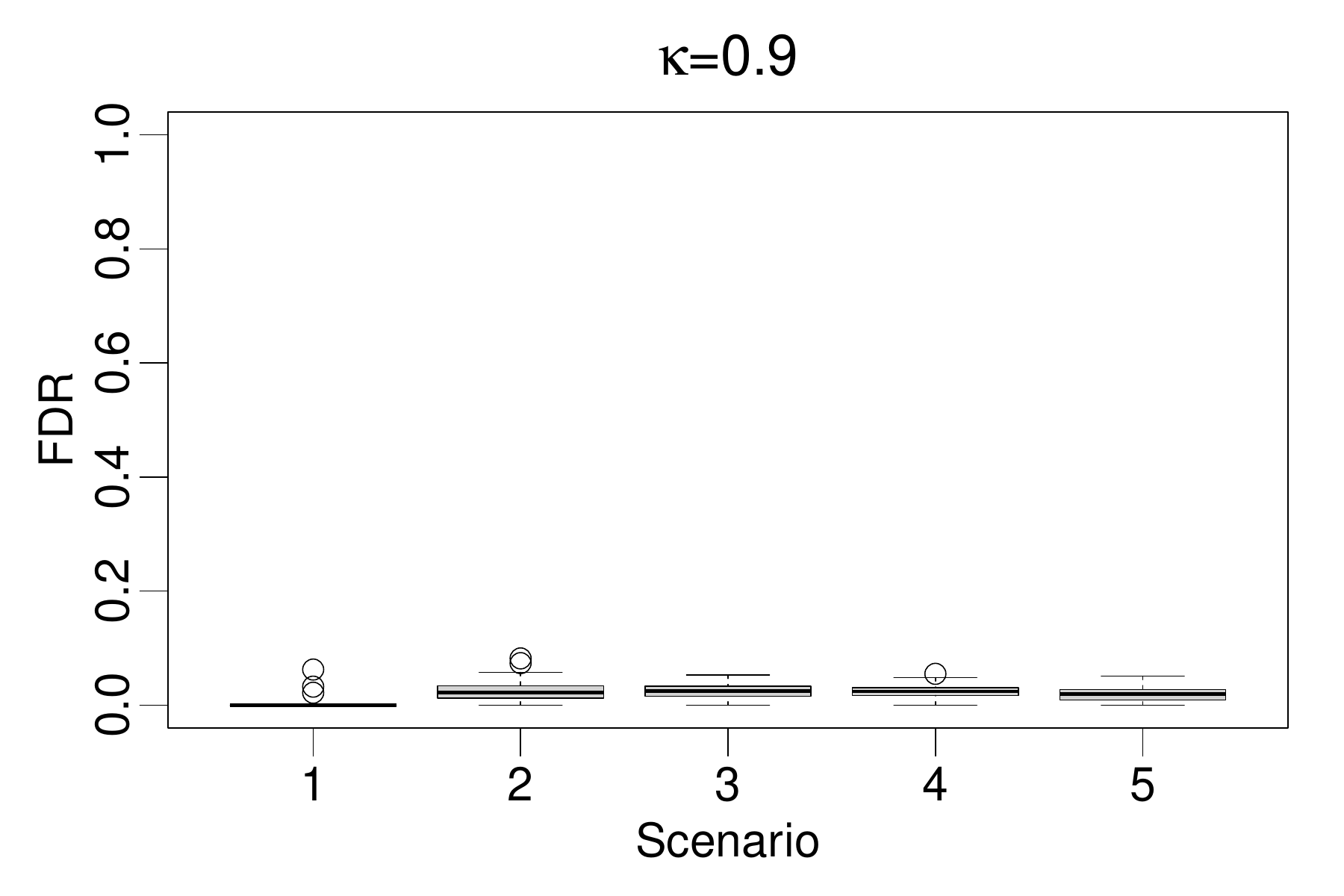}
		\includegraphics*[width=0.49\textwidth]{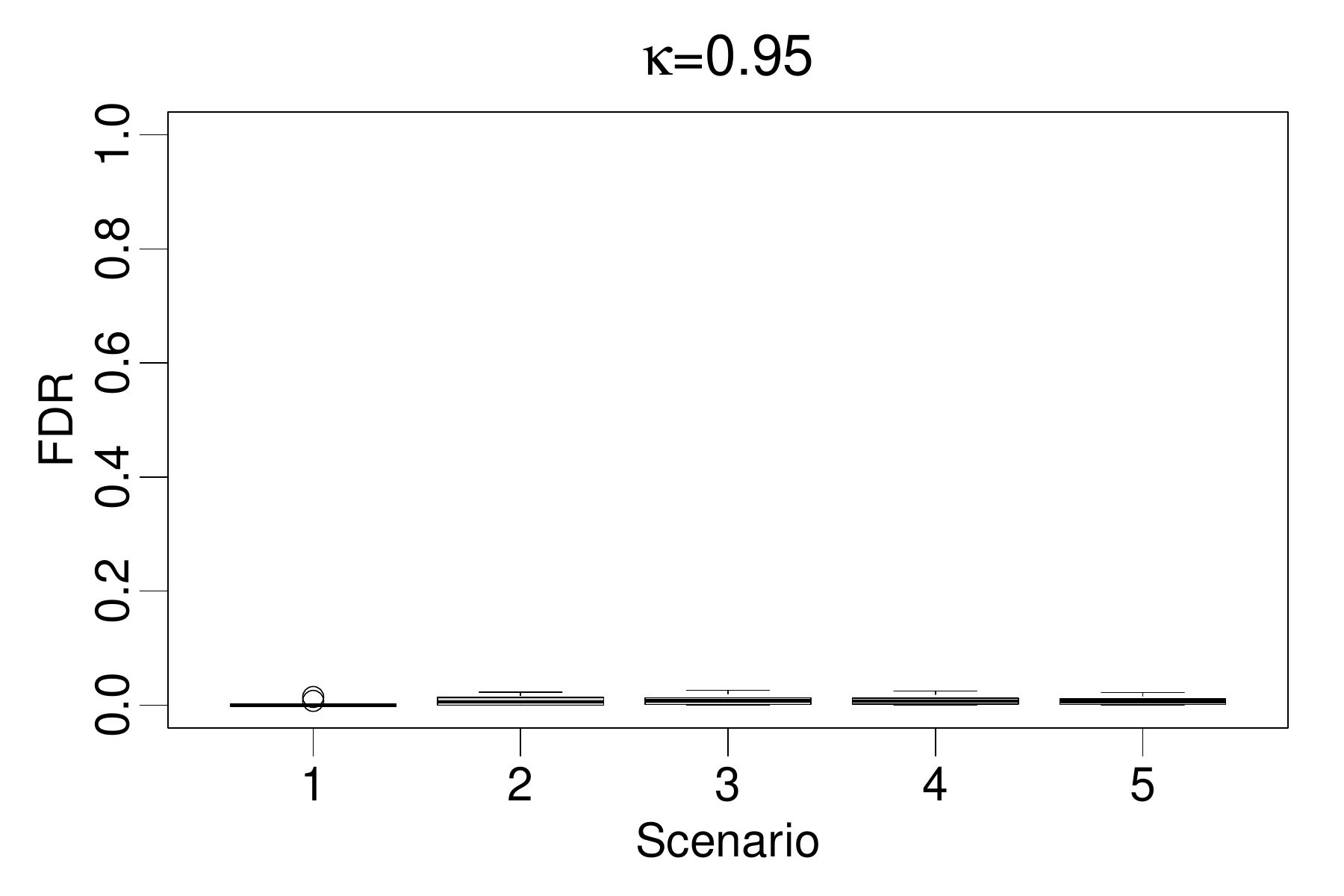}
		
		\caption{Expected false discovery rate for all scenarios with different thresholds}
		\label{FDR}
	\end{figure}
	
	
	We computed the expected FDR for the MFMDTM model with thresholds $\kappa$ equal to $0.5$, $0.6$, $0.7$, $0.8$, $0.9$ and $0.95$ for all five simulation scenarios. The boxplots given in Figure \ref{FDR} show that $\kappa\geq0.7$ controls the FDR under $0.2$ for all simulation scenarios. 
	
	\subsection{Impact of threshold for ``high abundance" OTUs}
	In reporting the feature selection performance of the MFMDM model on the simulated data (Section 4 of the main manuscript), 
	we focused on identifying ``high abundance" features, using a threshold of $0.001$,
	as microbiome data are highly zero-inflated and many OTUs have close to $0$ abundance. Those extremely rare OTUs do not contribute a lot of information in clustering and bring challenges to biological interpretation.
	To gauge the impact of this threshold, 
	we computed the AUC under different thresholds and show the mean AUC over 200 simulated datasets and the empirical $95\%$ confidence interval in Figure \ref{AUC}. 
	\begin{figure}[H]
		\includegraphics*[width=\textwidth]{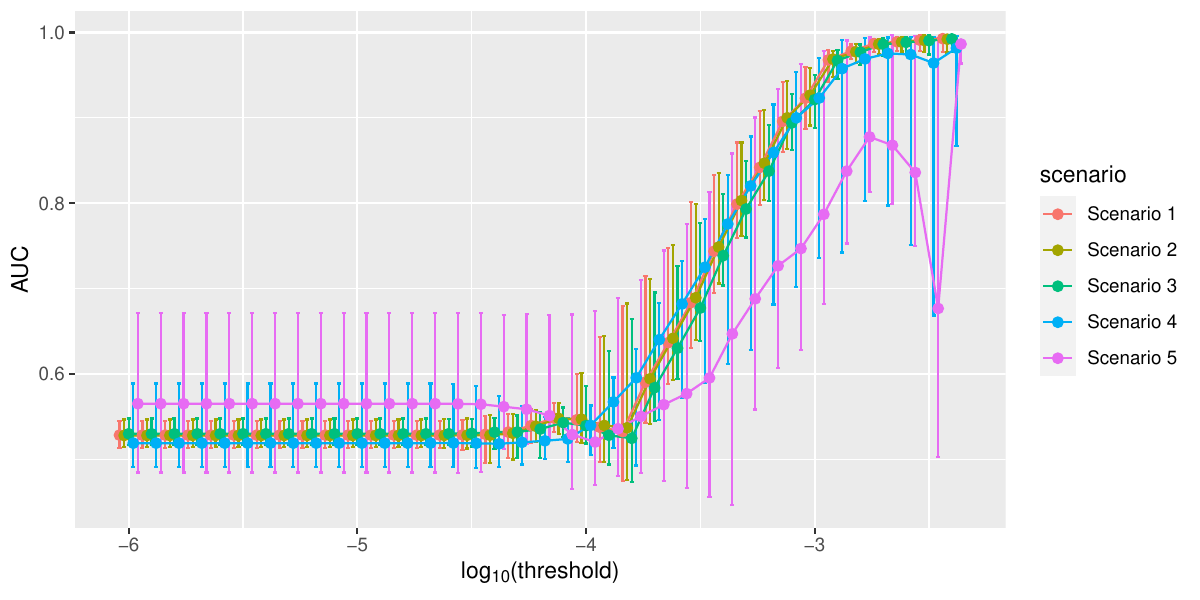}
		\caption{AUC under different thresholds}
		\label{AUC}
	\end{figure}
	
	Here, the x-axis shows the threshold on the common logarithm scale ($\log_{10}$). In general, the AUC increases with the threshold as the higher abundance OTUs tend to be more informative. The AUC curve for the easiest scenario (Scenario 5) is lower than other scenarios when the threshold is high, and it starts to decrease when the threshold passes $10^{-3.2}$. For the chosen threshold $0.001,$ we also plot the ROC curves for five scenarios. In Figure \ref{ROC}, each line represents an ROC curve from one dataset. For Scenarios 1-3, the ROC curves reflect consistently good performance. For Scenarios 4 and 5, some datasets show low specificity for reasonably high sensitivity values, which implies that the model selects some false positives. 
	\setcounter{figure}{10} 
	\begin{figure}[H]
		\includegraphics*[width=0.33\textwidth]{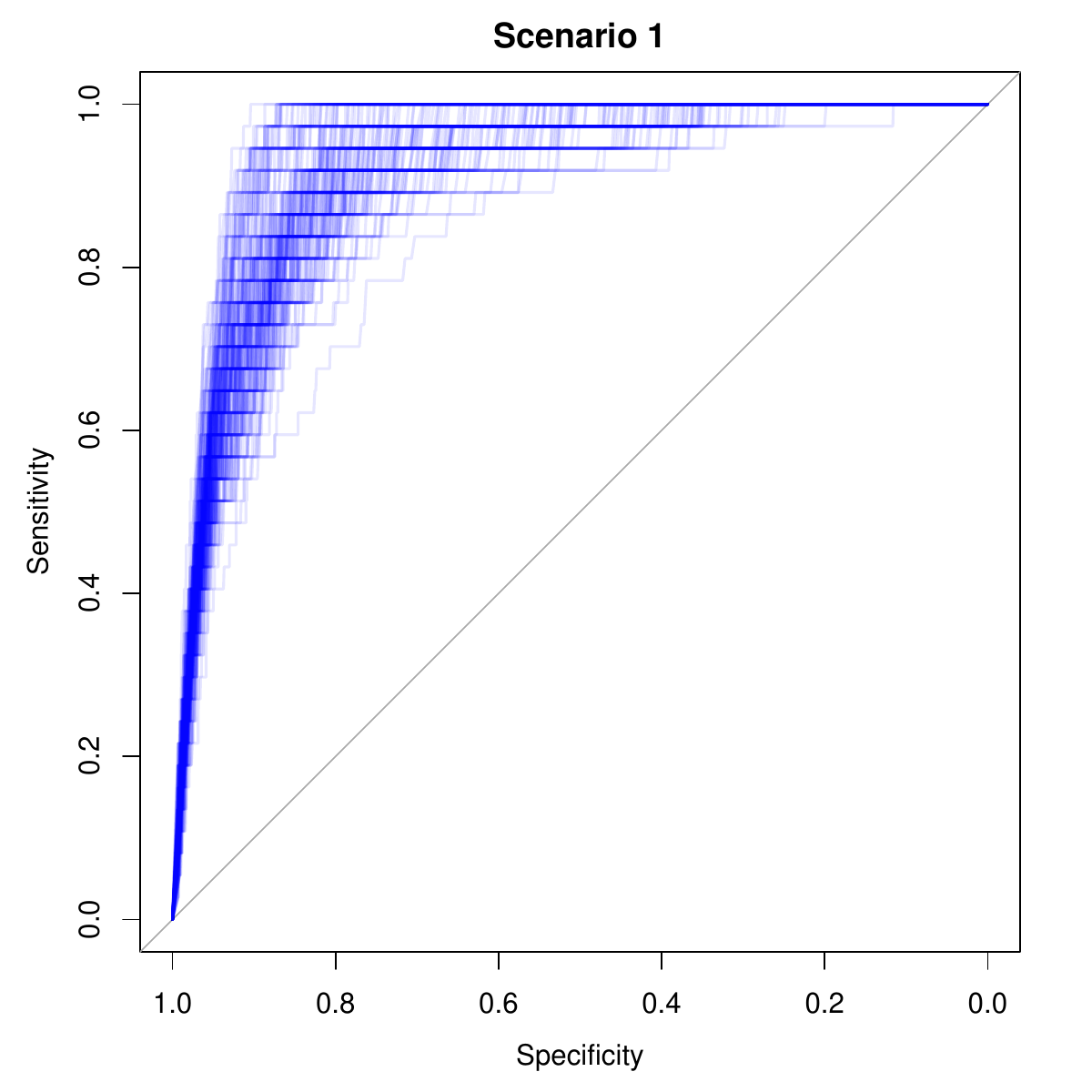}
		\includegraphics*[width=0.33\textwidth]{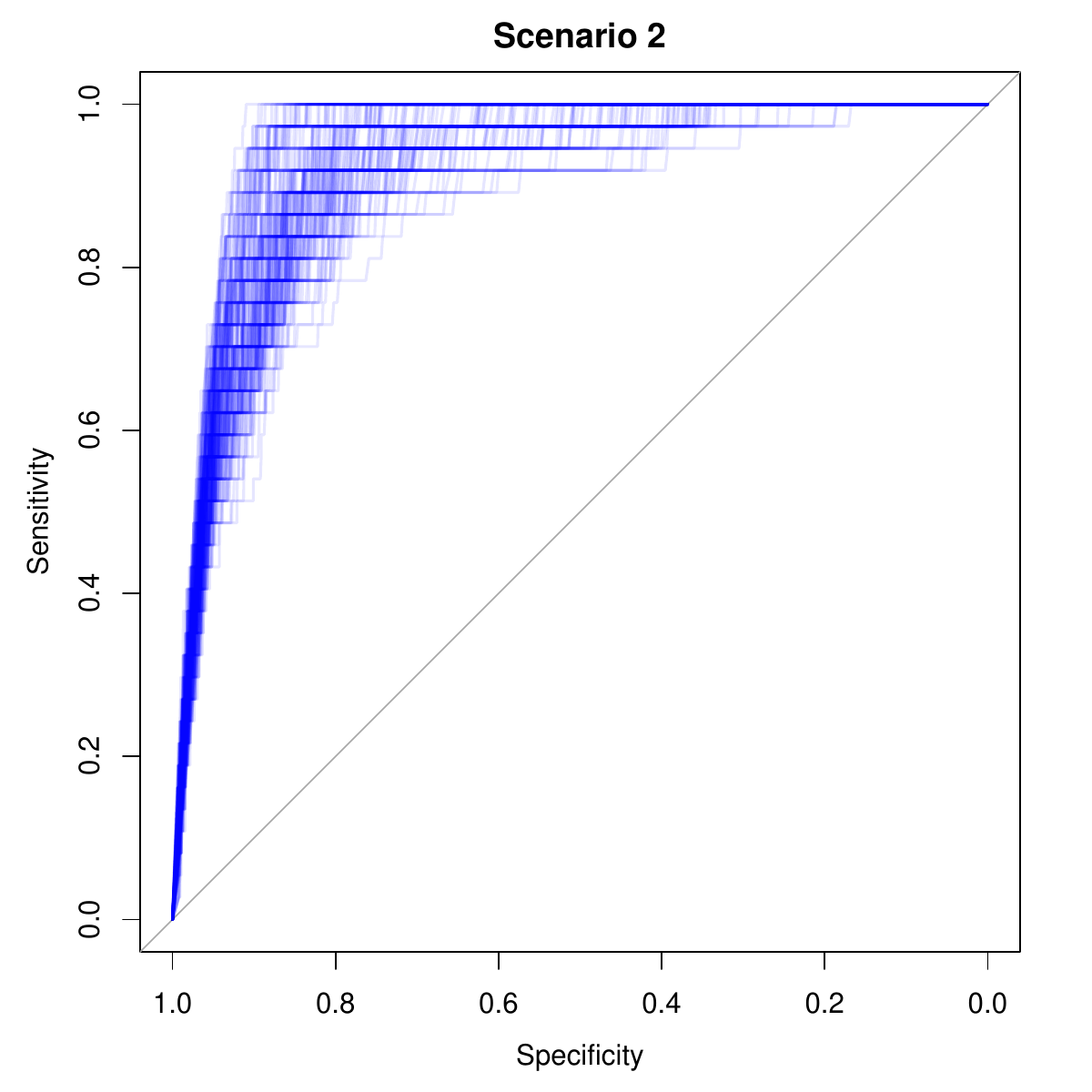}
		\includegraphics*[width=0.33\textwidth]{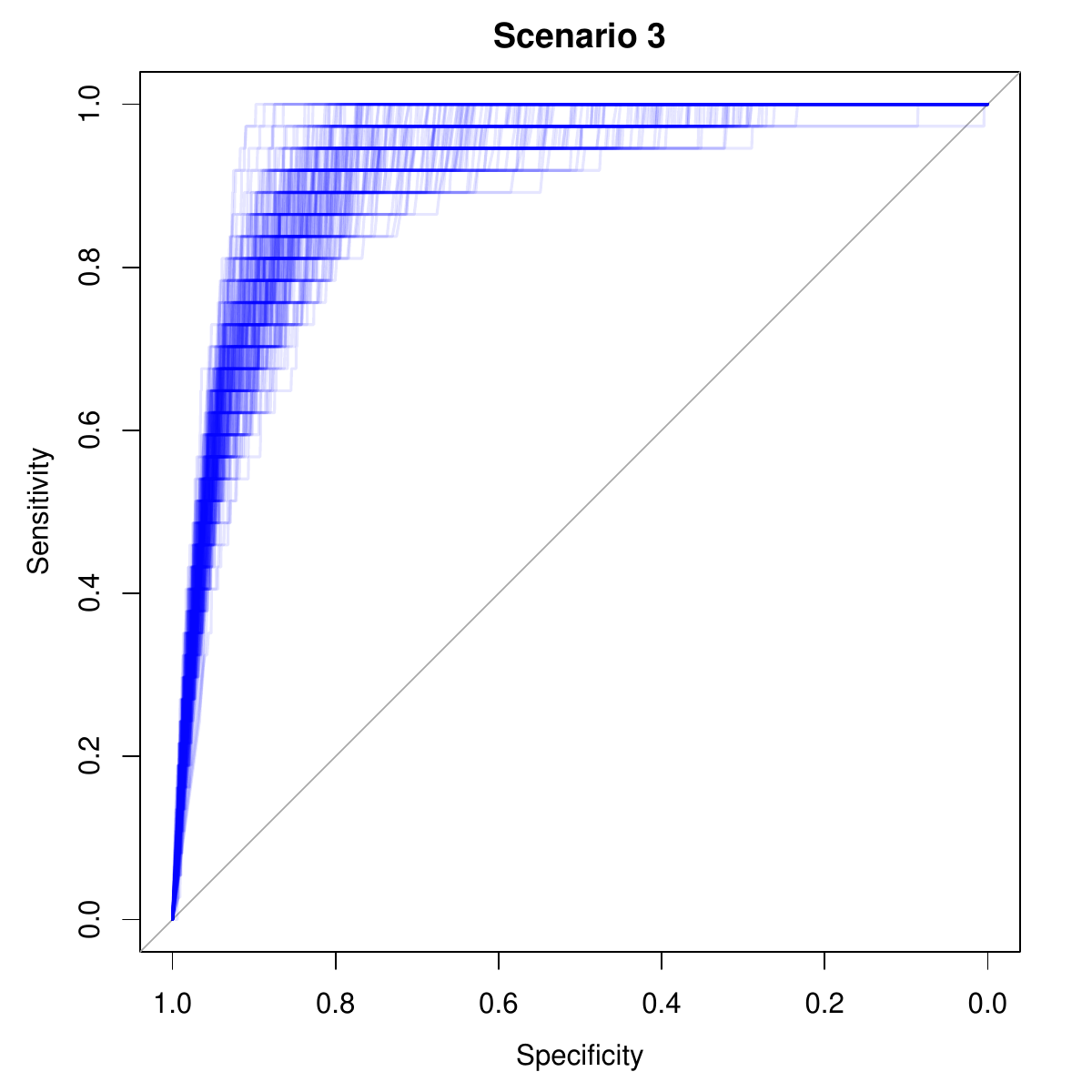}
		
	\end{figure}
	
	\begin{figure}[H]\ContinuedFloat
		\includegraphics*[width=0.33\textwidth]{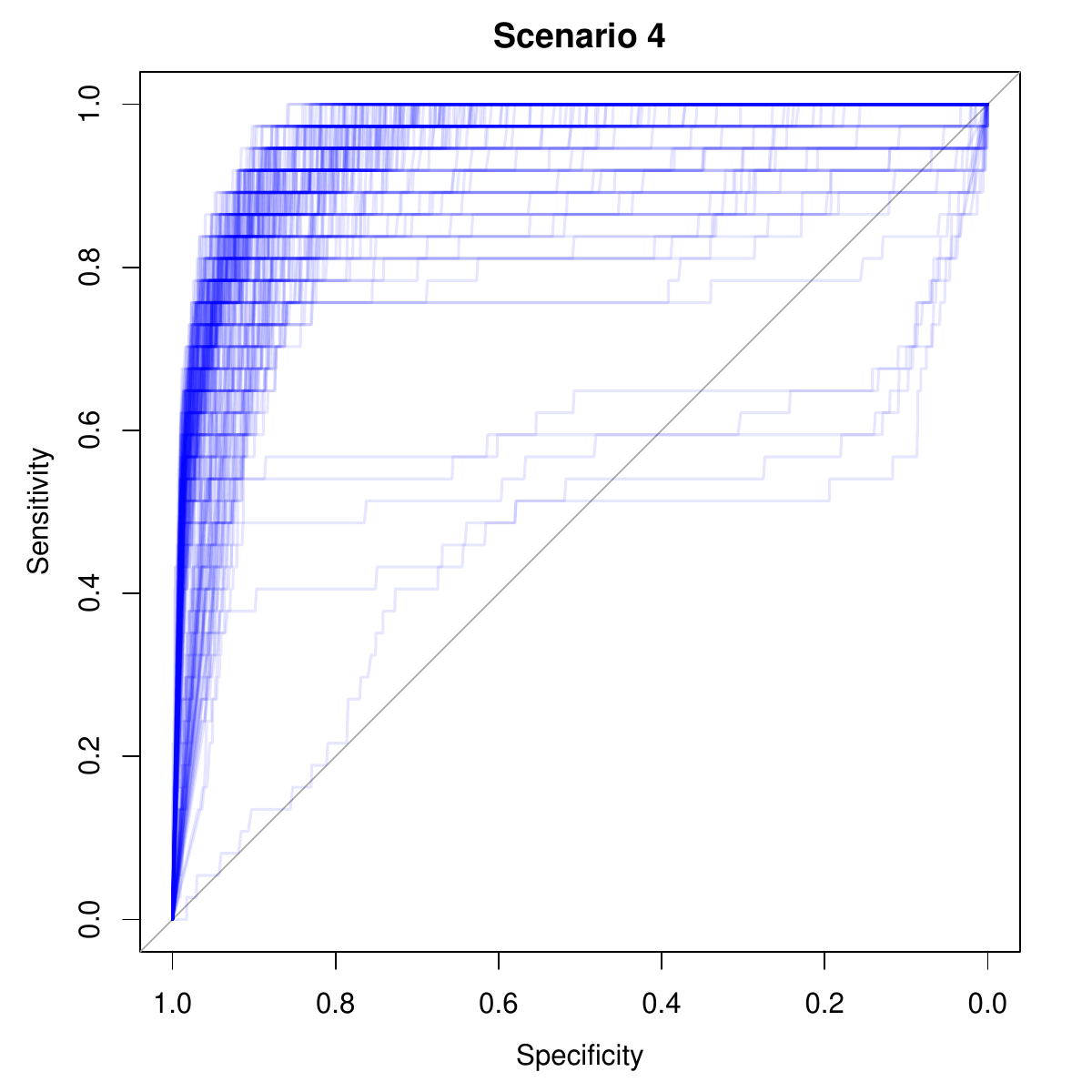}
		\includegraphics*[width=0.33\textwidth]{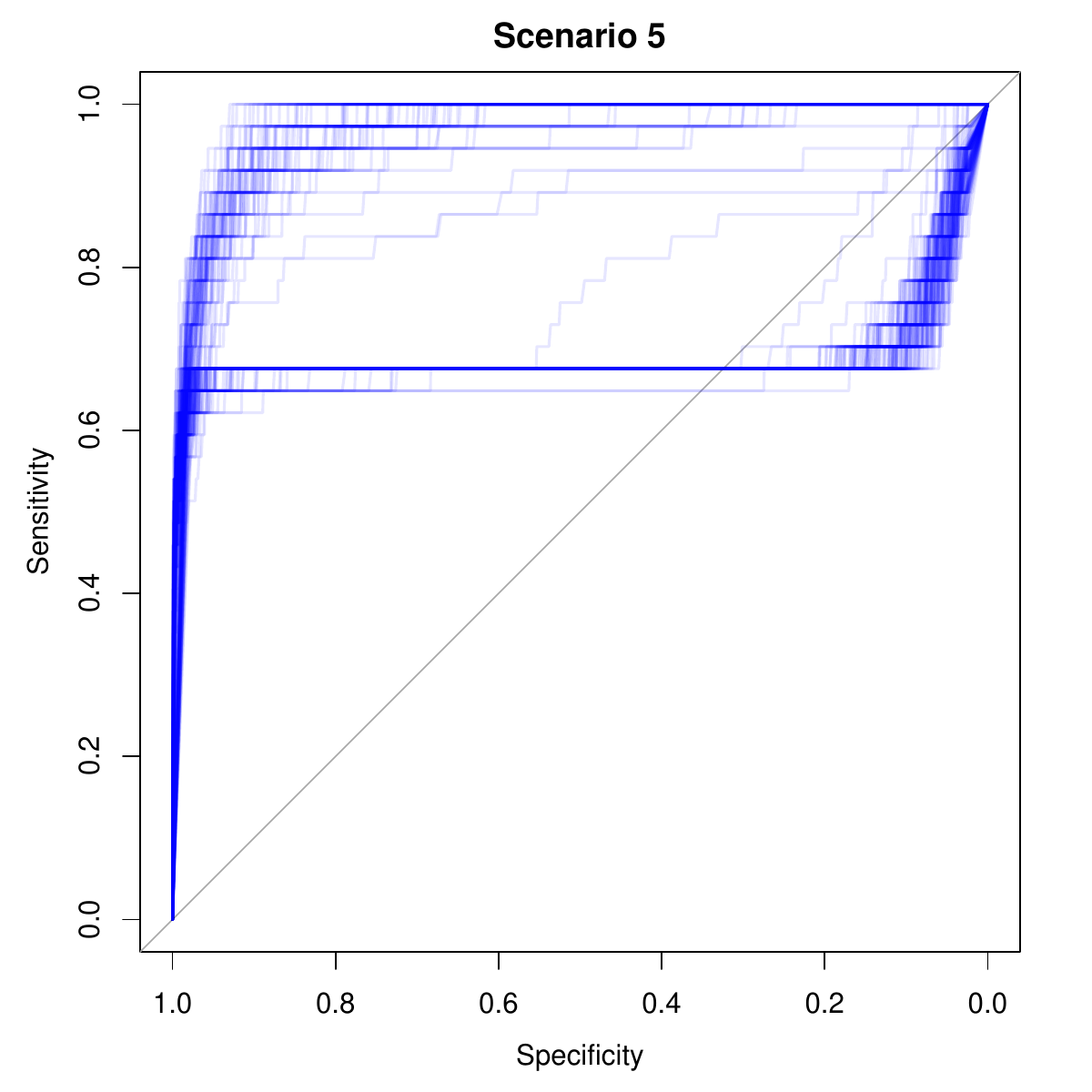}
		\caption{ROC curves of 200 simulated datasets for five simulation scenarios}
		\label{ROC}
	\end{figure}

}
\bibliographystyle{rss}
\bibliography{ref2}